\documentclass[fleqn,usenatbib]{mnras}

\usepackage{amsfonts}

\usepackage[T1]{fontenc}
\usepackage{ae,aecompl}
\usepackage[position=top]{subfig}
\usepackage{float}
\usepackage{comment}
\usepackage{xcolor}
\usepackage{upgreek}

\DeclareRobustCommand{\VAN}[3]{#2}
\let\VANthebibliography\thebibliography
\def\thebibliography{\DeclareRobustCommand{\VAN}[3]{##3}\VANthebibliography}

\usepackage{graphicx}	
\usepackage{amsmath}	
\usepackage{amssymb}	
\usepackage{newtxtext,newtxmath}
\usepackage{pifont}
\makeatletter 
  \patchcmd{\NAT@citex}
    {\@citea\NAT@hyper@{%
      \NAT@nmfmt{\NAT@nm}%
      \hyper@natlinkbreak{\NAT@aysep\NAT@spacechar}{\@citeb\@extra@b@citeb}%
      \NAT@date}}
    {\@citea\NAT@nmfmt{\NAT@nm}%
    \NAT@aysep\NAT@spacechar\NAT@hyper@{\NAT@date}}{}{}

  \patchcmd{\NAT@citex}
    {\@citea\NAT@hyper@{%
      \NAT@nmfmt{\NAT@nm}%
      \hyper@natlinkbreak{\NAT@spacechar\NAT@@open\if*#1*\else#1\NAT@spacechar\fi}%
        {\@citeb\@extra@b@citeb}%
      \NAT@date}}
    {\@citea\NAT@nmfmt{\NAT@nm}%
    \NAT@spacechar\NAT@@open\if*#1*\else#1\NAT@spacechar\fi\NAT@hyper@{\NAT@date}}
    {}{}
\makeatother

\hypersetup{allcolors=blue}

\usepackage{tikz}
\usetikzlibrary{shapes.geometric, arrows}
\tikzstyle{startstop} = [rectangle, rounded corners, minimum width=1cm, minimum height=0.5cm,text centered, draw=black, fill=red!30]
\tikzstyle{io} = [rectangle, minimum width=1cm, minimum height=0.5cm, text centered, draw=black, fill=blue!30]
\tikzstyle{process} = [rectangle, minimum width=1cm, minimum height=0.5cm, text centered, draw=black, fill=orange!30]
\tikzstyle{case} = [rectangle, minimum width=1cm, minimum height=0.5cm, text centered, draw=black]
\tikzstyle{quantity} = [rectangle, minimum width=1cm, minimum height=0.5cm, text centered, draw=black, fill=cyan!30]
\tikzstyle{decision} = [rectangle, minimum width=1cm, minimum height=0.5cm, text centered, draw=black, fill=green!30]
\tikzstyle{arrow} = [thick,->,>=stealth]
\tikzstyle{narrow} = [thick,-,>=stealth]
\tikzstyle{darrow} = [thick,<->,>=stealth]

\title[Mergers of Pop~III BBHs]{Gravitational waves from mergers of Population~III binary black holes: roles played by two evolution channels}
\author[B. Liu et al.]{Boyuan Liu\textsuperscript{\href{https://orcid.org/0000-0002-4966-7450}{\includegraphics[width=2.5mm]{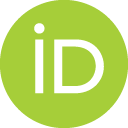}}\,}\thanks{E-mail: treibeis1995@gmail.com}$^{1,5,12}$, 
Tilman Hartwig\textsuperscript{\href{https://orcid.org/0000-0001-6742-8843}{\includegraphics[width=2.5mm]{orcid.png}}\,}$^{2,3,4}$, Nina S. Sartorio\textsuperscript{\href{https://orcid.org/0000-0003-2138-5192}{\includegraphics[width=2.5mm]{orcid.png}}\,}$^{1,6}$, Irina Dvorkin\textsuperscript{\href{https://orcid.org/0000-0002-2353-9194}{\includegraphics[width=2.5mm]{orcid.png}}\,}$^{7}$, 
Guglielmo Costa\textsuperscript{\href{https://orcid.org/0000-0002-6213-6988}{\includegraphics[width=2.5mm]{orcid.png}}\,}$^{8}$,  
\newauthor
Filippo Santoliquido\textsuperscript{\href{https://orcid.org/0000-0003-3752-1400}{\includegraphics[width=2.5mm]{orcid.png}}\,}$^{9,10}$, 
Anastasia Fialkov$^{1,11}$, 
Ralf S. Klessen\textsuperscript{\href{https://orcid.org/0000-0002-0560-3172}{\includegraphics[width=2.5mm]{orcid.png}}\,}$^{12,13}$, 
and 
Volker Bromm\textsuperscript{\href{https://orcid.org/0000-0003-0212-2979}{\includegraphics[width=2.5mm]{orcid.png}}\,}$^{5,14}$
\\
$^{1}$Institute of Astronomy, University of Cambridge, Madingley Road, Cambridge, CB3 0HA, UK\\
$^{2}$Institute for Physics of Intelligence, School of Science, The University of Tokyo, Bunkyo, Tokyo 113-0033, Japan\\ 
$^{3}$Department of Physics, School of Science, The University of Tokyo, Bunkyo, Tokyo 113-0033, Japan\\
$^{4}$AI-Lab, German Environment Agency, Alte Messe 6, 04103 Leipzig, Germany\\
$^{5}$Department of Astronomy, University of Texas, Austin, TX 78712, USA\\
$^{6}$Department of Physics and Astronomy, University of Ghent, Technologiepark 903, Ghent, 9052 Zwijnaarde, Belgium\\
$^{7}$Institut d’Astrophysique de Paris, Sorbonne Université \& CNRS, UMR 7095, 98 bis bd Arago, F-75014 Paris, France\\
$^{8}$Univ Lyon, Univ Lyon1, ENS de Lyon, CNRS, Centre de Recherche Astrophysique de Lyon UMR5574, F-69230 Saint-Genis-Laval, France\\
$^{9}$Gran Sasso Science Institute (GSSI), 67100 L’Aquila, Italy\\
$^{10}$INFN, Laboratori Nazionali del Gran Sasso, 67100 Assergi, Italy\\
$^{11}$Kavli Institute for Cosmology, Madingley Road, Cambridge, CB3 0HA, UK\\
$^{12}$Universit\"at Heidelberg, Zentrum fur Astronomie, Institut f\"ur Theoretische Astrophysik, D-69120 Heidelberg, Germany\\
$^{13}$Universit\"at Heidelberg, Interdiszipli\"ares Zentrum f\"ur Wissenschaftliches Rechnen, D-69120 Heidelberg, Germany\\
$^{14}$Weinberg Institute for Theoretical Physics, University of Texas, Austin, TX 78712, USA
}

\date{Accepted XXX. Received YYY; in original form ZZZ}

\pubyear{2024}

\begin{document}
\label{firstpage}
\pagerange{\pageref{firstpage}--\pageref{lastpage}}
\maketitle

\defcitealias{Sana2012}{S12}
\defcitealias{Stacy2013}{SB13}

\begin{abstract}
The gravitational wave (GW) signal from binary black hole (BBH) mergers is a promising probe of Population~III (Pop~III) stars. 
To fully unleash the power of the GW probe, one important step is to understand the relative importance and features of different BBH evolution channels. We model two channels, isolated binary stellar evolution (IBSE) and nuclear star cluster-dynamical hardening (NSC-DH), in one theoretical framework based on the semi-analytical code \textsc{a-sloth}, under various assumptions on Pop~III initial mass function (IMF), initial binary statistics and high-$z$ nuclear star clusters (NSCs). 
The NSC-DH channel contributes $\sim 8-95\%$ of Pop~III BBH mergers across cosmic history, with higher contributions achieved by initially wider binary stars, more top-heavy IMFs, and more abundant high-$z$ NSCs.  
The dimensionless stochastic GW background (SGWB) produced by Pop~III BBH mergers has peak values $\Omega^{\rm peak}_{\rm GW}\sim 10^{-11}-8\times 10^{-11}$ around observer-frame frequencies $\nu\sim 10-100\ \rm Hz$. The Pop~III contribution can be a non-negligible ($\sim 2-32\%$) component in the total SGWB at $\nu\lesssim 10\ \rm Hz$. 
The estimated detection rates of Pop~III BBH mergers by the Einstein Telescope are $\sim 6-230\ \rm yr^{-1}$ and $\sim 30-1230\ \rm yr^{-1}$ for the NSC-DH and IBSE channels, respectively. Pop~III BBH mergers in NSCs are more massive than those from IBSE, so they dominate the Pop~III SGWB below $20$ Hz in most cases. Besides, the detection rate of Pop~III BBH mergers involving at least one intermediate-mass BH above $100\ \rm M_\odot$ by the Einstein Telescope is $\sim 0.5-200\ \rm yr^{-1}$ in NSCs but remains below $0.1\ \rm yr^{-1}$ for IBSE. 

\end{abstract}




\section{Introduction}
\label{s1}

One of the major goals of modern astrophysics is to understand the onset of star and galaxy formation at Cosmic Dawn, during the first billion years after the Big Bang \citep{loeb2013}. In particular, the first generation of so-called Population~III (Pop~III) stars, formed in extremely metal-poor primordial gas via cooling by molecular hydrogen, are believed to have distinct features compared with present-day, Population~I/II (Pop~I/II), stars \citep[reviewed by e.g.,][]{Bromm2009,Bromm2013,Haemmerle2020,Klessen2023}. 

The first luminous objects play important roles in early cosmic history through their metal enrichment, radiation fields and cosmic ray production \citep[e.g.,][]{Pan2013,Salvador-Sole2017,Jaacks2018,Ohira2019,Yamaguchi2023,Sartorio2023,Gessey-Jones2023}, significantly impacting the chemical and thermal evolution of the intergalactic medium (IGM; see, e.g., \citealt{Karlsson2013,Barkana2016,Dayal2018} for reviews). This early feedback also establishes the conditions for the formation and evolution of subsequent populations of stars, galaxies and (supermassive) black hole (BH) seeds \citep[see, e.g.,][]{Bromm2011,Johnson2013,Pawlik2013,Jeon2015,Smith2015,Regan2017,Sakurai2017,Haemmerle2020,Inayoshi2020,Schauer2021,Chon2022,Sarmento2022,Chiaki2023,Sanati2023,Regan2024}, providing powerful diagnostics for early structure formation and even fundamental physics, such as the nature of dark matter \citep[e.g.,][]{Hirano2018,Hirano2018fuzzy,Sullivan2018,Liu2019,Liu2019bdms,Nebrin2019,Ilie2021,Cappelluti2022,Driskell2022,Liu2022,Liu2022apj,Liu2023pbh,Hibbard2022,Kulkarni2022,Hirano2024,Zhang2024,Zhang2024rad,Zhang2024ds}.

Although the importance of Pop~III stars is evident in theoretical predictions, their detailed properties are still unclear in the absence of direct observations, which are still challenging (especially for pure Pop~III systems in minihaloes at $z\gtrsim 15$) even with the James Webb Space Telescope (JWST) \citep{Gardner2006,Riaz2022,Katz2023,Nakajima2022,Bovill2024} and calls for extremely large ($\gtrsim 100\ \rm m$) extraterrestrial telescopes \citep{Angel2008,Rhodes2020,Schauer2020}. Detection of Pop~III features\footnote{See \citet{Windhorst2018}, \citet{Nakajima2022}, \citet{Trussler2023}, \citet{Katz2023}, \citet{Larkin2023}, \citet{Zackrisson2023}, and \citet{Venditti2024heii} for detailed discussions of the observability and characterization of Pop~III stars with JWST.} is possible in rare cases of magnification via extreme gravitational lensing \citep{Schauer2022,Welch2022,Vanzella2023,Zackrisson2023} or trace Pop~III populations\footnote{For instance, recently \citet{Maiolino2023} discover a broad ($\sim170$~\AA) HeII$\lambda$1640 emission feature from NIRSpec-IFU and NIRSpec-MSA observations of GN-z11 at $z=10.6$ that can be explained by a Pop~III starburst of $\sim 6\times 10^{5}\ \rm M_{\odot}$ \citep[for an interpretation, see, e.g.,][]{Venditti2023,Venditti2024heii}.} co-existing with Pop I/II stars and active galactic nuclei (AGN) at $z\lesssim 10$ \citep[e.g.,][]{Grisdale2021,Nanayakkara2022,Wang2024,Maiolino2023}. 
Another complementary approach to constrain Pop~III stars, in particular their initial mass function (IMF), with future telescopes, e.g., Roman Space Telescope, is to consider electro-magnetic transients, such as gamma-ray bursts and supernova (SN) explosions \citep[e.g.,][]{Fryer2022,Lazar2022,Hartwig2023,Venditti2024,Wiggins2024}. 
Currently, observational constraints on Pop~III stars are mainly derived from two indirect probes: stellar archaeology and the 21-cm signal\footnote{In general, intensity mapping of other lines such as HeII$\lambda$1640 and H$\alpha$ can also constrain early star formation \citep[e.g.,][]{Parsons2022}.} from neutral hydrogen at Cosmic Dawn. 

Stellar archaeology connects the statistics and chemical patterns of extremely metal-poor stars observed in the local Universe with the IMF and SN properties of Pop~III stars \citep[reviewed by, e.g.,][]{Frebel2015}. Such extremely metal-poor stars are expected to be the bona-fide second-generation (Pop~II) stars that preserve the chemical imprints of Pop~III metal enrichment \cite[such as carbon enhancement, see, e.g.,][]{Yoon2016,Yoon2018,Hansen2019,Dietz2021,Zepeda2023}. Recent progress
in this direction \citep[e.g.,][]{Ji2015,Hartwig2015,Salvadori2015,Salvadori2019,Sarmento2017,Sharma2018,Debennassuti2017,jeon2017,Jeon2021,Ishigaki2018,Magg2018,Magg2019,Chiaki2020,Komiya2020,Yuta2020,Rossi2021,Rossi2024,Liu2021wind,Lucey2022,Hartwig2023,Koutsouridou2023,Skuladottir2024} finds that although there is evidence for diverse explosion mechanisms of Pop~III SNe including scenarios of jet-induced hypernovae, `faint' SNe with fallback and mixing, as well as pair-instability SNe (PISNe) \citep[e.g.,][]{Heger2002,Heger2003,Maeda2003,Umeda2003,Umeda2005,Iwamoto2005,Kobayashi2006,Tominaga2009,Heger2010}, the chemical patterns of extremely metal-poor stars are mostly consistent with the enrichment by core-collapse SNe from Pop~III stars with initial masses $m_{\star}\sim 20-40\ \rm M_{\odot}$, and more massive stars 
are rather rare or collapse directly into BHs at the end of metal-free stellar evolution \citep[e.g.,][]{Heger2003,Fryer2012,Tanikawa2020}, which potentially contribute to metal enrichment via strong post main sequence (MS) winds \citep{Liu2021wind,Jeena2023,Nandal2024,Nandal2024vms,Nandal2024rotat,Tsiatsiou2024}. 
Similarly, the imprints of Pop~III metal enrichment can also be inferred from observations of high-$z$ extremely metal-poor absorption systems in the spectra of distant quasars \citep[e.g.,][]{Welsh2019,Welsh2022,Welsh2023,Yoshii2022,Saccardi2023,Salvadori2023,Vanni2024,Sodini2024,Zou2024} and metal-poor galaxies with strong metal emission lines \citep{D'Eugenio2023,Cameron2023,Senchyna2023,Ji2024jwst,Schaerer2024,Topping2024}. In general, stellar archaeology confirms the massive nature of Pop~III stars (with a broad, top-heavy IMF) as well as their formation in small clusters predicted by semi-analytical models and (magneto)hydrodynamic simulations of primordial star formation \citep[e.g.,][]{Greif2011,Greif2012,Susa2014,Hirano2014,Hirano2015,Hirano2018sc,Stacy2013,Stacy2016,Stacy2022,Hirano2017,Susa2019,Sugimura2020,Sugimura2023,McKee2020,Wollenberg2020,Park2021,Park2022,Park2024,Chon2021,Sharda2020,Sharda2021,Sharda2022,Latif2021,Latif2022,Riaz2018,Riaz2023,Prole2022res,Prole2022,Prole2023,Liu2024sms,Sharda2024,Sadanari2024}.


The 21-cm signal from the high-$z$ IGM is shaped by the UV and X-ray photons as well as cosmic rays from the first stars, SNe and galaxies \citep[reviewed by e.g.,][]{Barkana2016}, from which the timing, efficiency and mode of early star formation, as well as source properties (e.g., Pop~III IMF), can be constrained \citep[e.g.,][]{Fialkov2013,Fialkov2014rich,Fialkov2017,Ma2018,Madau2018,Mirocha2019,Schauer2019,Chatterjee2020,Qin2020,Gessey-Jones2022,Gessey-Jones2023,Kamran2022,Kaur2022,Kovlakas2022,Magg2022tr,Munoz2022,Bera2023,Bevins2024,Fialkov2023,Hassan2023,Ma202321,Mondal2023,Pochinda2023,Ventura2023,Lewis2024}. 
For instance, the average star formation efficiency 
of Pop~III stars is constrained to be below 5.5\% at 68\% confidence by the 21-cm data from the Hydrogen Epoch of Reionization Array and Shaped Antenna measurement of the background RAdio Spectrum \citep{Pochinda2023}. 

Another promising probe of the first stars that has just become possible thanks to the LIGO-Virgo-KAGRA (LVK) network \citep{LIGOScientificCollaboration2015,Acernese2015,KagraCollaboration2019} 
is provided by gravitational waves (GWs) from compact object mergers (see, e.g., \citealp{Mapelli2021,Mandel2022,Spera2022} for recent reviews). So far, the LVK network has reported about one hundred events of mergers between stellar-mass compact objects (mostly BHs with $\sim 5- 100\ \rm M_\odot$) within $z\lesssim 1$ in the third gravitational wave transient catalogue \citep[GWTC-3,][]{Abbott2023gwtc3}. Moreover, the 3rd-generation GW detectors such as the Einstein Telescope \citep[ET,][]{Punturo2010,Maggiore2020} and the {Cosmic Explorer Observatory \citep[CEO,][]{Reitze2019,Evans2023}} will observe thousands of compact object mergers per year up to $z\sim 30$. The mass, spin, and redshift distributions of these mergers, especially at high redshifts, can place novel constraints on Pop~III star formation and stellar evolution. 
Indeed, compared with Pop~I/II stars, Pop~III stars are more massive, compact, and experience little mass loss in the absence of metal-line-driven winds, so they are more likely to produce (massive) compact object remnants. Besides, recent hydrodynamic simulations converge on the picture that Pop~III stars are typically born in small clusters \citep{Haemmerle2020,Hartwig2023,Klessen2023} in which the formation of binary and multiple systems are common. Therefore, Pop~III stars are expected to be efficient progenitors of compact object binaries, in particular binary black holes (BBHs). Recent theoretical studies \citep[e.g.,][]{Kinugawa2020,Kinugawa2021,Hijikawa2021,Liu2020gw,Liu2021,Tanikawa2022} reveal the possibility that mergers of Pop~III BBH remnants are abundant and energetic GW sources, contributing a significant fraction ($\sim 0.1-10$\%) of BBH mergers across cosmic history, especially for massive BHs at high-$z$, much higher than the Pop~III fraction in the total mass budget of stars ever formed in the Universe ($\lesssim 10^{-4}$). A large fraction of the currently observed BBH merger events may originate from Pop~III stars \citep[e.g.,][]{Kinugawa2020,Kinugawa2021,Iwaya2023}, which can explain the high-mass regime ($\gtrsim 20\ \rm M_\odot$) of the chirp mass distribution for the detected events \citep{Abbott2021,Abbott2023gwtc3}. 

For instance, the recently detected event GW190521 at $z=0.82_{-0.34}^{+0.28}$ with unusual BH masses $85_{-14}^{+21}\ \rm M_{\odot}$ and $66_{-18}^{+17}\ \rm M_\odot$ (inferred with the NRSur7dq4 waveform model by \citealt{abbott2020gw190521}, see also \citealt{Abbott2020,O'Brien2021,Mehta2022}) 
is suspected to have a Pop~III origin\footnote{The origin and properties of GW190521 are still in debate \citep[see, e.g.,][]{Fishbach2020,Gayathri2020,Romero-Shaw2020}. Beyond the Pop~III scenario, other scenarios, such as IBSE of Pop~I/II stars with special stellar evolution models, hierarchical BH mergers and stellar collisions in star clusters and discs of AGN, {can also explain this event \citep[e.g.,][]{Belczynski2020,Fragione2020gw,DiCarlo2020,Kremer2020,Renzo2020sc,Arca-Sedda2021,Costa2021,Costa2022,Gayathri2021,Gerosa2021,Kimball2021,Liu2021gw,Mapelli2021hierarchical,Tagawa2021,Anagnostou2022,Antonini2024}}.}. GW190521 is special because the mass range $50-130\ \rm M_{\odot}$ is mostly forbidden for BHs born by Pop~I/II stars according to the standard PISN models unless under peculiar conditions (e.g., \citealt{Heger2003,Belczynski2016,Yoshida2016,Woosley2017,Woosley2019,Spera2017,Farmer2019,Farmer2020,Leung2019,Marchant2019,Marchant2020,Mapelli2020rotation,Renzo2020,Costa2021,Vink2021,Winch2024}), which is called the (standard) PISN mass gap. The lower edge of the mass gap can be shifted up to $\sim 100\ \rm M_\odot$ for Pop~III stars with their unique evolution tracks, 
so Pop~III stars can produce BBH mergers like GW190521 at a similar rate as observed under proper conditions \citep{Farrell2020,Kinugawa2021gw,Liu2020,Liu2021,Tanikawa2021,Tanikawa2022,Volpato2023,Tanikawa2024}. 

However, there are significant discrepancies in current theoretical predictions for Pop~III BBH mergers, making it difficult to derive accurate constraints on Pop~III stars from current GW data. 
The reason is that different studies consider different BBH formation and evolution channels and adopt different assumptions on a variety of physical processes over a broad range of scales. The most intensively studied channel to date is isolated binary stellar evolution (IBSE) of close binary stars \citep[with initial separations $a\lesssim 10\ \rm AU$, see, e.g.,][]{Kinugawa2014,Kinugawa2016,Kinugawa2017,Kinugawa2020,Hartwig2016,Belczynski2017,Inayoshi2017,Hijikawa2021,Tanikawa2021mrd,Tanikawa2022,Costa2023,Santoliquido2023}, which relies on binary interactions (e.g., mass transfer, tidal effects and common envelope evolution) and SN natal kicks to shrink the binary orbits and facilitate merging. The predicted merger efficiencies and properties are highly sensitive to the uncertain initial binary properties of Pop~III stars \citep[see, e.g.,][]{Stacy2013,Liu2021binary} and parameters of (binary) stellar evolution \citep[for broad discussions including Pop~I/II stars, see, e.g.,][]{deMink2015,Kinugawa2020,Bavera2021,Olejak2021,Marchant2021,Santoliquido2021,Santoliquido2023,Belczynski2022,Tanikawa2022,Stevenson2022,Willcox2023,Iorio2023,Dorozsmai2024}. Besides, it is implied by recent hydrodynamic simulations of Pop~III star formation \citep[e.g.,][]{Sugimura2020,Sugimura2023,Park2022,Park2024} that most binaries of Pop III stars have wide orbits ($a\gtrsim 100\ \rm AU$, due to outward migrations of Pop~III protostars and their circumstellar discs by accretion of gas with high angular momentum), potentially hampering the efficiency of the IBSE channel \citep{Liu2021binary,Costa2023,Santoliquido2023}. 

In light of this, \citet{Liu2020} and \citet{Liu2021} propose an alternative channel based on dynamical hardening (DH) in nuclear star clusters (NSCs). In this scenario, Pop~III BBHs fall into NSCs (made of Pop~I/II stars) by dynamical friction, in which the orbits of hard binaries are shrunk by {binary-single} encounters, so BBHs from initially wide binary stars can also merge within a Hubble time. The NSC-DH channel is expected to be less sensitive to initial binary properties and binary stellar evolution processes and can be as efficient as the IBSE channel under favourable conditions. 
Nevertheless, Pop~III BBH mergers from the NSC-DH channel are highly sensitive to the unknown properties of high-$z$ dwarf galaxies and their NSCs. {Another dynamical channel (in-situ) is given by $N$-body dynamics within massive ($\sim 10^{4}-10^{7}\ \rm M_{\odot}$) clusters of Pop~III stars and remnants themselves \citep[][]{Wang2022,Liu2023sc,Mestichelli2024}, which can also produce non-negligible merger rates. In particular, it is shown in \citet{Mestichelli2024} that massive BBH mergers involving intermediate-mass BHs (IMBHs, $\gtrsim 100\ \rm M_{\odot}$) are common in massive Pop~III clusters, while such mergers can be extremely rare from isolated evolution \citep{Costa2023,Santoliquido2023}.} However, massive Pop~III star clusters (SCs) are only expected to form under special conditions \citep[e.g., strong dynamical heating, streaming motion, and Lyman-Werner radiation,][]{Li2021,Lupi2021}, and their abundance and detailed properties are still uncertain\footnote{In our simulations, most Pop~III stars form in small clusters with total stellar masses of $\sim 10^{3}\ \rm M_{\odot}$, and the contribution of Pop~III stars in massive ($\sim 10^{4}-10^{5}\ \rm M_{\odot}$) clusters to the total number of Pop~III stars ever formed is very small ($\lesssim 3\%$). Therefore, we do not consider the pure Pop~III (in-situ) dynamical channel in this work.}. 

To 
fully unleash the power of the GW probe, deeper and more systematic investigations of the relevant physics of BBH formation and evolution are required to establish robust connections between properties of the first stars (e.g., IMF, binary statistics at birth, and star formation history) and GW observations. One important step towards this ultimate goal is \textit{to understand the relative importance of different BBH formation/evolution channels as well as the corresponding features of GW sources}. To do so, we implement the IBSE and NSC-DH channels in \textit{one} theoretical framework based on the public semi-analytical model \textsc{a-sloth}\footnote{\url{https://gitlab.com/thartwig/asloth}} \citep[Ancient Stars and Local Observables by Tracing haloes,][]{Hartwig2022,Magg2022}. 
Given the input of binary population synthesis (BPS) data, our framework for the first time follows the (external) dynamics (in host NSCs/galaxies/haloes and large-scale structure formation) as well as the (internal) orbital evolution of Pop~III BBHs (in NSCs) on-the-fly, together with the underlying star/galaxy/structure formation process. This enables us to model different BBH evolution channels self-consistently and characterize the resulting Pop~III BBH mergers and their host systems. 
In this paper, we explore the key properties of Pop~III BBH mergers 
from the NSC-DH and IBSE channels in 18 simulations combining different assumptions on the yet uncertain (initial) properties of Pop~III binary stars and high-$z$ NSCs, based on the BPS results produced by the \textsc{sevn} code \citep{Costa2023} and the merger trees from the cosmological simulation in \citep{Ishiyama2016}. 
We plan to investigate the properties of host haloes/galaxies/NSCs of Pop~III BBH mergers, the rate and spatial distribution of mergers in the host galaxy, and the dependence of merger properties on star formation, (binary) stellar evolution, and feedback parameters in future work \citep[see, e.g.,][]{Pacucci2017,Artale2019,Santoliquido2021,Santoliquido2022,Santoliquido2023,Iorio2023,Rauf2023,Srinivasan2023}.

\begin{table}
    \caption{List of key abbreviations.}
    \begin{tabular}{l|l}
    \hline
        Abbreviation & Full name \\
    \hline
        GW & Gravitational wave\\
        BH & Black hole\\
        BBH & Binary black hole\\
        SN & Supernova\\
        PISN & Pair-instability supernova\\
        MS & Main sequence\\
        ZAMS & Zero-age main sequence\\
        IMF & Initial mass function\\
        BPS & Binary population synthesis\\
        IBS & Initial binary statistics\\
        IBSE & Isolated binary stellar evolution\\
        NSC & Nuclear star cluster\\
        SC & Star cluster\\
        GC & Globular cluster\\
        DF & Dynamical friction\\
        DH & Dynamical hardening\\
        CE & Common envelope \\
        SMT & Stable mass transfer\\
        SFRD & Star formation rate density\\
        MRD ($\dot{n}$) & Merger rate density (co-moving)\\
        SGWB ($\Omega_{\rm GW}$) & Stochastic gravitational wave background\\
        SNR & Signal-to-noise ratio\\
        GWTC & Gravitational wave transient catalogue\\
        ET & The Einstein Telescope\\
        {CEO} & {The Cosmic Explorer Observatory}\\
        LVK & LIGO-Virgo-KAGRA \\
        JWST & James Webb Space Telescope\\
        AGN & Active galactic nuclei/nucleus \\
    \hline
    \end{tabular}
    \label{tab:abb}
\end{table}

\textsc{a-sloth} was calibrated based on six independent observational constraints. It has been refined and applied to a variety of topics, including GWs from mergers of Pop~III remnants \citep[but treating the IBSE and NSC-DH channels separately]{Hartwig2016,Liu2021}, stellar archaeology constraints on the Pop~III IMF \citep{Hartwig2015,Magg2018,Yuta2020} and local baryonic streaming velocity \citep{Uysal2023}, hypernova signatures of Pop~III stars in nearby dwarf satellite galaxies \citep{Lee2024}, impact of the transition from Pop~III to Pop~II star formation on the global 21-cm signal \citep{Magg2022tr}, rates of Pop~III SNe \citep{Magg2016}, formation of Milky Way (MW) satellites \citep{Chen2022mw,Chen2022}, Pop~III signatures and evolution of dust in high-$z$ galaxies \citep{Riaz2022,Tsuna2023,Dzieciol2024}, and the imprint of Pop~III stars on the faint end of the MW white dwarf luminosity function. This broad range of applications implies that our model can be easily used to explore the correlations between the GW signals from Pop~III (and Pop~I/II) BBH mergers and other observational properties of the first stars and galaxies \citep[see, e.g.,][]{Dvorkin2016,Inayoshi2016,Santoliquido2022,Tanikawa2022met,Veronesi2024}, although this is not the focus of this paper. Considering such broader applications, we make our code public\footnote{\url{https://gitlab.com/Treibeis/a-sloth-cob} }.

This paper is organized as follows. In Sec.~\ref{sec:binary}-\ref{sec:extra} we describe our implementation of the relevant physics in \textsc{a-sloth}, covering the formation of Pop~III BBHs (Sec.~\ref{sec:sample}), their subsequent evolution in different environments (Sec.~\ref{sec:evo}), the formation and evolution of NSCs (Sec.~\ref{sec:nsc}), 
and an extension of \textsc{a-sloth} to follow the evolution of Pop~III BBHs and their host galaxies/haloes in late epoch ($z\lesssim 4.5$) that is not covered by merger trees from cosmological simulations (Sec.~\ref{sec:extra}). In Sec.~\ref{sec:res}, we present the predicted properties of Pop~III BBH mergers (e.g., merger rate, contribution to the stochastic GW background, mass distribution, and simple demography of host systems) 
from the IBSE and NSC-DH channels, starting with the fiducial model (Sec.~\ref{sec:fid}), and then exploring the dependence on Pop~III IMF, initial binary statistics (IBS), and abundances of (high-$z$) NSCs (Sec.~\ref{sec:comp}). Finally, in Sec.~\ref{sec:dis}, we summarize our results and discuss their implications for the GW signatures of the first stars. A list of the key abbreviations used in this paper is given in Table~\ref{tab:abb}.




\section{Pop~III binary stars and BBHs}
\label{sec:binary}
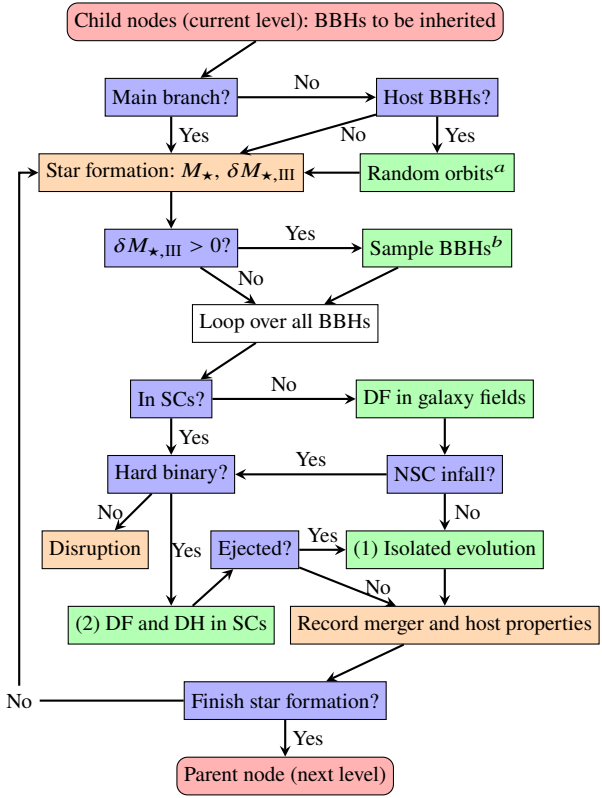
\begin{figure}
    \centering
\begin{tikzpicture}[node distance=1cm]
\node (cnode) [startstop] {Child nodes (current level): BBHs to be inherited};
\node (pre) [io, below of=cnode,xshift=-1.5cm] {Main branch?};
\node (asloth) [process, below of=pre, xshift=0cm] {Star formation: $M_{\star}$, $\delta M_{\star,\rm III}$};
\node (pop3sf) [io, below of=asloth,xshift=-0cm] {$\delta M_{\star,\rm III}>0$?};
\node (bbh) [io, below of=cnode, xshift=2cm] {Host BBHs?};
\node (orbit) [decision, below of=bbh,xshift=0cm]{Random orbits$^a$};
\node (sample) [decision, below of=orbit,xshift=0cm]{Sample BBHs$^b$};
\node (loop) [case, below of=sample, xshift=-2cm]{Loop over all BBHs};
\node (nsc) [io, below of=loop, xshift=-1.5cm]{In SCs?};
\node (df) [decision, right of=nsc, xshift=2.6cm]{DF in galaxy fields};
\node (hard) [io, below of=nsc]{Hard binary?};
\node (eject) [io, below of=hard,xshift=1.1cm]{Ejected?};
\node (infall) [io, below of=df]{NSC infall?};
\node (ibse) [decision, below of=infall]{{(1) Isolated evolution}};
\node (dh) [decision, below of=eject,xshift=-1.1cm]{{(2) DF and DH in SCs}};
\node (dis) [process, below of=hard, xshift=-1cm]{Disruption};
\node (sf) [io, below of=dh,xshift=1.5cm]{Finish star formation?};
\node (record) [process, below of=ibse]{Record merger and host properties};
\node (mid0) [left of=sf,xshift=-2.5cm]{No};
\node (pnode) [startstop, below of=sf]{Parent node (next level)};
\draw [arrow] (cnode) -- (pre);
\draw [arrow] (pre) -- node[xshift=0.3cm]{Yes} (asloth);
\draw [arrow] (pre) -- node[yshift=0.2cm]{No} (bbh);
\draw [arrow] (bbh) -- node[xshift=0.3cm]{Yes} (orbit);
\draw [arrow] (bbh) -- node[xshift=0.6cm]{No} (asloth);
\draw [arrow] (asloth) -- (pop3sf);
\draw [arrow] (pop3sf) -- node[yshift=0.2cm]{Yes} (sample);
\draw [arrow] (pop3sf) -- node[xshift=0.3cm,yshift=0.1cm]{No} (loop);
\draw [arrow] (sample) -- (loop);
\draw [arrow] (loop) --  (nsc); 
\draw [arrow] (nsc) -- node[yshift=0.2cm]{No} (df);
\draw [arrow] (nsc) -- node[xshift=0.3cm]{Yes} (hard);
\draw [arrow] (hard) -- node[xshift=0.2cm]{Yes} (dh);
\draw [arrow] (hard) -- node[xshift=-0.3cm]{No} (dis);
\draw [arrow] (eject) -- node[yshift=0.2cm]{Yes} (ibse);
\draw [arrow] (eject) -- node[xshift=0.4cm]{No} (record);
\draw [arrow] (dh) -- (eject);
\draw [arrow] (df) -- (infall); 
\draw [arrow] (infall) -- node[yshift=0.2cm]{Yes} (hard);
\draw [arrow] (infall) -- node[xshift=0.3cm]{No} (ibse);
\draw [arrow] (ibse) -- (record);
\draw [arrow] (record) -- (sf);
\draw [arrow] (orbit) -- (asloth);
\draw [narrow] (sf) -- (mid0);
\draw [arrow] (mid0) |- (asloth);
\draw [arrow] (sf) -- node[xshift=0.3cm]{Yes} (pnode);
\end{tikzpicture}
    \caption{Overview of the Pop~III BBH routine (Sec.~\ref{sec:binary}) within \textsc{a-sloth}. 
    Red rounded boxes: interfaces with the merger tree. Blue boxes: decision-making steps. Orange boxes: other (sub-)routines that contain or are called by the BBH routine. Here, the star formation routine of \textsc{a-sloth} predicts the expected mass $\delta M_{\star,\rm III}$ of Pop~III stars formed at each timestep and updates the galaxy mass $M_{\star}$ (total mass of Pop~I/II stars) that is crucial for DF of BBHs in galaxy fields. Green boxes: key processes for BBH formation (Sec.~\ref{sec:sample}) and evolution (Sec.~\ref{sec:evo}) via the (1) IBSE and (2) NSC-DH channels.  
    \\ \small$^a$ The external (galactic) orbits of the BBHs from the smaller halo during a halo merger event are randomized according to Eq.~\ref{pop3ic}.
    \\ \small$^b$ Newly born BBHs are sampled from the input \textsc{sevn} catalogue (see Sec.~\ref{sec:sample}). Their galactic orbits are initialized with Eq.~\ref{pop3ic}. }
    \label{bbh_process}
\end{figure}

We implement the formation, dynamics and internal orbital evolution of Pop~III BBHs in galaxy fields and NSCs, as well as the formation and evolution of NSCs in the public semi-analytical model \textsc{a-sloth} \citep
{Hartwig2022} for cosmic star/structure formation along halo merger trees. 
The reader is referred to \citet{Hartwig2022} and \citet{Magg2022} for a detailed introduction of \textsc{a-sloth}. Our implementation is based on the earlier work by \citet{Liu2021} with updated treatments for the sampling/formation of Pop~III BBHs, formation and evolution of galaxies and NSCs. 
Below we explain our BBH, galaxy and NSC models, focusing on the new features with respect to \citet{Liu2021}. We start with the BBH model in this section covering their formation (Sec.~\ref{sec:sample}) and evolution (Sec.~\ref{sec:evo}) in the context of external (galactic) dynamics (Sec.~\ref{sec:gd}) and internal orbital dynamics (Sec.~\ref{sec:bevo}). 
The flow chart of our BBH routine is shown in Fig.~\ref{bbh_process}. 
The model of NSCs will be described in the next section.

\subsection{Formation of Pop~III BBHs}
\label{sec:sample}
We couple the star formation routine in \textsc{a-sloth} with input catalogues of Pop~III BBHs from BPS simulations of \textsc{sevn} \citep{Costa2023} to sample Pop~III BBHs from Pop~III star formation events. Each simulation follows the evolution of a large number ($\sim 1-2\times 10^{7}$) of Pop~III binary stars randomly sampled from the given IMF (for the primary star) and distributions of initial binary parameters (mass ratio, orbital period and eccentricity), based on the stellar evolution tracks of effectively metal-free stars (with an absolute metallicity of $Z=10^{-11}$) generated by the \textsc{parsec} code \citep{Bressan2012,Costa2021,Costa2022,Nguyen2022}. 
The masses and orbital parameters of the resulting Pop~III BBHs at the moment when both stars become BHs (and the binary remains bound), together with the initial masses, orbital parameters, and lifetimes of their progenitor stellar binaries are recorded in a catalogue. Here, the lifetime $\tau_{\rm B}$ of a BBH progenitor stellar binary is the time taken for both stars to evolve from zero-age main sequence (ZAMS) to a BH. In this way, the simulation also predicts the mass fraction of binaries that become BBHs $f_{\rm BBH}$ and the average initial mass of BBH progenitor stellar binaries $\bar{m}_{\rm p,BBH}$, which are crucial parameters for our sampling process (see below).

In each Pop~III star-forming halo, \textsc{a-sloth} predicts the expected mass $\delta M_{\star,\rm III}$ of Pop~III stars formed at the current timestep $t$. We assume that on average a fraction $f_{\rm B}=0.7$ of the mass of newly-born Pop~III stars is made up by binaries based on the $N$-body simulations of Pop~III SCs in \citet{Liu2021binary}. 
Treating Pop~III star formation as a stochastic\footnote{It is likely that star formation is self-regulated and not fully stochastic \citep{Lewis2023,Yan2023,Liu2024sms}, especially considering Pop~III systems made of small numbers of massive stars {\citep[but see also][]{Grudic2023imf}}. Since we are only concerned with the global population of Pop~III BBH mergers in this paper, a detailed investigation of IMF sampling schemes \citep[see, e.g.,][]{Kroupa2013} is deferred to future work.} process at the scales resolved by \textsc{a-sloth}, the number of Pop~III BBHs (as well as the corresponding progenitor stellar binaries) to be sampled at this timestep $N_{\rm B}$ is drawn from a Poisson distribution with parameter $f_{\rm boost}f_{\rm B}f_{\rm BBH}\delta M_{\star,\rm III}/\bar{m}_{\rm p, BBH}$, where $f_{\rm boost}=50$ is a boost factor introduced to obtain better statistics. If $N_{\rm B}>0$, we randomly draw $N_{\rm B}$ BBH progenitor binary stars from the input BPS catalogue, which will become BBHs at $t+\tau_{\rm B}$. 

Since the (initial/ZAMS) properties of Pop~III binary stars are largely unknown in the lack of direct observations, to explore the parameter space, we consider six \textsc{sevn} simulations (LOG1, TOP1, KRO1, LOG5, TOP5, and KRO5) with diverse initial conditions from \citet[see their Sec. 2 for details]{Costa2023}, combining three IMF models and two sets of initial binary parameter distributions. 
In particular, we fix the mass range of primary stars as $m_{\star,1}\in [5,550]\ \rm M_\odot$ and consider three models for (the shape of) the primary star IMF as follows \citep[see also fig.~4 of][]{Costa2023}.
\begin{itemize}
    \item A log-flat distribution $p(m_{\star,1})\equiv dN/dm_{\star,1}\propto m_{\star,1}^{-1}$ (labelled by `LOG'), which is motivated by the mass distributions of Pop~III protostars in (magneto)hydrodynamic simulations \citep[see, e.g., fig 6 in][and references therein]{Klessen2023} and chosen as the fiducial model.
    \item A \citet{Kroupa2001} IMF $p(m_{\star,1})\propto m_{\star,1}^{-2.3}$ (labelled by `KRO'), which is consistent with the IMF of Pop I/II stars (for $m_{\star,1}>0.5\ \rm M_\odot$) in local observations.
    \item A top-heavy distribution $p(m_{\star,1})\propto m_{\star,1}^{-0.17}e^{-m^{2}_{\rm cut}/m_{\star,1}^{2}}$ given $m_{\rm cut}^{2}=20\ \rm M_\odot^{2}$ (labelled by `TOP') as an extreme case, which is adopted in some cosmological simulations \citep{Jaacks2018,Jaacks2019,Liu2020,Liu2020did}.
\end{itemize}
In all cases, the IMF follows a power-law at the high-mass end. Therefore, each IMF model can also be characterized by the (asymptotic) power-law slope $\alpha$, with $\alpha=1$ for LOG, 2.3 for KRO and 0.17 for TOP. 

For IBS, we consider two distinct sets of distributions of the mass ratio $q_{\star}\equiv m_{\star,2}/m_{\star,1}$, orbital period $P$, which is characterized by the variable $\pi\equiv \log(P/{\rm day})$ in practice, and eccentricity $e$ \citep[see also fig.~3 of][]{Costa2023}. A lower bound on the secondary mass $m_{\star,2}>2.2\ \rm M_\odot$ is also imposed in both cases.
\begin{itemize}
    \item The first model is based on observations of present-day massive binary stars \citep[henceforth \citetalias{Sana2012}]{Sana2012} with $p(q)\propto p^{-0.1}$ for $q\in [0.1,1]$, $p(\pi)\propto \pi^{-0.55}$ for $\pi\in [0.15,5.5]$, and $p(e)\propto e^{-0.42}$ for $e\in [0,1)$. This model favours initially close binary stars and is labelled by the number `1' following the convention in \citet{Costa2023}. We call it the close IBS model from now on.
    \item The second model adopts the power-law and log-normal fits to the mass ratio and orbital period distributions of Pop~III protostars in the hydrodynamic simulations by \citet[hereafter \citetalias{Stacy2013}]{Stacy2013}: $p(q)\propto q^{-0.55}$ for $q\in [0.1,1]$, and $p(\pi)\propto\exp[-(\pi-\mu)^{2}/(2\sigma^{2})]$, given $\mu=5.5$ and $\sigma=0.85$. The eccentricity distribution is assumed to be thermal, i.e., $p(e)=2e$ for $e\in [0,1)$, following previous BPS studies of Pop~III BBHs \citep[e.g.,][]{Kinugawa2014,Hartwig2016,Tanikawa2021mrd}. This model is dominated by initially wide binaries and labelled by the number `5'. We refer to it as the wide IBS model henceforth. 
\end{itemize}
As it is still challenging to simulate the entire Pop~III star formation process with well-resolved disc fragmentation, protostellar feedback, and $N$-body dynamics, the properties of Pop~III SCs are still unclear in current simulations, and so is the statistics of binary stars \citep[see, e.g.,][]{Liu2021binary}. We hope to bracket the reality with these two models, which are supported by different simulations with different setups (numerical scheme of hydrodynamics, resolution, modelling of feedback, and cosmological context). 
For instance, the simulations by \citet{Hirano2018sc} show that it is possible to form highly eccentric, close binaries 
via violent $N$-body dynamics in compact Pop~III SCs formed by mergers of multiple star-forming gas clumps in relatively large haloes ($M_{\rm h}\sim 10^{7}\ \rm M_{\odot}$) under strong streaming motions between gas and dark matter. The formation of close binaries 
by dynamical friction is also favoured in dense gas structures with collision-induced emission cooling \citep{Riaz2023}, at least in the initial stage (a few thousand years after the formation of the first protostar) when protostellar feedback is expected to be unimportant \citep{Jaura2022}. On the other hand, the model dominated by wide binaries is motivated by the trend that Pop~III protostar clusters/binaries formed in typical minihaloes ($M_{\rm h}\sim 10^{5-6}\ \rm M_{\odot}$) 
tend to expand due to accretion/inflow of gas with high angular momentum in (magneto-)hydrodynamic simulations 
of hot, thick star-forming discs \citep[e.g.,][]{Chon2019,Heath2020,Sugimura2020,Sugimura2023,Mignon-Risse2023,Franchini2023,Park2022,Park2024}. 

Given $\delta M_{\star,\rm III}$, \textsc{a-sloth} also samples individual Pop~III stars from an input power-law IMF $p(m_{\star})\propto m_{\star}^{-\alpha_{\rm s}}$, which are treated as single stars to model stellar feedback, i.e., photoheating, ionization, SN-driven outflows and metal enrichment \citep[see sec. 2 of][]{Hartwig2023}. In principle, one can pair a fraction of the sampled stars into binaries and follow their evolution on the fly to predict the formation of BBHs (and meanwhile model the feedback from binary stars). 
However, as described above, we instead adopt a simple approach 
by sampling BBHs directly from the input \textsc{sevn} catalogue to explore the parameter space of the poorly constrained properties of Pop~III binary stars. In this way, the progenitor stellar binaries of Pop~III BBHs sampled separately by our scheme in a star formation event are not consistent with the (underlying) population of Pop~III stars sampled by the stellar feedback routine of \textsc{a-sloth}, and the impact of binary interactions on stellar feedback is also ignored. 
To approximately capture the variations of stellar feedback and BBH formation with IMF spontaneously, we set the power-law slope and mass range of the single-star IMF $p(m_{\star})$ for stellar feedback to those of the primary star IMF $p(m_{\star,1})$ adopted by the input BPS catalogue, i.e., $\alpha_{\rm s}=\alpha=1$ for LOG (fiducial), 2.3 for KRO, and 0.17 for TOP\footnote{For the TOP model with $p(m_{\star,1})\propto m_{\star,1}^{-0.17}e^{-m^{2}_{\rm cut}/m_{\star,1}^{2}}$, we do not include the exponential cutoff term in the IMF used to sample single stars for stellar feedback for simplicity. 
We have verified by numerical experiments that the exact shape of the IMF at the low-mass ($\lesssim 20\ \rm M_\odot$) end has negligible impact on the star formation history in this case since stellar feedback is dominated by massive stars for such a top-heavy IMF.}, with $m_{\star}\in [5,550]\ \rm M_\odot$. 
We defer a more complete modelling of binary stars, particularly their feedback features such as enhanced production of (ionizing) UV photons and different metal yields \citep[e.g.,][]{Sansom2009,Chen2015,Gotberg2019,Secunda2020,Ma2022,Tsai2023,Lecroq2024,Liu2024,Yates2024}, to future work. 
We expect the effects of binary interactions to be comparable to those of varying the IMF (See Sec.~\ref{sec:imf}), 
which will be overwhelmed by the uncertainties in other aspects. For instance, the star formation history of Pop~III stars is still highly uncertain in theory, showing orders of magnitude of discrepancies between different studies with different feedback models, resolution, and simulation volumes \citep[see, e.g., fig.~13 in][]{Hartwig2022}. The poorly understood initial binary properties (captured by the six \textsc{sevn} models considered here) also lead to order-of-magnitude uncertainties in the merger rate of Pop~III BBHs \citep[see Sec.~\ref{sec:comp} and, e.g.,][]{Santoliquido2023}.

\subsection{Evolution of Pop~III BBHs}
\label{sec:evo}
Once a Pop~III BH binary is sampled, we follow on-the-fly the evolution of its orbit in the host galaxy or NSC as well as the internal evolution of the binary orbit. 


\subsubsection{Galactic dynamics}
\label{sec:gd}
The dynamics of Pop~III BBHs in host galaxies is driven by galaxy mergers, which randomize the (galactic/external) orbits of Pop~III BBHs coming from the smaller progenitor halo, and dynamical friction (DF) by field stars and dark matter, 
through which massive Pop~III BBHs sink towards galaxy centres where they can further fall into NSCs. 
We use the same method as in \citet[see their sec.~2.2 for details]{Liu2021} to model such dynamics but further include the effects of dark matter in addition to stars. Here, we only consider the DF of stars and dark matter, ignoring the effects of gas. It is shown in simulations \citep[e.g.][]{Chen2022df} that DF is dominated ($\gtrsim 99\%$) by collisionless particles. Besides, for Pop~III BBHs, which are typically much lighter than supermassive BH seeds and SCs, the DF timescale is only shorter than the Hubble time in the central region ($r\lesssim 300\ \rm pc$) dominated by stars in most cases. 

\paragraph{Initial distribution} 
When a Pop~III BBH is born or falls into a larger halo during a halo merger event, given the virial radius $R_{\rm vir}$ of the (post-merger) host halo, its apocentre distance $r$ is drawn randomly from a (cumulative) probability distribution of $x\equiv r/R_{\rm vir}$: 
\begin{align}
    F(<x)\simeq
    \begin{cases}
        10^{-4}(1000x)^{2.23}\ ,& x<0.03\ ,\\
        0.2(100x/3)^{0.88}\ ,& 0.03\le x<0.1\ ,\\
        0.58(10x)^{0.24}\ ,&0.1\le x<0.95\ ,
    \end{cases}\label{pop3ic}
\end{align}
which is a broken-power-law fit of the spatial distribution of Pop~III remnants in high-$z$ haloes predicted by the cosmological simulation FDbox\_Lseed in \citet{Liu2020gw,Liu2020did}. 
The eccentricity $e_{\star}$ of a (galactic) orbit is initially drawn from a uniform distribution in $[0,1)$, following \citet{Arca-Sedda2014}. Here we do not consider the effects of SN natal kicks during the formation of Pop~III BBHs on their initial galactic orbits. We expect the initial orbits at birth to be unimportant for the inspirial of Pop~III BBHs in galaxies hosting NSCs, as such galaxies reside in relatively massive haloes with $M_{\rm h}\gtrsim 10^{9}\ \rm M_{\odot}$ in which most Pop~III BBHs come from cosmic accretion flows and halo mergers, so their dynamics are mainly determined by halo assembly processes 
captured by the distribution of $x\equiv r/R_{\rm vir}$ in cosmological simulations \citep{Liu2020gw,Liu2020did}. 
Besides, we assume that during a halo merger event, all Pop~III BBHs in the smaller halo will be immediately stripped and redistributed in the post-merger halo according to Eq.~\ref{pop3ic}.

\paragraph{Inspiral by dynamical friction} 
Within the host galaxy, we follow the inspiral of each binary with 
\begin{align}
\begin{split}
    \frac{dr}{dt}&\simeq -r\left[\frac{1}{\tau_{\rm DF}(r,e_{\star},m|\vec{\lambda}_{\star})}+\frac{1}{\tau_{\rm DF}(r,e_{\star},m|\vec{\lambda}_{\chi})}\right]\ ,\\ 
    \frac{de_{\star}}{dt}&\simeq -e_{\star}\left[\frac{1}{\tau_{\rm DF}(r,e_{\star},m|\vec{\lambda}_{\star})}+\frac{1}{\tau_{\rm DF}(r,e_{\star},m|\vec{\lambda}_{\chi})}\right]\ ,
\end{split}\label{df}
\end{align}
where $m\equiv m_{1}+m_{2}$ is the total mass of the binary given the primary BH mass $m_{1}$ and secondary BH mass $m_{2}$, $\tau_{\rm DF}$ is the DF timescale formula (see below), $\vec{\lambda}_{\chi}$ captures the dark matter halo properties ($M_{\rm vir}$ and $R_{\rm vir}$), and $\vec{\lambda}_{\star}$ denotes parameters of the background stellar system, which are the mass $M_{\star}$, size $R_{\star}$ and inner slope of density profile $\gamma_{\star}$ of the host galaxy in our case. Here $M_{\star}$ is set to the total mass of Pop~I/II stars predicted by the baryon cycle in \textsc{a-sloth} or an extrapolation scheme for the low-$z$ regime not covered by the merger trees (see Sec.~\ref{sec:extra}). We update $r$ and $e_{\star}$ at each star formation timestep of the host halo. 
Note that Pop~III stars typically form in small clusters, which, if they remain bound, can sink into the galaxy centre more efficiently. The reason is that in this case, $m$ in Eq.~\ref{df} should be the total mass of the cluster ($\sim 10^{3}\ \rm M_\odot$), and $\tau_{\rm DF}$ is shorter for higher $m$ (see below). By using the masses of individual binaries in Eq.~\ref{df}, we make the conservative assumption that Pop~III SCs have been dissolved by internal relaxation, mass loss from stellar evolution, and tidal disruption 
during halo/galaxy mergers before falling into galaxies hosting NSCs\footnote{Most Pop~III stars form in small clusters with total masses of $\sim 100-10^{4}\ \rm M_\odot$ in our simulations. Assuming a half-mass radius of $\sim 1~\rm pc$ and an average stellar mass of $\sim 10\ \rm M_{\odot}$, the dissolution timescale of such clusters by internal relaxation is $\sim 100$~Myr, similar to 
the timescale $\sim 100\ \rm Myr$ in which minihaloes hosting Pop~III stars/remnants experience major mergers that bring Pop~III remnants to large haloes ($M_{\rm h}\gtrsim 10^{9}\ \rm M_\odot$) with NSCs.}.

We use the minimum of the values predicted by two DF timescale formulae to estimate $\tau_{\rm DF}$. The first is the Chandrasekhar formula \citep{BT2011}
\begin{align}
    \frac{\tau_{\rm DF,C}}{\mathrm{Myr}}=\frac{342}{\ln\Lambda}\left(\frac{r}{\rm3\ pc}\right)^{2}\left(\frac{v}{\rm 10\ km\ s^{-1}}\right)\left(\frac{m}{100\ \rm M_{\odot}}\right)^{-1}\ , \label{tdf1}
\end{align}
where $\ln\Lambda\sim\ln[M_{\star}r/(0.8mR_{\star})]$ is the Coulomb logarithm, and $v\sim\sigma_{\star}\sim\sqrt{GM_{\star}/(0.8R_{\star})}$. The second formula is a generalization of Eq.~\ref{tdf1} and can be applied to both cusped and cored density profiles \citep{Arca-Sedda2015,Arca-Sedda2016}: 
\begin{align}
\begin{split}
    \frac{\tau_{\rm DF,A}}{\mathrm{Myr}}=0.3g\sqrt{\left(\frac{R_{\star}^{3}}{\rm 1\ kpc^{3}}\right)M_{\star,11}^{-1}}\left(\frac{m}{M_{\star}}\right)^{\alpha}\left(\frac{r}{R_{\star}}\right)^{\beta}\ ,\\
    g=(2-\gamma_{\star})\left\{a_{1}\left[\frac{1}{(2-\gamma_{\star})^{a_{2}}}+a_{3}\right](1-e_{\star})+e_{\rm if}\right\}\ ,
\end{split}\label{tdf2}
\end{align}
given $\alpha=-0.67$, $\beta=1.76$, $a_{1}=2.63\pm0.17$, $a_{2}=2.26\pm 0.08$, and $a_{3}=0.9\pm 0.1$. This formula is only valid for $0<\gamma_{\star}<2$ and Eq.~\ref{tdf1} is used otherwise\footnote{We have $\tau_{\rm DF,A}<\tau_{\rm DF,C}$ in most cases.}. 

\paragraph{Host galaxy/halo model} 
To evaluate the first dynamical friction term driven by stars, we assume that the galaxy (stellar) density profile is described by a Dehnen sphere \citep[see also Eq.~\ref{dehnen}]{Dehnen1993}, and the galaxy size (scale length) $R_{\star}$ is derived from the empirical relation between galaxy mass and half-mass radius $R_{1/2}$ in local observations \citep[see their eq.~18 and fig.~1]{Arca-Sedda2014}:
\begin{align}
\begin{split}
    R_{\star}&=[2^{1/(3-\gamma_{\star})}-1]R_{1/2}\ ,\\
    R_{\rm 1/2}&=\max(0.02R_{\rm vir},\ 2.37M_{\star,11}^{0.14}\ \rm kpc)\ ,
\end{split}
\label{rstar}
\end{align}
where $M_{\star,11}\equiv M_{\star}/(10^{11}\ \rm M_{\odot})$, and the inner slope $\gamma_{\star}$ is generated from a uniform distribution in the ranges of $0-2$. 
We also impose a lower limit of $0.02R_{\rm vir}$ on $R_{1/2}$ according to the results (for median galaxy half-mass radii) from 
sub-halo abundance matching \citep{Somerville2018} for galaxies with $M_{\star}\sim 10^8-10^{11}\ \rm M_{\odot}$ at $z\le 3$. 

In reality, galaxy size/morphology is regulated by a variety of processes, such as gas accretion, galaxy mergers, stellar and AGN feedback \citep[see the discussions in e.g.,][]{Hopkins2023,Karmakar2023}, which introduce dispersion to the scaling relation. To capture such dispersion, we perturb the value of $\log(R_{\star})$ predicted by Eq.~\ref{rstar} with a random number (drawn once\footnote{It is found in cosmological simulations that most galaxies oscillate around the median size several times in their lifetimes \citep[see their fig.~8 and 9]{Karmakar2023}. We ignore such oscillations for simplicity.} 
and inherited along the main branch\footnote{In a merger tree, starting with the root the main branch goes through the most massive child halo in each halo merger event (branching).}) following a uniform distribution in [-0.36, 0.36]~dex, motivated by the scatter in observations \citep{Leigh2012}. Besides, additional uncertainties may arise when extrapolating the local empirical scaling relation to the high-$z$ regime ($z\sim 3-10$) that is important for our purposes. For instance, it is found in recent cosmological simulations \citep{Roper2023} that galaxies at $z\gtrsim 5$ are expected to follow a different size-mass relation with a negative slope, and are more compact, embedded in central concentration of non-star forming gas (i.e., `inside-out' galaxy formation), which is also supported by JWST observations \citep[e.g.,][]{Baggen2023,Langeroodi2023,Ito2023,Ormerod2024,Ji2024}. Our $z$-independent estimate of $R_{\star}$ can be regarded as an upper limit. It will be shown below (Eq.~\ref{tdf2}) that the dependence of DF timescale $\tau_{\rm DF}$ on galaxy size is weak ($\tau_{\rm DF}\propto R_{\star}^{-0.26}$), which, combined with the weak mass and redshift dependence of $R_{\star}$ in Eq.~\ref{rstar}, implies that our results are insensitive to the modelling of $R_{\star}(M_{\star})$. Actually, we find by numerical experiments that changing $R_{\star}$ by a factor of 10 can only alter the number of Pop~III BBHs (and mergers) in NSCs by up to $\sim 40$ percent.

For the second term in Eq.~\ref{df} driven by dark matter, which is only important for Pop~III BBHs at galaxy outskirts ($r\gtrsim 300$\ \rm pc), we only use the Chandrasekhar formula (Eq.~\ref{tdf1}) with $\ln\Lambda\sim\ln[M_{\rm vir}r/(0.8mR_{\rm vir})]$ and $v\sim\sqrt{GM_{\rm vir}/(0.8R_{\rm vir})}$, implicitly assuming a power-law slope $\gamma_{\chi}>2$ for the dark matter density profile\footnote{Apparently, this simple assumption ($\gamma_{\chi}>2$) does not hold in the central region, where the dynamical friction by dark matter can be important for the dynamics of (normal) SCs (see Sec.~\ref{sec:sc_evo}). Nevertheless, we find that our results are insensitive to the dynamics of SCs.}, which describes the asymptotic behaviour of the NFW profile at large radii. 
During a merger between two haloes, the orbits of binaries from the smaller halo are randomized with the same distributions adopted for initialization\footnote{For simplicity, we ignore possible enhancement of inspiral by DF during galaxy mergers implied by recent observations \citep{Roman2023}.}. 
 
We further introduce a free parameter $M_{\star,\min}$, as the (Pop~I/II stellar) mass threshold above which galaxies have regular morphology and well-defined dynamical centres to which massive objects can sink. DF is turned off in galaxies below this threshold. The morphology and dynamics of high-$z$ galaxies are still poorly constrained observationally. The recent numerical experiments by \citet{Hopkins2023} found that the formation of a centrally concentrated density profile that defines the dynamical centre drives the transition from irregular to disc morphology, which mostly occurs around $M_{\star}\sim 10^{10}\ \rm M_{\odot}$, although in principle the transition can happen at any mass scale and less massive disc galaxies are also seen in simulations. 
In this study, we adopt $M_{\star,\min}=10^{6}\ \rm M_{\odot}$ by default as an optimistic choice, which is approximately the minimum mass of galaxies hosting NSCs observed in the local Universe \citep{Neumayer2020nuclear}. The motivation is that the observed properties of NSCs in local low-mass galaxies ($M_{\star}\lesssim 10^{9}\ \rm M_{\odot}$), including both early- and late-type, can be well explained by the globular cluster (GC) accretion scenario \citep[see, e.g.,][]{Fahrion2022,Leaman2022,Roman2023}, in which multiple GCs inspiral to the centre by DF and build up the NSC \citep[e.g.,][]{Arca-Sedda2014,Tsatsi2017,Arca-Sedda2018,Fahrion2022gc,Wang2023gc}, implying that DF must work in 
such NSC host galaxies. 

\subsubsection{Binary orbital evolution}
\label{sec:bevo}
The internal evolution of a BBH orbit starts at $\tau_{\rm B}$ after the initial formation of the progenitor binary stars. The earlier stage ($t<\tau_{\rm B}$) is governed by binary stellar evolution whose outcome is given by the input BPS catalogue and serves as the initial condition for BBH evolution in \textsc{a-sloth}. 
We use the same set of differential equations detailed in \citet[see their sec.~2.3 and below]{Liu2021} to follow the internal evolution of BBHs in the field and NSCs as well as the inspiral of BBHs in NSCs. Here we integrate the equations for evolution inside SCs on the fly with the explicit Euler method in \textsc{a-sloth} and adopt an approximation based on the merger timescale formula from \citet{Iorio2023} to facilitate the computation for the internal evolution of binaries in the field. When numerical integration\footnote{At each timestep $\Delta t_{i}$, we first update $a$ by $\Delta t_{i}/2$: $a_{i+1/2}=a_{i}+\dot{a}(a_{i},e_{i})\Delta t_{i}/2$. Then we evolve $e$ for the whole step: $e_{i+1}=e_{i}+\dot{e}(a_{i+1/2},e_{i})\Delta t_{i}$. Finally, we assign $a_{i+1}=a_{i+1/2}+\dot{a}(a_{i+1/2},e_{i+1})\Delta t_{i}/2$. } is used to evolve $a$ and $e$ in SCs, we further do sub-cycling with timesteps $\delta t_{\rm binary}\le 0.01 a/(da/dt)$. A merger happens when the expected time taken to merge has passed since BBH formation or $a<6G(m_{1}+m_{2})/c^{2}$ according to numerical integration. Then the merger properties are recorded and the binary is removed from the data structure in \textsc{a-sloth}. 
This approach is more accurate and flexible than post-processing since the evolution of host SCs is self-consistently taken into account (see Sec.~\ref{sec:nsc}).

\paragraph{Isolated evolution in galaxy fields} 
In the field, we assume that binary evolution is only driven by GW emission\footnote{BH binaries with highly eccentric ($e\gtrsim 0.99$), wide ($a\sim 10^{4}\ \rm AU$) orbits (and wide BH triples) can also be driven to merge by perturbations from fly-by field stars \citep{Michaely2019,Michaely2020}. 
This channel can be important for Pop~III BBH mergers if the Pop~III IBS is dominated by wide binaries and Pop~III stellar/BH triples are common \citep[see, e.g., fig.~8 of][]{Sharda2020}. 
We plan to consider this channel in the future.} \citep{Peters1964}:
\begin{align}
\begin{split}
     \dot{a}(a,e)\equiv\frac{da}{at}=\left.\frac{da}{dt}\right|_{\rm GW}&=-\frac{64}{5}\frac{A}{a^{3}(1-e^{2})^{7/2}}f_{1}(e)\ ,\\
     \dot{e}(a,e)\equiv\frac{de}{dt}=\left.\frac{de}{dt}\right|_{\rm GW}&=-\frac{304}{15}e\frac{A}{a^{4}(1-e^{2})^{5/2}}f_{2}(e)\ ,
\end{split}\label{begw}
\end{align}
where $A=G^{3}m_{1}m_{2}(m_{1}+m_{2})/c^{5}$, and
\begin{align}
\begin{split}
    f_{1}(e)=1+\frac{73}{24}e^{2}+\frac{37}{96}e^{4}\ ,\quad
    f_{2}(e)=1+\frac{121}{304}e^{2}\ .
\end{split}
\end{align}

An approximated solution of these equations is given by \citet[see their appendix C]{Iorio2023} based on \citet{Zwick2020} that predicts the time taken to merge from the initial BBH condition ($a_{i}$ and $e_{i}$) as
\begin{align}
\begin{split}
    &t_{\rm GW}(a_{i},e_{i})=\begin{cases}
        t_{\rm GW,0}(a_{i})/[1+f_{\rm corr}(e_{i})]\ ,& e_{i}<0.999\ ,\\
        t_{\rm peters}(a_{i})8^{1-\sqrt{1-e_{i}}}\ ,& e_{i}\ge 0.999\ ,
    \end{cases}\\
    &t_{\rm GW,0}(a_{i})=\left[5a_{i}^{4}(1-e^{2}_{i})^{7/2}\right]/(256A)\ ,\\
    &t_{\rm peters}(a_{i})=t_{\rm GW,0}/f_{1}(e_{i})\ ,\\
    &f_{\rm corr}=e_{i}^{2}\left[-0.443+0.58(1-e_{i}^{3.074})^{f_{3}(e_{i})}\right]\ ,\\
    &f_{3}(e_{i})=1.105-0.807e_{i}+0.193e_{i}^{2}\label{tgw_approx}\ ,
\end{split}
\end{align}
which is accurate enough for our purposes. Therefore, instead of integrating Eqs.~\ref{begw}, we set a clock for each BBH at formation and trigger the merger after $t_{\rm GW}$ has passed. If a BBH falls into a NSC and is ejected later (either by binary-single encounters or disruption of the SC, see below), we update the clock on an individual basis with a new merger timescale $t_{\rm GW,ej}(a_{\rm ej},e_{\rm ej})$ calculated from the orbital parameters ($a_{\rm ej}$ and $e_{\rm ej}$) at the moment of ejection, which have evolved inside the SC (see below). 

\paragraph{Dynamical evolution in NSCs}
A binary falls into a NSC when the apocentre distance of its (galactic) orbit is smaller than the characteristic size (i.e., scale length) of the NSC in the host galaxy (i.e., $r<R_{\rm SC}$). Once inside SCs, soft binaries will be destroyed and hard binaries will be hardened by {binary-single} encounters. The criterion for hard binaries is $a<R_{\rm SC}$ and $a<a_{\rm HDB}=Gm_{1}m_{2}/\left[m_{\star,\rm SC}\sigma_{\star}(r)^{2}\right]$ \citep{Mapelli2021hierarchical}, given the SC size $R_{\rm SC}$, local velocity dispersion $\sigma_{\star}(r)$ and typical mass of objects in the cluster. Throughout this study, we adopt $m_{\star,\rm SC}=0.5\ \rm M_{\odot}$ for simplicity, although $m_{\star,\rm SC}$ may vary from cluster to cluster and evolve with redshift in reality. 
The hard binary criterion is checked at each (star formation) timestep and soft binaries are immediately removed. For simplicity, here we have ignored the hardening of BBHs by interactions with gas in NSCs, which can even shrink the orbits of initially soft binaries with $a>a_{\rm HDB}$ to make them hard, shifting the soft-hard boundary to larger separations \citep{Rozner2024}.

For the surviving hard binaries, we include additional {binary-single} encounter terms in the binary evolution equations \citep{Sesana2015,Mapelli2021hierarchical,ArcaSedda2020}:
\begin{align}
\begin{split}
    \frac{da}{dt}=\left.\frac{da}{dt}\right|_{\rm \star}+\left.\frac{da}{dt}\right|_{\rm GW}\ ,\quad
    \frac{de}{dt}=\left.\frac{de}{dt}\right|_{\star}+\left.\frac{de}{dt}\right|_{\rm GW}\ ,
\end{split}\label{be}\\
\begin{split}
    \left.\frac{da}{dt}\right|_{\rm \star}=-\frac{GH\rho_{\star}(r)}{\sigma_{\star}(r)}a^{2}\ ,\quad
    \left.\frac{de}{dt}\right|_{\star}=\kappa\frac{G \mathcal{H}\rho_{\star}(r)}{\sigma_{\star}(r)}a\ ,
\end{split}\label{be3b}
\end{align}
where $\rho_{\star}(r)$ is the local stellar density of the cluster, $\mathcal{H}\sim 1-20$ and $\kappa\sim 0.01-0.1$ \citep{Sesana2006,Sesana2015,Mapelli2021hierarchical} are two dimensionless factors. We adopt $\mathcal{H}=20$ following \citet{Liu2021} and fix $\kappa=0.01$ as a conservative assumption. We also impose a lower limit of $10^{-8}$ on $1-e$ under the optimistic assumption that the binary will merge immediately when driven to this limit by {binary-single} interactions. 

As in \citet{ArcaSedda2020}, here a SC of mass $M_{\rm SC}$ is characterized by a cored Dehnen sphere \citep{Dehnen1993} with
\begin{align}
\begin{split}
    \rho_{\star}(r)&=\begin{cases}
    \frac{(3-\gamma_{\rm SC})M_{\rm SC}/(4\pi R_{\rm SC}^{3})}{x^{\gamma_{\rm SC}}(1+x)^{4-\gamma_{\rm SC}}}\ ,\quad &r\ge r_{\rm c}\ ,\\
    \rho_\star(r_{\rm c})\ ,\quad &r<r_{\rm c}\ ,
    \end{cases}\\
    \sigma_{\star}(r)&=\begin{cases} 
    \sqrt{\frac{GM_{\rm SC}}{R_{\rm SC}}}\left(\frac{r}{R_{\rm SC}}\right)^{\delta/2}\ ,\quad &r_{\rm c}\le r< R_{\rm SC}\ ,\\
    \sigma_{\star}(r_{\rm c})\ ,\quad &r<r_{\rm c}\ ,
    \end{cases}
\end{split}\label{dehnen}
\end{align}
where $x\equiv r/R_{\rm SC}$, $\delta=\gamma_{\rm SC}$ for $\gamma_{\rm SC}<1$ and $\delta=2-\gamma_{\rm SC}$ for $\gamma_{\rm SC}\ge 1$, given the scale length $R_{\rm SC}$ and inner slope $\gamma_{\rm SC}$ of the Dehnen sphere, which are derived from $M_{\rm SC}$ using the empirical scaling relations in \citet[][]{Liu2021} based on local observations \citep{Neumayer2020nuclear,Pechetti2020}. 
In this model, most hardening happens in the core, and the hardening efficiency is sensitive to the core overdensity parameter $\Delta_{\rm c}$ that determines the core size  
$r_{\rm c}=\max(R_{\rm SC}/\Delta_{\rm c}^{1/\gamma_{\rm SC}}, r_{\bullet})$, which is no smaller than the influence radius of the BBH $r_{\bullet}=R_{\rm SC}(2m/M_{\rm SC})^{1/(3-\gamma_{\rm SC})}$. We adopt an optimistic core overdensity $\Delta_{\rm c}=100$.  
The external orbit of the binary also evolves by DF within the cluster as described by Eq.~\ref{df} in which the background stellar system now becomes the SC. 
In our case, $\rho_{\star}(r)$, $\sigma_{\star}(r)$ and $\tau_{\rm DF}(r)$ are all updated at each star formation timestep to capture the evolution of the host SC (see Sec.~\ref{sec:nsc}).

Similar to \citet{Liu2021}, we assume spherical symmetry and ignore higher-order processes involving tidal fields, general relativity effects, and interactions with binary Pop~I/II BHs, central massive BHs, and gas, such as relativistic phase space diffusion, tides-driven eccentricity excitation, the Kozai–Lidov mechanism, hierarchical and repeated mergers, dynamical hardening/softening and accretion in AGN discs \citep[see, e.g.,][]{Petrovich2017,Hoang2018,Antonini2019,Yang2019,ArcaSedda2020,McKernan2020,Tagawa2020,Mapelli2021hierarchical,Mapelli2022,Zhang2021,Fragione2022,Fragione2023,Kritos2022,Atallah2023,DeLaurentiis2023,Chattopadhyay2023,Chen2023,Dall'Amico2023,Li2023,Rozner2023,Rowan2023,Trani2023,Vaccaro2023,Wang2023agn,Winter-Granic2023,Hamilton2024,Balberg2024,Barber2024,Gangardt2024,Fabj2024,Purohit2024,Rozner2024,Torniamenti2024}. Here, interactions with existing Pop~I/II BBHs are expected to be important in high-$z$ metal-poor SCs \citep{Barber2024}, which we plan to take into account in future work. 
For completeness, we also consider ejections of Pop~III binaries by {binary-single} encounters, which occur when $a_{\rm GW}<a<a_{\rm ej}$, given the critical separations for GW-dominance $a_{\rm GW}$ and ejection $a_{\rm ej}$ \citep{Miller2009,Antonini2016,fragione2020repeated,Mapelli2021hierarchical}:
\begin{align}
    a_{\rm GW}&=\left[\frac{64}{5H}\frac{A\sigma_{\star}f_{1}(e)}{G\rho_{\star}(1-e^{2})^{7/2}}\right]^{1/5}\ ,\label{agw}\\
    a_{\rm ej}&=\frac{Hm_{\star,\rm SC}^{2}}{\pi(m_{1}+m_{2})^{3}}\frac{Gm_{1}m_{2}}{v_{\rm esc}^{2}}\ ,\label{aej}\\
    \frac{v_{\rm esc}}{\mathrm{km\ s^{-1}}}&=40\left(\frac{M_{\rm SC}}{10^{5}\ \rm M_{\odot}}\right)^{1/3}\left[\frac{\rho_{\star}(R_{\rm SC})}{10^{5}\ \rm M_{\odot}\ pc^{-3}}\right]^{1/6}\ ,
\end{align}
where $\rho_{\star}(R_{\rm SC})$ is the typical stellar density evaluated at the scale length $R_{\rm SC}$ of the cluster. 
Once a binary is ejected from a cluster, its subsequent evolution is only driven by GWs. Besides, an ejected binary can no longer sink towards the galaxy centre by DF until its host halo merges into a larger halo and a new orbit is assigned to this binary in the post-merger system. The ejection criterion can only be satisfied by low-mass BBHs ($m\lesssim 20\ \rm M_{\odot}$) in low-mass ($M_{\rm SC}\lesssim 10^{5}\ \rm M_{\odot}$) SCs 
where $a_{\rm ej}>a_{\rm GW}$. In our case, 
the dynamical hardening timescale in such low-mass ($M_{\rm SC}\lesssim 10^{5}\ \rm M_{\odot}$) SCs is typically comparable to the Hubble time, and light BBHs in low-mass SCs are rare, so ejection by {binary-single} encounters is unimportant. 

Since Eqs.~\ref{begw} are not integrated explicitly for BBHs in galaxy fields, 
integration of Eqs.~\ref{be} starts from the initial conditions at the moment of BBH formation rather than infall. That is to say, we make an approximation for the initial condition of evolution inside SCs, which is justified by the fact that the progenitors of most mergers in SCs seldom evolve in fields and the evolution inside SCs is insensitive to the initial condition. This approximation does not affect the fate of binaries in SC, since the maximum separation for a BBH to be a hard binary inside SCs ($a\sim 10^{2}-10^{4}\ \rm AU$) is much larger than that required for efficient evolution in fields by GW emission ($a\lesssim 0.5\ \rm AU$). We have verified with numerical experiments that the difference between the results from explicit integration of Eqs.~\ref{begw} and the approximated merger timescale solution (Eqs.~\ref{tgw_approx}) is negligible. 


\section{Nuclear star clusters}
\label{sec:nsc}

\begin{figure}
    \centering
\begin{tikzpicture}[node distance=1cm]
\node (cnode) [startstop] {Child nodes (current level): SCs to be inherited};
\node (pre) [io, below of=cnode,xshift=-0.8cm] {Main branch?};
\node (asloth) [process, below of=pre, xshift=0cm] {Star formation: $M_{\star}$, $p_{i}$};
\node (focc) [io, below of=asloth,xshift=-1cm] {$p_{i}<f_{\rm occ}(M_{\star})$?};
\node (insc) [io, below of=cnode, xshift=2cm] {Host SCs?};
\node (insc0) [io, right of=focc, xshift=1.2cm] {Host SCs?};
\node (orbit) [decision, below of=insc,xshift=0cm]{Random orbits$^a$};
\node (insc1) [io, below of=focc, xshift=-0cm] {Host SCs?};
\node (old) [io, right of=insc1, xshift=1.2cm] {Oldest SC?};
\node (nnsc) [case, below of=insc1, xshift=0cm] {NSC formation};
\node (ensc) [case, below of=old,xshift=-0cm] {NSC evolution};
\node (sat) [case, below of=old, xshift=2cm]{Normal SCs};
\node (scale) [decision, below of=nnsc, xshift=0cm]{Empirical scaling relations};
\node (nsc) [quantity, below of=scale, xshift=-0.5cm]{$M_{\rm SC}$, $R_{\rm SC}$, $\gamma_{\rm SC}$};
\node (dyn) [decision, below of=sat, xshift=-0.cm]{DF, expansion, mass loss};
\node (sc) [quantity, below of=dyn, xshift=0.cm]{$M_{\rm SC}$, $R_{\rm SC}$, $r$, $e_{\star}$};
\node (nvp) [io, right of=nsc,xshift=1.2cm]{$M_{\rm SC}<100\ \rm M_{\odot}$?};
\node (insc3) [io, below of=nvp, xshift=2cm]{Host NSC?};
\node (nmg) [io, below of=insc3,xshift=-0cm]{$r<R_{\rm NSC}$?};
\node (evap) [process, below of=nvp,xshift=-0.cm]{Evaporation};
\node (merge) [process, below of=evap, xshift=-1cm]{NSC-SC merger};
\node (final) [process, below of=nmg, xshift=-2cm]{Binary evolution in SCs};
\node (sf) [io, below of=final]{Finish star formation?};
\node (pnode) [startstop, below of=sf]{Parent node (next level)};
\draw [arrow] (asloth) -- (focc);
\draw [arrow] (insc1) -- node[xshift=0.3cm]{No} (nnsc);
\draw [arrow] (focc) -- node[yshift=0.2cm]{No} (insc0);
\draw [arrow] (focc) -- node[xshift=0.3cm]{Yes} (insc1);
\draw [arrow] (insc) -- node[xshift=0.3cm]{Yes} (orbit);
\draw [arrow] (insc1) -- node[yshift=0.2cm]{Yes} (old);
\draw [arrow] (orbit) -- (asloth);
\draw [arrow] (old) -- node[xshift=0.3cm]{Yes}(ensc);
\draw [arrow] (old) -| node[yshift=0.2cm,xshift=-0.8cm]{No} (sat);
\draw [arrow] (insc0) -| node[yshift=-0.5cm,xshift=0.3cm]{Yes} (sat);
\draw [arrow] (nnsc) -- (scale);
\draw [arrow] (ensc) -- (scale);
\draw [arrow] (sat) -- node[xshift=0.4cm]{Loop} (dyn);
\draw [arrow] (dyn) -- (sc);
\draw [arrow] (scale) -- (nsc);
\draw [arrow] (nsc) |- (final);
\draw [arrow] (sc) -- (nvp);
\draw [arrow] (nvp) -- node[xshift=-0.3cm]{Yes} (evap);
\draw [arrow] (nvp) -- node[xshift=0.5cm]{No} (insc3);
\draw [arrow] (insc3) -- node[xshift=0.3cm]{Yes} (nmg);
\node (mid0) [right of=final,xshift=2cm]{No};
\node (mid1) [right of=sf,xshift=3.3cm]{No};
\draw [narrow] (insc3) -| (mid0);
\draw [arrow] (mid0) -- (final);
\draw [arrow] (nmg) -- node[yshift=0.2cm]{Yes} (merge);
\draw [arrow] (merge) -- (nsc);
\draw [arrow] (nmg) -- node[xshift=0.5cm]{No} (final);
\draw [arrow] (final) -- (sf);
\draw [arrow] (cnode) -- (pre);
\draw [arrow] (pre) -- node[xshift=0.3cm]{Yes} (asloth);
\node (mid) [left of=sf, xshift=-2.8cm]{No};
\draw [narrow] (sf) --  (mid);
\draw [arrow] (mid) |- (asloth);
\draw [arrow] (sf) -- node[xshift=0.3cm]{Yes} (pnode);
\draw [arrow] (insc) -- node[xshift=0.6cm]{No} (asloth);
\draw [arrow] (pre) -- node[yshift=0.2cm]{No} (insc);
\draw [narrow] (insc0) -| (mid1);
\draw [arrow] (mid1) -- (sf);
\end{tikzpicture}
    \caption{Same as Fig.~\ref{bbh_process} but for the NSC routine. 
    Here, $M_{\rm SC}$, $R_{\rm SC}$, and $\gamma_{\rm SC}$ denote the mass, size, and density profile inner slope of the SC characterized by a cored Dehnen sphere (Eq.~\ref{dehnen}). The orbit of a normal SC in the host halo is described by the apocentre distance $r$ and eccentricity $e_{\star}$. 
    The green boxes denote the key input physics that evolve the SC properties (cyan boxes), through two pathways: `NSC formation/evolution' and `Normal SCs' (white boxes), detailed in Sec.~\ref{sec:nsc_form} and \ref{sec:sc_evo}, respectively.
    \\ \small$^a$ The orbits of the SCs from the smaller halo during a halo merger event are randomized according to Eq.~\ref{scdis}.}
    \label{nsc_process}
\end{figure}
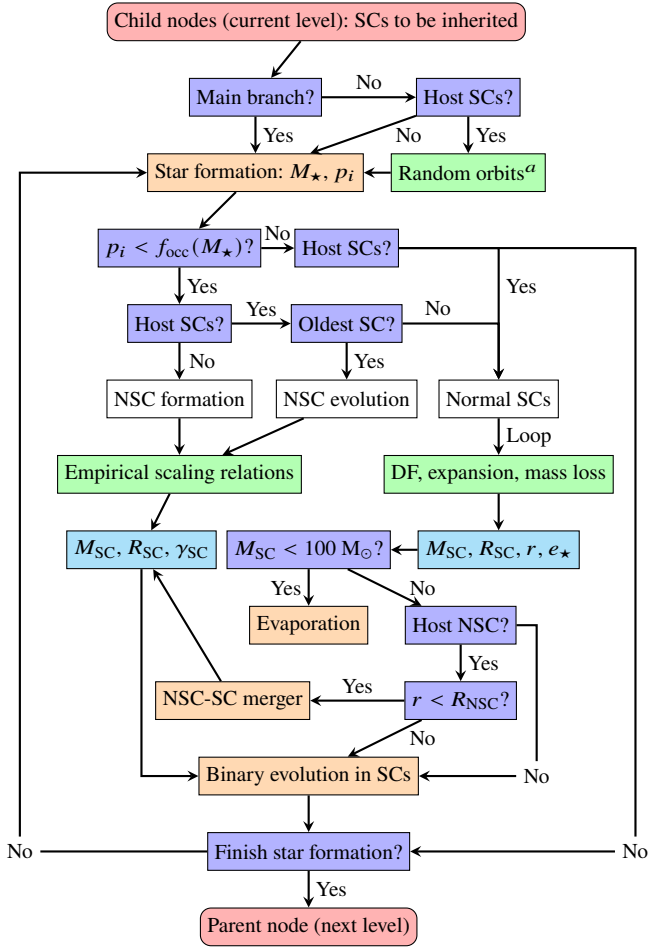

For simplicity, we only consider NSCs and their descendants (e.g., from disrupted satellite galaxies during galaxy mergers) as the potential sites that can provide efficient dynamical hardening of Pop~III binaries. All SCs considered in this work are initially NSCs although later on some of them become normal SCs (see Sec.~\ref{sec:sc_evo}) either as remnant NSCs from stripped satellite galaxies or NSCs disrupted by internal processes captured by the NSC occupation fraction (Eq.~\ref{focc_fit}). 

Since Pop~III star formation typically peaks at $z\sim 10-20$ as predicted by cosmological simulations \citep[e.g.,][]{Tornatore2007,Wise2012,Johnson2013,Muratov2013,Pallottini2014,Xu2016,Sarmento2017,Liu2020did}, we have to consider NSCs across the entire cosmic history of galaxy formation (particularly at $z\gtrsim 1$) to model their interactions with Pop~III BBHs. However, the properties of NSCs (and SCs in general) beyond the local volume and nearby galaxy clusters are poorly constrained in observations. Although the formation and evolution of (high-$z$) SCs (in dwarf galaxies) has been intensively studied with semi-analytical models, $N$-body and (cosmological) hydrodynamic (zoom-in) simulations \citep[e.g.,][]{Devecchi2009,Devecchi2010,Devecchi2012,Renaud2015,Guillard2016,Safranek-Shrader2016,Li2017,Li2018,Li2022sc,Brown2018,Howard2018,Pfeffer2018,Li2019,Ma2020,Lahen2020,Chen2021,McKenzie2021,Fahrion2022,Hislop2022,vanDonkelaar2022,Grudic2023,Lahen2023,Livernois2023,Sameie2023,vanDonkelaar2023,ArcaSedda2024i,ArcaSedda2024iii,Chen2024gc,Gao2024,Gray2024}, 
our understanding of high-$z$ SCs is still limited due to theoretical and numerical problems \citep[see, e.g.,][]{Chen2021,Hislop2022}. 

In the absence of a robust and universal theoretical model for NSCs in the broad redshift range ($z\sim 0-30$) involved in our simulations, we extrapolate the empirical scaling relations and occupation fraction of NSCs in local observations \citep{Neumayer2020nuclear,Pechetti2020} to high redshifts, as detailed in Sec.~2.3 of \citet{Liu2021}. In this work, we incorporate their NSC model into the merger trees of \textsc{a-sloth} to keep track of NSCs on the fly (Sec.~\ref{sec:nsc_form}). We also model the dynamics, mergers and evaporation of normal SCs as NSC descendants (Sec.~\ref{sec:sc_evo}). An illustration of our NSC routine is shown in Fig.~\ref{nsc_process}. 
For simplicity, we have ignored the effects of Pop~III BBHs on SC evolution (e.g., heating via {binary-single} encounters), which are expected to be small because Pop~III BBHs are completely sub-dominant in the NSCs made of Pop~I/II stars considered here (see Sec.~\ref{sec:fd_progenitor}). 
Building a physically motivated model for NSCs, other populations of SCs (e.g., GCs and young star clusters) and SMBHs that are important for NSC formation and evolution in \textsc{a-sloth} is an intriguing direction for future research \citep[see, e.g.,][]{El-Badry2019,Fahrion2022,Fahrion2022gc,DeLucia2023,Park2022gc,Polkas2023,Chen2024gc,Gao2024,Kaur2024}. 

\subsection{Formation and evolution of nuclear star clusters}
\label{sec:nsc_form}
In our model, formation of NSCs (in galaxies with $M_{\star}>M_{\star,\rm min}$) is governed by the input occupation fraction of NSCs as a function of galaxy mass $f_{\rm occ}(M_{\star})\equiv f_{\rm occ}(\tilde{M}_{\star})$, $\tilde{M}_{\star}\equiv\log(M_{\star}/\rm M_{\odot})$. We adopt the fit formula in \citet[see their fig.~8]{Liu2021} 
\begin{align}
\hat{f}_{\rm occ}&\simeq
\begin{cases}
    0.15\ , &\tilde{M}_{\star}< 6.6\ ,\\
    0.15+0.36(\tilde{M}_{\star}-6.6)\ ,  &6.6\le\tilde{M}_{\star}<8.2\ ,\\
    0.72\ , &8.2\le\tilde{M}_{\star}<9.5\ ,\\
    0.72-0.185(\tilde{M}_{\star}-9.5)\ , &9.5\le\tilde{M}_{\star}\ ,
\end{cases}\label{focc_fit}    
\end{align}
for our fiducial model (labelled by `\_obs'.), based on observations of nearby galaxies in all environments\footnote{$f_{\rm occ}$ is generally higher in denser environments \citep{Neumayer2020nuclear,Hoyer2021,Zanatta2024}. For instance, the $f_{\rm occ}$ measured in the Virgo cluster is higher than that in the local ($\lesssim 11\ \rm Mpc$) volume by $\sim 20\%$ at $M_{\star}\sim 10^{7}-10^{9}\ \rm M_{\star}$ \citep{Hoyer2021}. Our fiducial model $\hat{f}_{\rm occ}$ may slightly overestimate the cosmic average by including observations of galaxies in all environments.} \citep{Neumayer2020nuclear}. We further impose a low-mass truncation at $M_{\star,\min}=10^{6}\ \rm M_\odot$, i.e., $f_{\rm occ}=0$ for $M_{\star}\le M_{\star,\min}$, consistent with our modeling of DF. Here we also ignore the redshift evolution of $f_{\rm occ}$ for simplicity. 

In the GC accretion scenario of NSC formation relevant for low-mass ($M_{\star}\lesssim 10^{9}\ \rm M_{\odot}$) galaxies, 
$f_{\rm occ}$ is sensitive to galaxy sizes, so the occupation fraction can be higher if the majority of NSC formation happens before the dominant epoch of size growth at $z\sim 2$ \citep[see their fig.~6]{Leaman2022}. 
In fact, galaxies tend to be more compact at higher $z$ according to recent JWST observations \citep[e.g.,][]{Baggen2023,Langeroodi2023,Ito2023,Ji2024,Ormerod2024} and cosmological (zoom-in) simulations \citep[e.g.,][]{Roper2023}, which support the inside-out scenario of galaxy formation. 
Besides, it is found in the Romulus cosmological simulations that dwarf galaxies formed earlier are more likely to host massive BHs \citep{Tremmel2023}, which is consistent with the enhanced NSC occupation observed in galaxy clusters. These outcomes imply that the dense environments at high-$z$ favour BH and NSC formation. On the other hand, it is also likely that high-$z$ galaxies are typically irregular and clumpy where DF is ineffective, as predicted by e.g., the simulations in \citet{Ma2021} for galaxies with $M_{\star}\lesssim 10^{10}\ \rm M_{\odot}$ at $z\gtrsim 5$ \citep[see also][]{Biernacki2017,Pfister2019,Bortolas2020}\footnote{
Recent JWST observations find that (purely) irregular galaxies do not dominate at $z\gtrsim 3$ (especially for $z\lesssim 5$) as previously thought \citep{Ferreira2023,Kartaltepe2023}. In particular, disc galaxies dominate the galaxy populations with $M_{\star}\gtrsim 10^{9}\ \rm M_{\odot}$ at $z\sim 0.6-8$, while the fraction of irregular galaxies remains below $\lesssim 20\%$, consistent with cosmological simulations \citep{Lee2021,Lee2023,Park2022gal}. However, one should be cautious with the interpretations of observational data in comparison with galaxy formation simulations \citep[see, e.g.,][]{Vega-Ferrero2024}.}. 

In light of this, in addition to the fiducial model, we also consider two extreme cases with higher and lower NSC occupation fractions. The three NSC models are summarized as follows.
\begin{itemize}
    \item In the fiducial model (labelled by `obs'), we have $f_{\rm occ}=\hat{f}_{\rm occ}$ and $M_{\star,\min}=10^{6}\ \rm M_{\odot}$ based on local observations.
    \item In the optimistic model (labelled by `full'), we consider full occupation with $f_{\rm occ}(M_{\star})\equiv 1$ for $M_{\star,\min}=10^{6}\ \rm M_{\odot}$.
    \item In the conservative model (labelled by `low'), we reduce the efficiency of DF\footnote{In this scenario, NSCs can still form without DF in `low-mass' galaxies ($M_{\star}\lesssim 10^{10}\ \rm M_{\odot}$) at high $z$ via in-situ star formation \citep[e.g.,][]{Guillard2016,Brown2018,Howard2018,Gray2024}. However, without DF, Pop~III BBHs can hardly fall into such NSCs, so they are irrelevant to our work. Therefore, we do not track NSCs in galaxies with $M_{\star}<M_{\star,\min}$.} (as well as NSC formation) with $f_{\rm occ}=\hat{f}_{\rm occ}$ and $M_{\star,\min}=\min\{1,[(1+z)/6]^{5.14}\}\times 10^{10}\ \rm M_{\odot}$, assuming that $M_{\star,\min}$ evolves from $10^{10}\ \rm M_{\odot}$ at $z=5$ to $10^{6}\ \rm M_{\odot}$ at $z=0$ following a power-law of $(1+z)$.
\end{itemize}

Given the input $f_{\rm occ}(M_{\star})$, we first generate a random number from a uniform distribution in $[0,1]$ for each \textit{leaf} of the merger tree. This random number is then inherited along the main branch such that every node $i$ has a random number $p_{i}$. At each (star formation) timestep, we check the criterion $p_{i}<f_{\rm occ}(M_{\star,i})$, which indicates the presence of a NSC, where $M_{\star,i}$ is the current galaxy mass of node $i$. Given $p_{i}<f_{\rm occ}(M_{\star,i})$, if this node contains no SCs, 
we create a new SC object as its NSC, whose properties (i.e., mass $M_{\rm SC}$, size $R_{\rm SC}$, and inner slope of density profile $\gamma_{\rm SC}$, see Eq.~\ref{dehnen}) are derived from the empirical NSC-galaxy scaling relations\footnote{More recent results for the (local) NSC-galaxy scaling relations are available in \citet{Hannah2024}, which show higher NSC occupation fractions ($f_{\rm occ}\simeq 1$ for $M_{\star}\sim 10^{6.5}-10^{9.5}\ \rm M_\odot$) than those adopted in our fiducial model based on \citet{Neumayer2020nuclear}. Their impact on our final results is expected to be within the range covered by the three NSC models considered here. We plan to consider these new relations in future work.}, detailed in Sec.~2.3 of \citet{Liu2021} based on the observational data compiled in \citet{Neumayer2020nuclear} and \citet{Pechetti2020}, given the host galaxy stellar mass $M_{\star,i}$ as the input\footnote{The scatter in NSC-galaxy scaling relations are also implemented with random variables inherited along main branches, similar to the NSC occupation and scatter of galaxy size. }. 
If the node already contains one or multiple SC(s) but does not host a NSC in the previous timestep, we convert the oldest SC (remnant) along the main branch into the NSC with properties also specified by the NSC-galaxy scaling relations, regardless of the status of this SC (see Sec.~\ref{sec:sc_evo} below). If the node already hosts a NSC, we simply update the NSC properties according to the scaling relations that reflect the variation of $M_{\star,i}$, \textit{assuming adiabatic evolution of NSCs regulated solely by host galaxy mass}. A NSC host galaxy will lose its NSC if $p_{i}>f_{\rm occ}(M_{\star,i})$, which will only happen for $M_{\star}> 10^{9.5}\ \rm M_{\odot}$. In this case, the original NSC is not removed but turned into a normal SC that no longer follows the NSC-galaxy relations and will be evaporated and redistributed during galaxy mergers (see below). In reality, such internal disruption of NSCs can be caused by merging with supermassive black holes, which is more common in more massive (elliptical) galaxies \citep{Neumayer2020nuclear}, as reflected in the decrease of $\hat{f}_{\rm occ}$ with $M_{\star}$ for $M_{\star}> 10^{9.5}\ \rm M_{\odot}$. 

In our model, any snapshot of the merger tree(s) satisfies (statistically) the input occupation fraction and NSC-galaxy scaling relations. During a halo/galaxy merger, the parent node inherits the SCs from child nodes and only one of them can be the NSC of the post-merger galaxy. 
The rest are normal SCs, i.e., NSCs from stripped satellite galaxies \citep{Wang2023}, which are subject to orbital redistribution, DF, tidal forces, and internal relaxation heating, so they can merge with NSCs or evaporate, as discussed below.

\subsection{Dynamics and evolution of normal star clusters}
\label{sec:sc_evo}
Similar to Pop~III BBHs, the orbit of a normal SC of mass $M_{\rm SC}$ also shrinks and its eccentricity $e_{\star}$ decays under DF according to Eq.~\ref{df}, now replacing $m$ with $M_{\rm SC}$. To integrate these equations in \textsc{a-sloth}, we must specify the conditions after galaxy mergers. We assume that the orbits of SCs coming from the larger (target) halo remain unchanged by the merger while the SCs from the smaller (satellite/infalling) halo are redistributed 
in the post-merger halo following a cored power-law profile:
\begin{align}
 n_{\rm SC}(r)=
\begin{cases}
    n_{0}\ , &r\le R_{\rm core}\ ,\\
    n_{0}(r/R_{\rm core})^{-3}\ ,  &r>R_{\rm core}\ ,
\end{cases}\label{scdis}
\end{align}
where $n_{0}$ is fixed by $4\pi\int_{0}^{R_{\rm vir}}n_{\rm SC}(r)r^{2}dr=N_{\rm SC}$, given the halo virial radius $R_{\rm vir}$ and number of (satellite) SCs $N_{\rm SC}$. Also, $R_{\rm core}=\min(31.62R_{\star}M_{\star, 11}^{1/6}, R_{\rm vir})$ is the core size, which is the minimum of the halo virial radius and total galaxy radius given by the size-mass relation in \citet[see their eqs.~18-19]{Arca-Sedda2014} with a scatter of 0.36 dex. An inner core of SCs with a uniform distribution is supported by the GC accretion model in \citet{Arca-Sedda2014} that can reproduce the observed relations between NSC mass, galaxy mass and velocity dispersion \citep{Erwin2012,Leigh2012,Scott2013}, while the power-law form of the outer region with a slope of $-3$ is motivated by the asymptotic feature of the NFW profile at $r\rightarrow \infty$, assuming that SCs and dark matter follow similar spatial distributions at halo outskirts. 

Now for each SC from the smaller halo, we draw its apocentre distance $r$ in the post-merger halo from a probability distribution $p(r)=4\pi n_{\rm SC}(r)r^{2}/N_{\rm SC}$. The eccentricity $e_{\star}$ is again generated from a uniform distribution in $[0,1)$ \citep{Arca-Sedda2014}. This orbit randomization process for infalling SCs is supported by $N$-body simulations \citep[e.g.,][]{Pagnini2023}. Here we assume instantaneous disruption of infalling galaxies so it is always the SC mass that is used to evaluate Eq.~\ref{df}, even for the NSC of the infalling galaxy. 

Since our modelling is highly simplified under spherical symmetry, we only consider mergers between NSCs and normal SCs, not among normal SCs themselves. A merger happens when a SC gets too close to a NSC with $r<R_{\rm SC}$. 
Immediately after the merger, the Pop~III BBHs contained in the normal SC are placed at $r=R_{\rm SC}$ and their external orbital eccentricities $e_{\star}$ are set to that of the infalling SC. For simplicity, we do not model the effects of individual merger events on NSC properties. We assume that the long-term effects of mergers are already captured by the empirical scaling relations \citep{Neumayer2020nuclear} that govern NSC evolution in our model. 

In addition to dynamics, we also calculate the mass loss and expansion of normal SCs under external tidal fields and internal relaxation by\footnote{It is possible to calculate the mass loss of SCs more accurately based on simulation results \citep[see, e.g.,][]{Madrid2017,Gieles2023}, which is particularly important for the formation and growth of NSCs by SC mergers. Since modelling such processes is beyond the scope of our study, we adopt this simple equation to capture the long-term decrease of dynamical hardening efficiency during SC relaxation. }
\begin{align}
    \frac{dM_{\rm SC}}{dt}&\simeq-\frac{M_{\rm SC}}{t_{\rm dis}}\ ,\label{dmsc}\\
    \frac{d R_{\rm SC}}{dt}&\simeq\frac{2}{3}\frac{R_{\rm SC}}{t_{\rm rh}}\ ,\label{drsc}
\end{align}
given the (half-mass) two-body relaxation timescale $t_{\rm th}$, and the overall dissolution timescale $t_{\rm dis}$ which is estimated with the minimum of the two timescales of internal and external effects \citep{PortegiesZwart2010}:
\begin{align}
\begin{split}
    &t_{\rm rh}=200\ {\rm Myr}\left(\frac{M_{\rm SC}}{10^{6}\ \rm M_{\odot}}\right)^{1/2}\left(\frac{R_{\rm SC}}{\rm pc}\right)^{3/2}\left(\frac{m_{\star,\rm SC}}{\rm M_\odot}\right)^{-1}\frac{10}{\ln\Lambda'}\ , 
\end{split}\label{tdisi}\\
\begin{split}
    &t_{\rm dis}=\min(t_{\rm dis, relax},t_{\rm dis,tidal})\ ,\quad t_{\rm dis, relax}\approx 30t_{\rm rh}\ ,\\
    &t_{\rm dis, tidal}\approx 2\ {\rm Myr}\\
    &\quad\quad\times \left(\frac{N}{\ln\Lambda'}\right)^{3/4}\left(\frac{r_{\rm g}}{\rm kpc}\right)\left(\frac{v_{\rm g}}{220\ \rm km\ s^{-1}}\right)^{-1}(1-e_{\star})\ ,
\end{split}\label{tdise}
\end{align}
where $\ln\Lambda'=0.11 N$, $r_{\rm g}=\max(r, R_{\rm SC})$, 
$v_{\rm g}=\sqrt{GM_{\star}/(0.8R_{\star})}$ is the typical circular velocity of the host galaxy of mass $M_{\star}$ and scale length $R_{\star}$ \citep[see their eq.~18]{Arca-Sedda2014}, and $N=M_{\rm SC}/m_{\star,\rm SC}$ is the total number of stars in the cluster. In Eq.~\ref{drsc}, we only consider long-term expansion driven by internal heating, which leads to an asymptotic solution $R_{\rm SC}\propto t^{2/3}$ at $t\rightarrow\infty$ \citep{PortegiesZwart2010}, if ignoring mass loss, given $t_{\rm th}\propto R_{\rm SC}^{3/2}$. 
For simplicity, we assume that $\gamma_{\rm SC}$ remains constant. A SC evaporates 
when $M_{\rm SC}$ drops below $100\ \rm M_{\odot}$. At this moment, the Pop~III remnants contained in it 
inherit the orbit of the SC in the galaxy.
All NSC properties are updated in timesteps no larger than $\delta t_{\rm NSC}=1\ \rm Myr$, and sub-cycling is introduced within star formation timesteps if needed. 


\section{Low-redshift extrapolation}
\label{sec:extra}

Cosmological simulations that resolve Pop~III star-forming minihaloes in large representative volumes down to $z=0$ are still prohibitively expensive. 
Therefore, cosmological simulations resolving minihaloes typically have small volumes and are only representative at high redshifts. For instance, the simulation data from \citet{Ishiyama2016} adopted in our \textsc{a-sloth} runs provides merger trees for every halo in a co-moving volume of $(8\ h^{-1}\rm Mpc)^{3}$ with a mass resolution of $5000\ h^{-1}\rm M_{\odot}$ down to $z_{\rm f}\simeq 4.5$ when the box is marginally representative. We design an extrapolation scheme to follow the evolution of Pop~III BBHs and their host galaxies/haloes further down to $z=0$ from the last simulation snapshot ($z=z_{\rm f}$), based on the halo growth model in \citet{Fakhouri2010} and the stellar-halo mass relation (SHMR) in \citet{Behroozi2019}.


We assume that all haloes, except for significantly stripped (satellite) ones, remain quasi-isolated from $z_{\rm f}$ and grow smoothly at the cosmic average rate as a function of halo mass $M_{\rm h}$ and redshift $z$, inferred from the Millennium cosmological simulations \citep{Boylan-Kolchin2009} for $\Lambda\rm CDM$ \citep{Fakhouri2010}:
\begin{align}
\begin{split}
    \frac{dM_{\rm h}}{dt}&=46.1\ \mathrm{M_{\odot}\ yr^{-1}}\left(\frac{M_{\rm h}}{10^{12}\ \rm M_{\odot}}\right)^{1.1}\\
    & \times (1+1.11z)\sqrt{\Omega_{m}(1+z)^{3}+\Omega_{\Lambda}}\ .
\end{split}\label{dmhalo}
\end{align}
Once the halo mass $M_{\rm h}(z)$ is known, we can derive the galaxy (stellar) mass with
\begin{align}
M_{\star}(z)=(B+1)M_{\star,\rm SHMR}(M_{\rm h}(z),z)\ ,\label{ms_post}
\end{align}
where $M_{\star,\rm SHMR}(M_{\rm h},z)$ is given by the SHMR fitting formula for the true stellar mass values including both star-forming and quiescent galaxies and excluding intrahalo light in \citet[see column 8 of their table J1]{Behroozi2019}, and $B$ is a normalization factor. Here, we extrapolate the fitting formula to low-mass haloes with a minimum star formation efficiency of $\eta\equiv M_{\star}\Omega_{m}/(\Omega_{b}M_{\rm h})=10^{-3}$. In this way, we ignore the fluctuations of the star formation rate with respect to the cosmic average caused by different halo assembly histories and baryon cycles\footnote{There is evidence from recent observations that fluctuations of star formation rates are weak for intermediate-mass galaxies ($M_{\star}\sim 10^{9}-10^{10}\ \rm M_{\odot}$) at $z\sim 0.3-0.4$ \citep{Patel2023}.} under the same halo mass. Such stochastic effects are captured by \textsc{a-sloth} and can cause an offset between $M_{\star,\rm SHMR}$ and $M_{\star}(z_{\rm f})$ predicted by \textsc{a-sloth} at $z_{\rm f}$. Assuming that this offset decays after cosmic noon at $z_{\rm noon}=2$, for each halo we evolve $B$ from the initial condition $\left[(B+1)M_{\star,\rm SHMR})\right]|_{z=z_{\rm f}}=M_{\star}(z_{\rm f})$ with
\begin{align}
\frac{dB}{dt}=
\begin{cases}
    0\ , &z>z_{\rm noon}\ ,\\
    -\xi\frac{B}{t_{H}}\ , &z\le z_{\rm noon}\ ,
\end{cases}\label{bsf}
\end{align}
where $t_{H}=1/H(t)$ given the Hubble parameter at $H(t)$, and $\xi=2$ is a parameter that governs the decay rate. A satellite halo is regarded as significantly stripped when $B+1>B_{\max}+1=25$, whose halo/galaxy mass remains constant\footnote{The parameter $B+1$ measures the mass fraction of stars in the halo, and $B\sim 25$ means that stars make up the entire halo mass under a cosmic average star formation efficiency $\eta\equiv M_{\star,\rm SHMR}\Omega_{m}/(\Omega_{b}M_{\rm h})\sim 0.2$, or at least the entire baryonic mass for $\eta\gtrsim 0.03$. Here $\eta\sim 0.03-0.2$ is typical for galaxies with $M_{\star}\gtrsim 10^{9}\ \rm M_{\odot}$ at $z\lesssim 4$ \citep{Behroozi2019}.}. 
We also consider that in the post-reionization epoch ($z<6$), haloes with $M_{\rm h}<6.7\times 10^{8}\ \mathrm{ M_{\odot}}\ [(1+z)/5]^{-3/2}$ cannot form new stars according to the reionization models in \citet{Pawlik2015,Pawlik2017}, \citet{Benitez-Llambay2020} and \citet{Hutter2021}. These haloes still grow with Eq.~\ref{dmhalo} but their stellar masses are no longer updated with Eq.~\ref{ms_post}. Besides, we impose an upper bound to galaxy mass (growth) $M_{\star}<\eta_{\max}(\Omega_{b}/\Omega_{m})M_{\rm peak}$, given $M_{\rm peak}$ the maximum mass the halo has ever reached and $\eta_{\max}=0.5$ the maximum star formation efficiency. 

\begin{figure}
    \centering
    \includegraphics[width=1\columnwidth]{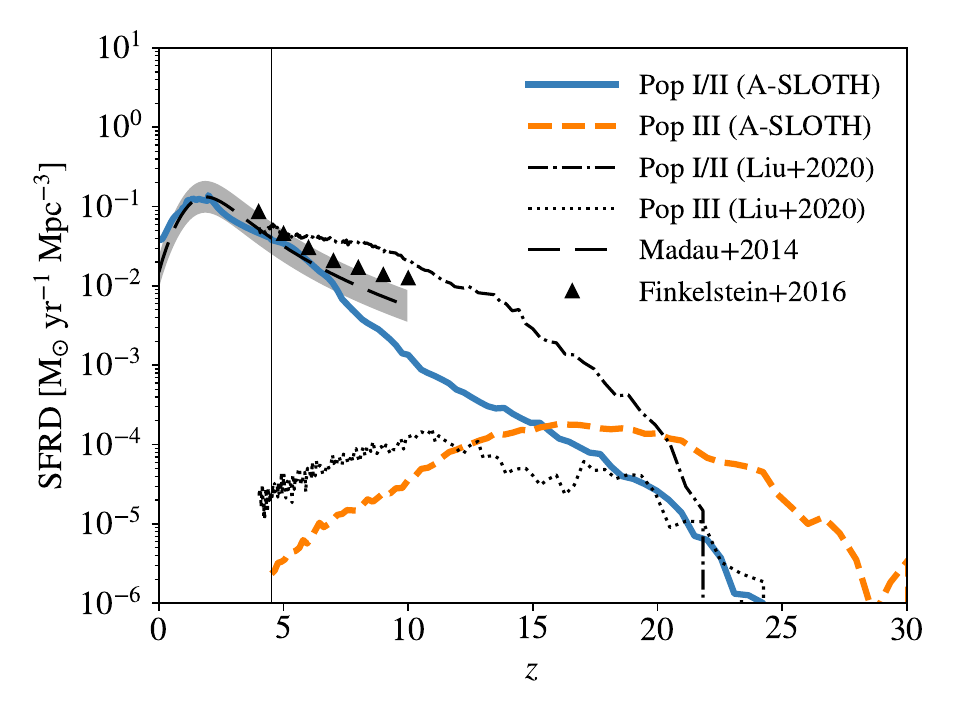}
    \vspace{-20pt}
    \caption{Co-moving SFRDs of Pop~I/II (solid) and Pop~III stars (dashed), predicted by the fiducial Pop~III IMF ($\alpha=1$) and the best-fit star formation and stellar feedback parameters in \textsc{a-sloth} \citep[see their table~3]{Hartwig2023} coupled with our extrapolation scheme, applied on the merger trees from the cosmological simulation by \citet{Ishiyama2016}. The observational results in \citet{Madau2014}, inferred from UV and IR galaxy surveys such as \citet[data points]{Finkelstein2016}, are plotted as the long-dashed curve (with a scatter of 0.2 dex embodied by the shaded region). For comparison, we also show the Pop~I/II and Pop~III SFRDs from \citet{Liu2020gw} with the dash-dotted and dotted curves, respectively. The thin vertical line denotes the final redshift $z_{\rm f}\simeq 4.5$ above which merger trees are constructed from the cosmological simulation. }
    \label{sfrd}
\end{figure}

We adopt fixed (star formation) timesteps of $\delta t_{\rm post}=1\ \rm Myr$ for the low-$z$ evolution of galaxies/haloes with the above formalism (Eqs.~\ref{dmhalo}-\ref{bsf}) where the values of $z_{\rm noon}$, $\xi$, $B_{\max}$ and $\eta_{\max}$ are chosen to reproduce the 
cosmic star formation rate density (SFRD) at $z\lesssim 6$ in observations \citep[e.g.,][]{Madau2014,Finkelstein2016}. 
For instance, Fig.~\ref{sfrd} shows the star formation histories of Pop~III and Pop~I/II stars predicted by the default star formation and stellar feedback parameters in \textsc{a-sloth} \citep[see their table~3]{Hartwig2022} coupled with our extrapolation scheme (Eqs.~\ref{dmhalo} and \ref{ms_post}), based on the merger trees from the cosmological simulation in \citet{Ishiyama2016}, for the fiducial Pop~III IMF ($\alpha=1$, see Sec.~\ref{sec:sample}). 
The Pop~I/II SFRD, which dominates the total SFRD at $z\lesssim 15$ in our model, is slightly overestimated 
at $z\rightarrow 0$ compared with observations. The reason is that we did not include explicitly galaxy mergers and mechanisms of quenching (e.g., AGN feedback and environmental quenching) in the low-$z$ regime. Nevertheless, by experimenting with other star formation models, we find that varying the late-time ($z\lesssim 1$) SFRD by up to a factor of $\sim 10$ has little impact on Pop~III BBH mergers. 

Given the star formation histories $M_{\star}(t)$ of individual haloes, all galaxy and NSC properties can be derived, which set the background for the evolution of SCs and Pop~III binaries as explained in the previous subsections and \citet{Liu2021}. 
In reality, haloes are not isolated and will merge into larger haloes. This process disrupts the inspiral of Pop~III BBHs in satellite galaxies but may enhance DF during galaxy mergers \citep{Roman2023}. Galaxy mergers also strip satellite galaxies, so their NSCs can no longer grow and become vulnerable to tidal forces from the central galaxy, reducing their ability to facilitate Pop~III mergers. Our results for the NSC infall rate and merger rate in NSCs at $z<z_{\rm f}$ are likely rather optimistic.

To better characterize the host galaxies of Pop~III mergers at $z<z_{\rm f}$, we also estimate the average gas-phase metallicity $Z$ of each galaxy by extrapolating from the value at $z=z_{\rm f}$ predicted by \textsc{a-sloth} with the redshift-dependent mass-metallicity relation $Z\propto M_{\star}^{0.3}(1+z)^{-0.9}$ based on \citet{Langeroodi2022}.

\section{Results} 
\label{sec:res}

We apply \textsc{a-sloth} to the merger trees constructed from the cosmological simulation by \citet{Ishiyama2016} in a co-moving volume of $V_{\rm com}=(8\ h^{-1}\rm Mpc)^{3}\simeq 1650\ Mpc^{3}$ with a dark matter mass resolution of $5000\ h^{-1}\rm M_{\odot}$, which only cover $z \ge z_{\rm f}\simeq 4.5$ where the box size is large enough to be marginally representative. We use the extrapolation scheme described in Sec.~\ref{sec:extra} to follow the subsequent evolution down to $z=0$. 
Throughout this work, we adopt the cosmological parameters from \citet{planck}\footnote{$\Omega_{m}=0.3089$, $\Omega_{b}=0.0486$, 
and $H_{0}=100h\ \rm km\ s^{-1}\ Mpc^{-1}$ with $h=0.6774$} and the default \textsc{a-sloth} star formation and feedback parameters \citep[see their table~3]{Hartwig2022}. Following \citet{Liu2021binary}, \citet{Liu2021} and \citet{Hartwig2023}, the mass fraction of Pop~III stars in binaries is fixed as $f_{\rm B}=0.7$. 
Since various stochastic processes are involved in our modelling and the simulation volume underlying the merger trees is not very large\footnote{The total mass of Pop~III stars formed in the original simulation volume is $\sim 7.6-9.4\times 10^{7}\ \rm M_{\odot}$, which only corresponds to $\sim 2.4-8.4\times 10^{4}$ BBHs.}, we boost the number of Pop~III binaries to be sampled in \textsc{a-sloth} by a factor of $f_{\rm boost}=50$ to achieve better statistics and reduce numerical noise in the merger history, so the effective simulation volume $\tilde{V}_{\rm com}=50 V_{\rm com}$ is used to calculate the cosmic average quantities from our merger populations, assuming that the actual simulation volume in \citet{Ishiyama2016} is cosmologically representative. 

We consider 18 models combining different choices of the Pop~III IMF, IBS, and NSC parameters (defined in Sec.~\ref{sec:sample} and \ref{sec:nsc_form}), as summarised in Table~\ref{runlist} (see also Table~\ref{tobs} and \ref{tmass} for the key properties of Pop~III mergers in these models). We divide Pop~III mergers into two groups based on their sites: galaxy fields and NSCs. Here and henceforth, we use the term `NSC' to denote all SCs including both NSCs and NSC descendants (i.e., normal SCs) defined in Sec.~\ref{sec:nsc} if not specially clarified. 
Most mergers in galaxy fields come from BBHs that evolve in isolation throughout their lifetimes, representative for the IBSE channel, except for a tiny ($\lesssim 2\%$) fraction of mergers from BBHs returned to galaxy fields from NSCs (via ejection by {binary-single} encounters or SC evaporation/disruption), which are influenced by the NSC-DH channel. We do not exclude these special field mergers when discussing the features of the IBSE channel. 
In this section, we first present the results for the fiducial model and discuss the general properties of Pop~III BBH mergers (Sec.~\ref{sec:fid}), in comparison with those of mergers of Pop~I/II BHs and primordial back holes (PBHs) predicted by \citet{Franciolini2022} and \citet{Bavera2022} based on low-$z$ observations \citep{Abbott2021,Abbott2023}. 
Next, we explore the dependence of our results on the parameters of Pop~III IMF, IBS, and NSCs (Sec.~\ref{sec:comp}). Finally, in Sec.~\ref{sec:et}, we briefly discuss the observational perspective of Pop~III BBH mergers, focusing on the detection rates of Pop~III BBH mergers by ET\footnote{We use the ET-D-sum sensitivity model from \url{https://www.et-gw.eu/index.php/etsensitivities} \citep[see also][]{Hild2008,Hild2010,Hild_2011,Branchesi2023}.} and the LVK network during O4\footnote{We use the sensitivity model for advanced LIGO from \url{https://dcc.ligo.org/LIGO-T2000012-v1/public} (\texttt{aligo\_O4high.txt}) to represent the maximum capacity of the LVK network during O4.} 
using the \texttt{python} package \textsc{gwtoolbox}\footnote{\url{https://bitbucket.org/radboudradiolab/gwtoolbox/src/master/}} \citep{Yi2022a,Yi2022b}, and current observations of BBH mergers involving massive BHs (like GW190521). 

Following \citet{Bavera2022}, we adopt the best-fit model in \citet{Franciolini2022} as the reference to evaluate the Pop~III contributions to the GW signals of BBH mergers. This model combines PBH mergers with the Pop~I/II BBH mergers formed in the IBSE channel via common envelope (CE) and stable mass transfer (SMT), as well as dynamically in GCs, with branching ratios inferred from the events in the second gravitational wave transient catalogue \citep[GWTC-2,][]{Abbott2021}. 
According to this model, most ($\sim 94$\%) events in GWTC-2 are explained by Pop~I/II BBH mergers, and the remaining ones (especially the special event GW190521) are attributed to PBHs that follow a log-normal distribution with a characteristic mass of $\sim 35\ \rm M_\odot$ and a width of $\sigma=0.41$, and make up $f_{\rm PBH}\sim 2\times 10^{-4}$ of dark matter \citep[see their fig.~4]{Franciolini2022}. Here, the contribution of PBHs to the observed events is statistically insignificant and subject to uncertainties in the modelling of astrophysical populations. 
Considering every possible origin of BBH mergers for comparison is beyond the scope of this work, and we do not expect the reference model to be the most representative model for the populations of BBH mergers other than that from Pop~III stars. 
In fact, large uncertainties remain in current theoretical predictions on the properties of BBH mergers, 
so inferences about the origins of BBH mergers and their branching ratios from the limited sample of detected events are highly model-dependent and tentative. 
The reference model considered here produces relatively conservative predictions for the merger rate and SGWB of BBH mergers involving Pop~I/II BHs and PBHs compared with other models in the literature. 
Therefore, our results on the relative contributions of Pop~III mergers should be regarded as heuristic 
optimistic estimates (given the specific Pop~III star formation histories predicted by \textsc{a-sloth}) meant to illustrate the special features of Pop~III BBH mergers, and the exact values can be off by a factor of a few. For simplicity, we focus on the intrinsic properties of entire BBH merger populations at $z>0$ when comparing our results with those in the reference model \citep{Franciolini2022,Bavera2022}. 
We plan to investigate the detailed (redshift-dependent) observational signatures of BBH mergers of different origins and evolution channels in future work fully taking into account source detectability and characterization \citep[see, e.g.,][]{Tanikawa2022,Santoliquido2023}. 



\begin{table*}
\caption{Summary of models. Column 1 gives the model name. 
Columns 2, 3, 4, and 5 define the initial condition model underlying the input \textsc{sevn} catalogue, in terms of the initial distributions of primary stellar mass, mass ratio, orbital period, and eccentricity, respectively (see Sec.~\ref{sec:sample}). Here the primary star IMF is specified with the power-law slope at the high-mass end (given the fixed ZAMS mass range $m_{\star,1}\in [5,550]\ \rm M_\odot$), which is identical to that of the Pop~III IMF adopted in \textsc{a-sloth} to model stellar feedback. Columns 6 and 7 define the NSC (and DF) model with the NSC occupation fraction as a function of galaxy (stellar) mass and the minimum galaxy mass for DF and NSC formation at $z>5$. See Sec.~\ref{sec:nsc_form} for a detailed description of how these parameters are used to determine whether Pop~III BBHs can fall into the NSC by DF in a galaxy. Column 8 shows the BBH formation efficiency $\epsilon_{\rm BBH}$, i.e., the number of BBHs formed per unit stellar mass. Similarly, Column 9 shows the BBH merger efficiency $\epsilon_{\rm GW}^{\rm all\ (field/NSC)}$, i.e., the number of BBH mergers at $z>0$ per unit stellar mass, for all (field/NSC) mergers. Column 10 shows the fraction of mergers in NSCs $f_{\rm NSC}=\epsilon_{\rm GW}^{\rm NSC}/\epsilon_{\rm GW}^{\rm all}$. Column 11 shows the fraction $f_{\rm GW}^{\rm field/NSC}$ of BBHs that merge at $z>0$ in galaxy fields/NSCs. The last column shows the fraction $f_{\rm infall}$ of Pop~III BBHs that fall into NSCs.} 
    \centering
    \begin{tabular}{ccccccccccccccc}
    \hline
    Model & $\alpha$ & $p(q)$ & $p(\pi)$ & $p(e)$ & $f_{\rm occ}$ & $M_{\star,\min}$ & $\epsilon_{\rm BBH}$ & $\epsilon_{\rm GW}^{\rm all\ (field/NSC)}$ & $f_{\rm NSC}$ & $f_{\rm GW}^{\rm field/NSC}$ & $f_{\rm infall}$ & \\
    & & & & & & $[\rm M_{\odot}]$ & $[10^{-4}\ \rm M_{\odot}^{-1}]$ & $[10^{-5}\ \rm M_{\odot}^{-1}]$ & & \\
    \hline 
LOG1\_obs & 1 & \citetalias{Sana2012} & \citetalias{Sana2012} & \citetalias{Sana2012} & $\hat{f}_{\rm occ}$ & $10^{6}$ & 4.78 & 6.49 (5.33/1.16) & 17.9\% & 11.6\%/62.8\% & 3.87\% \\
TOP1\_obs & 0.17 & \citetalias{Sana2012} & \citetalias{Sana2012} & \citetalias{Sana2012} & $\hat{f}_{\rm occ}$ & $10^{6}$ & 2.58 & 2.38 (1.66/0.722) & 30.3\% & 6.73\%/64.1\% & 4.36\% \\
KRO1\_obs & 2.3 & \citetalias{Sana2012} & \citetalias{Sana2012} & \citetalias{Sana2012} & $\hat{f}_{\rm occ}$ & $10^{6}$ & 7.05 & 11.8 (10.2/1.57) & 13.3\% & 15.1\%/60.6\% & 3.67\% \\
LOG5\_obs & 1 & \citetalias{Stacy2013} & \citetalias{Stacy2013} & $2e$ & $\hat{f}_{\rm occ}$ & $10^{6}$ & 10 & 3.64 (0.685/2.95) & 81.2\% & 0.716\%/61\% & 4.82\% \\
TOP5\_obs & 0.17 & \citetalias{Stacy2013} & \citetalias{Stacy2013} & $2e$ & $\hat{f}_{\rm occ}$ & $10^{6}$ & 7.94 & 2.76 (0.283/2.48) & 89.8\% & 0.376\%/62.5\% & 4.99\% \\
KRO5\_obs & 2.3 & \citetalias{Stacy2013} & \citetalias{Stacy2013} & $2e$ & $\hat{f}_{\rm occ}$ & $10^{6}$ & 8.9 & 3.14 (0.755/2.38) & 75.9\% & 0.888\%/58.6\% & 4.56\% \\
\hline 
LOG1\_full & 1 & \citetalias{Sana2012} & \citetalias{Sana2012} & \citetalias{Sana2012} & 1 & $10^{6}$ & 4.78 & 7.98 (5.27/2.71) & 34\% & 12.2\%/58.9\% & 9.65\% \\
TOP1\_full & 0.17 & \citetalias{Sana2012} & \citetalias{Sana2012} & \citetalias{Sana2012} & 1 & $10^{6}$ & 2.58 & 3.33 (1.64/1.68) & 50.6\% & 7.14\%/60.3\% & 10.8\% \\
KRO1\_full & 2.3 & \citetalias{Sana2012} & \citetalias{Sana2012} & \citetalias{Sana2012} & 1 & $10^{6}$ & 7.05 & 13.7 (10.1/3.64) & 26.5\% & 15.7\%/56.4\% & 9.14\% \\
LOG5\_full & 1 & \citetalias{Stacy2013} & \citetalias{Stacy2013} & $2e$ & 1 & $10^{6}$ & 10 & 7.29 (0.674/6.62) & 90.8\% & 0.761\%/55.6\% & 11.9\% \\
TOP5\_full & 0.17 & \citetalias{Stacy2013} & \citetalias{Stacy2013} & $2e$ & 1 & $10^{6}$ & 7.94 & 5.8 (0.276/5.53) & 95.2\% & 0.396\%/56.7\% & 12.3\% \\
KRO5\_full & 2.3 & \citetalias{Stacy2013} & \citetalias{Stacy2013} & $2e$ & 1 & $10^{6}$ & 8.9 & 6.01 (0.741/5.27) & 87.7\% & 0.937\%/52.9\% & 11.2\% \\
\hline 
LOG1\_low & 1 & \citetalias{Sana2012} & \citetalias{Sana2012} & \citetalias{Sana2012} & $\hat{f}_{\rm occ}$ & $10^{10}$ & 4.78 & 6.07 (5.34/0.73) & 12\% & 11.5\%/50.2\% & 3.04\% \\
TOP1\_low & 0.17 & \citetalias{Sana2012} & \citetalias{Sana2012} & \citetalias{Sana2012} & $\hat{f}_{\rm occ}$ & $10^{10}$ & 2.58 & 2.16 (1.66/0.495) & 22.9\% & 6.69\%/53.8\% & 3.57\% \\
KRO1\_low & 2.3 & \citetalias{Sana2012} & \citetalias{Sana2012} & \citetalias{Sana2012} & $\hat{f}_{\rm occ}$ & $10^{10}$ & 7.05 & 11.2 (10.3/0.894) & 8\% & 15\%/45.8\% & 2.77\% \\
LOG5\_low & 1 & \citetalias{Stacy2013} & \citetalias{Stacy2013} & $2e$ & $\hat{f}_{\rm occ}$ & $10^{10}$ & 10 & 2.71 (0.686/2.02) & 74.7\% & 0.711\%/51.9\% & 3.89\% \\
TOP5\_low & 0.17 & \citetalias{Stacy2013} & \citetalias{Stacy2013} & $2e$ & $\hat{f}_{\rm occ}$ & $10^{10}$ & 7.94 & 2.09 (0.28/1.81) & 86.6\% & 0.369\%/55.2\% & 4.13\% \\
KRO5\_low & 2.3 & \citetalias{Stacy2013} & \citetalias{Stacy2013} & $2e$ & $\hat{f}_{\rm occ}$ & $10^{10}$ & 8.9 & 2.18 (0.755/1.42) & 65.4\% & 0.879\%/45.4\% & 3.52\% \\
    \hline
    \end{tabular}
    \label{runlist}
\end{table*}

\begin{table*}
\caption{Basic observational signatures of Pop~III BBH mergers for the 18 models listed in Table~\ref{runlist}. Column 1: model name. Columns 2: all-sky merger rate $\dot{\mathcal{N}}$ (Eq.~\ref{rate}) for all mergers within $z<1$. Column 3: fraction $\mathcal{F}_{\rm NSC}$ of all-sky merger rate contributed by NSC mergers for $z<1$. Column 4: fraction $\mathcal{F}_{\rm PopIII}$ of all-sky merger rate of BBH mergers contributed by Pop~III stars for $z<1$. Column 5: all-sky merger rate $\dot{\mathcal{N}}_{\rm IMBH}$ of Pop~III BBHs with at least one IMBH (i.e., $m_{1}>100\ \rm M_\odot$) for $z<1$. Columns 6-9: same as Columns 2-5 but for $z<20$. Columns 10 and 11: peak value $\Omega_{\rm GW}^{\rm peak}$ and location $\nu_{\rm peak}$ of the SGWB energy spectrum from Pop~III BBH mergers. Column 12: maximum ratio $f_{\rm SGWB}^{\rm PopIII}$ between the SGWB produced by Pop~III BBH mergers and the conservative prediction of the SGWB from Pop~I/II and PBH mergers by \citet{Bavera2022}. Given the optimistic results for the total SGWB inferred from GWTC-3 events \citep{Abbott2023} as the reference, the maximum contribution of Pop~III BBH mergers to the total SGWB is approximately $f_{\rm SGWB}^{\rm PopIII}/5$. }
    \centering
    \begin{tabular}{cccccccccccc}
    \hline
    Model & $\dot{\mathcal{N}}\ [\rm yr^{-1}]$ & $\mathcal{F}_{\rm NSC}$ & $\mathcal{F}_{\rm PopIII}$ & $\dot{\mathcal{N}}_{\rm IMBH}$ & $\dot{\mathcal{N}}\ [\rm yr^{-1}]$ & $\mathcal{F}_{\rm NSC}$ & $\mathcal{F}_{\rm PopIII}$ & $\dot{\mathcal{N}}_{\rm IMBH}$ & $\Omega_{\rm GW}^{\rm peak}$ & $\nu_{\rm peak}$ & $f_{\rm SGWB}^{\rm PopIII}$ \\ 
    & $(z<1)$ & $(z<1)$ & $(z<1)$ & $(z<1)$ & $(z<20)$ & $(z<20)$ & $(z<20)$ & $(z<20)$ & $[10^{-12}]$ & [Hz] & \\ 
    \hline
LOG1\_obs & 8.95 & 41.6\% & 0.423\% &1.34 & 650 & 4.75\% & 2.93\% & 11.6 & 10.4 & 14 & 13.5\% \\
TOP1\_obs & 4.87 & 63\% & 0.231\% &2.18 & 264 & 8.41\% & 1.21\% & 16.2 & 13.8 & 12.6 & 18.7\% \\
KRO1\_obs & 21.2 & 30.9\% & 0.998\% &0.304 & 1400 & 3.35\% & 6.12\% & 2.56 & 11.8 & 105 & 9.05\% \\
LOG5\_obs & 10.1 & 91.9\% & 0.476\% &5.35 & 151 & 54\% & 0.697\% & 48.3 & 28.7 & 13.1 & 40.6\% \\
TOP5\_obs & 10.6 & 96\% & 0.499\% &8.53 & 117 & 70.4\% & 0.539\% & 70.7 & 44.8 & 12.6 & 64.8\% \\
KRO5\_obs & 10.5 & 88.9\% & 0.494\% &1.19 & 174 & 44.3\% & 0.799\% & 11.2 & 10.8 & 92.8 & 11.7\% \\
\hline
LOG1\_full & 13.4 & 62.4\% & 0.631\% &2.8 & 707 & 12.7\% & 3.17\% & 32.5 & 17.6 & 13.1 & 26.4\% \\
TOP1\_full & 8.1 & 78.7\% & 0.383\% &4.33 & 309 & 21.9\% & 1.41\% & 48.2 & 24.6 & 11.6 & 39.2\% \\
KRO1\_full & 28.2 & 50.2\% & 1.32\% &0.615 & 1490 & 9.56\% & 6.48\% & 6.91 & 16.7 & 98.6 & 13\% \\
LOG5\_full & 20.7 & 96.2\% & 0.973\% &10.9 & 310 & 77.7\% & 1.42\% & 140 & 50.4 & 11.6 & 83.6\% \\
TOP5\_full & 20.7 & 98.2\% & 0.974\% &16.7 & 279 & 87.7\% & 1.28\% & 207 & 78.9 & 11.2 & 134\% \\
KRO5\_full & 21 & 94.7\% & 0.988\% &2.38 & 320 & 69.9\% & 1.46\% & 30 & 19.3 & 83.9 & 23.3\% \\
\hline
LOG1\_low & 8 & 33.9\% & 0.379\% &1.08 & 626 & 1.05\% & 2.82\% & 2.76 & 11 & 14.8 & 13.3\% \\
TOP1\_low & 4.02 & 54.9\% & 0.191\% &1.61 & 248 & 2.3\% & 1.14\% & 4.32 & 14.7 & 14.5 & 17.9\% \\
KRO1\_low & 19 & 21.6\% & 0.893\% &0.235 & 1370 & 0.66\% & 5.97\% & 0.56 & 10.7 & 116 & 8.37\% \\
LOG5\_low & 8 & 89.7\% & 0.378\% &4.47 & 87.4 & 20.2\% & 0.404\% & 11.7 & 33.2 & 14.5 & 40.3\% \\
TOP5\_low & 8.22 & 95.1\% & 0.389\% &6.81 & 54.8 & 36.9\% & 0.253\% & 17.9 & 52.3 & 14.2 & 64.1\% \\
KRO5\_low & 7.26 & 83.9\% & 0.343\% &0.945 & 110 & 12.2\% & 0.508\% & 2.44 & 9.47 & 101 & 10.5\% \\
    \hline
    \end{tabular}
    \label{tobs}
\end{table*}

\subsection{Fiducial model}
\label{sec:fid}

The fiducial model LOG1\_obs is defined with a log-flat IMF for primary stars ($\alpha=1$), the IBS from local observations (\citetalias{Sana2012}), and the empirical NSC occupation fraction $f_{\rm occ}=\hat{f}_{\rm occ}$ (see Eq.~\ref{focc_fit}) with $M_{\star,\min}=10^{6}\ \rm M_{\odot}$. The basic statistics of Pop~III BBHs and their mergers in the fiducial model are summarized as follows: The mass fraction of binary stars that become BBHs is $f_{\rm BBH}\simeq 12.6\%$, which (given $f_{\rm B}=0.7$ and $\bar{m}_{\rm p,BBH}\simeq 193\ \rm M_\odot$) corresponding to a BBH formation efficiency (number of BBHs formed per unit stellar mass) $\epsilon_{\rm BBH}\simeq 4.78\times 10^{-4}\ \rm M_\odot^{-1}$. 
3.55\% of Pop~III BBHs fall into NSCs, among which 21.3\% are disrupted as soft binaries, and 37\% have reached the galaxy centre by DF before the formation of the NSC. 
In galaxy fields (NSCs), 11.6\% (62.8\%) of Pop~III BBHs have merged at $z>0$. The overall merger efficiency (number of mergers at $z>0$ per unit stellar mass) is $\epsilon_{\rm GW}^{\rm all}=6.49\times 10^{-5}\ \rm M_{\odot}^{-1}$. 17.9\% of Pop~III mergers occur in NSCs across cosmic history, among which only $\sim 2\%$ happen in NSC descendants (i.e., normal SCs). The small fraction of NSC mergers in NSC descendants is an artificial outcome of our low-$z$ extrapolation scheme that ignores halo mergers. Since mergers between galaxies hosting NSCs are rare at high-$z$, NSC descendants (produced by galaxy mergers) are expected to be important after cosmic noon ($z\lesssim 2$), which is not covered by the merger trees from \citet{Ishiyama2016}. 
In fact, for merger trees targeting Milky-Way-mass galaxies (reaching $z=0$) from the Caterpillar project \citep{Griffen2016}, we find that NSC descendants can host up to $\sim 60$\% of Pop~III BBH mergers in NSCs. 

\subsubsection{Merger rate}
\label{sec:fd_mrd}

\begin{figure}
    \centering
    \includegraphics[width=1\columnwidth]{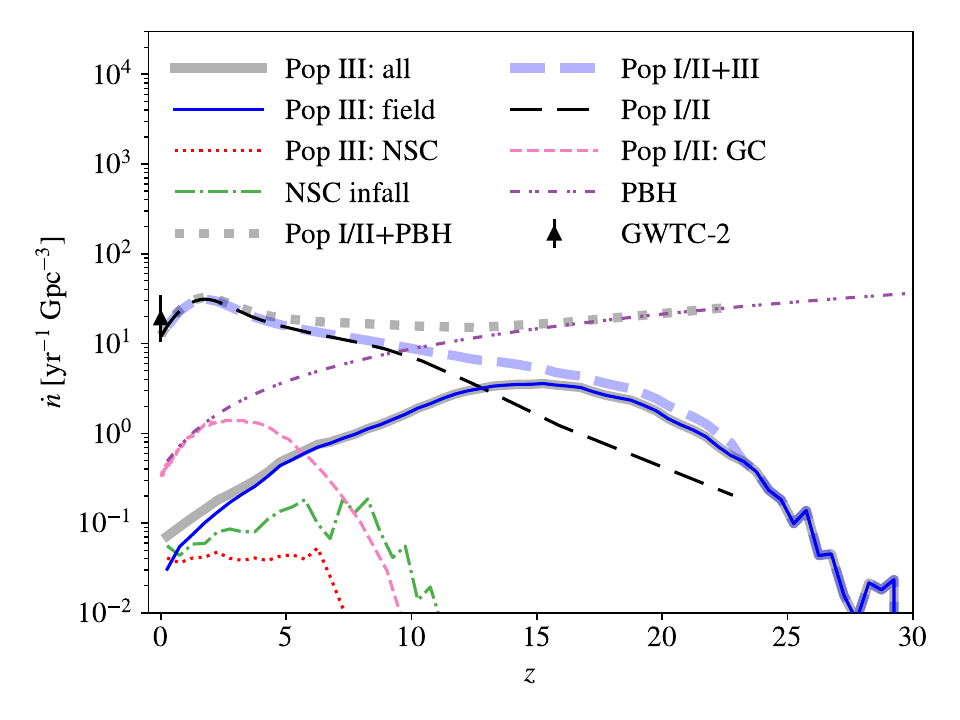}
    \vspace{-20pt}
    \caption{Co-moving MRD of Pop~III BBHs as a function of redshift in the fiducial model LOG1\_obs. The results for mergers in galaxy fields, NSCs and combined are shown with the thin solid, dotted and thick solid curves, respectively. The rate density of NSC infall is also shown with the dash-dotted curve. For comparison, we plot the MRDs of all Pop~I/II BBHs (long-dashed) in both galaxy fields (formed via CE and SMT) and GCs and only those in GCs (dashed) as well as that of PBH binaries (dash-dot-dotted), from \citet[see their fig.~3]{Bavera2022}, normalised to the model selection results in \citet{Franciolini2022}. {The thick dotted curve shows the total MRD combining Pop~I/II and PBH mergers, while the thick dashed curve shows the MRD of all stellar BBHs (Pop~I/II+III).} We find that Pop~III mergers account for $\sim 0.3\%$ of the (redshift-dependent) local BBH MRD $\dot{n}_{\rm obs}(z= 0)=19.3_{-9}^{+15.1}\ \rm yr^{-1}\ Gpc^{-3}$ inferred from the GWTC-2 events \citep[triangle data point,][]{Abbott2021}. }
    \label{fd_dndz}
\end{figure}

Figure~\ref{fd_dndz} shows the (co-moving) merger rate density (MRD) $\dot{n}$ as a function of redshift for Pop~III BBHs in comparison with the merger histories of Pop~I/II BBHs and PBHs from \citet{Franciolini2022} and the (redshift-dependent) local BBH MRD $\dot{n}_{\rm obs}(z= 0)=19.3_{-9}^{+15.1}\ \rm yr^{-1}\ Gpc^{-3}$ inferred from the GWTC-2 events \citep{Abbott2021}. 
Here, the MRD at a redshift bin $z_{j}$ is derived by counting the number of mergers $N_{j}$ in the redshift range $z\in [z_{j}-0.5\Delta z_{j},z_{j}+0.5\Delta z_{j})$, i.e., $\dot{n}(z_{j})=N_{j}/(\tilde{V}_{\rm com}\Delta t_{j})$, where $\Delta t_{j}$ is the cosmic age evolution across the redshift bin. Sixty evenly spaced bins are used for $z\sim 0-30$ with $\Delta z_{j}=0.5$. 

\begin{figure}
    \centering
    \includegraphics[width=1\columnwidth]{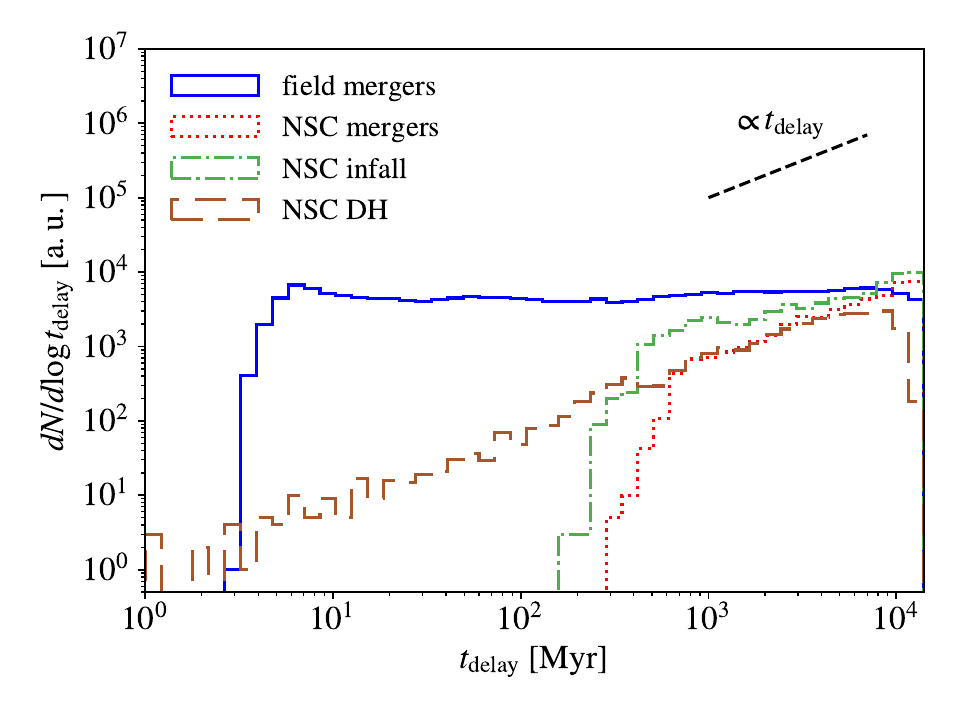}
    \vspace{-20pt}
    \caption{Delay time distribution for Pop~III BBH mergers in galaxy fields (solid) and NSCs (dotted), as well as NSC infall (dash-dotted) and DH (long-dashed) of BBHs in the fiducial model LOG1\_obs. }
    \label{fd_tgw}
\end{figure}

We find that the Pop~III MRD is dominated by field mergers at $z\gtrsim 1$ and peaks at $z \sim 15$ with $\dot{n}_{\rm peak}\sim 3.6\ \rm yr^{-1} Gpc^{-3}$, which is close to the peak of Pop~III star formation (Fig.~\ref{sfrd}). The MRD of field mergers is consistent with that derived in \citet{Santoliquido2023} for the IBSE channel using the same Pop~III star formation history predicted by the fiducial \textsc{a-sloth} model. The consistency also holds for the other 17 models considered in Table~\ref{runlist}. 
The MRD from Pop~III mergers in NSCs increases towards lower redshift and reaches a plateau at $z\lesssim 6$, and it slightly exceeds the MRD of field mergers at $z=0$. It is also shown in Fig.~\ref{fd_dndz} that NSC infall only starts at $z\sim 11$, because massive enough galaxies that host NSCs ($M_{\star}>10^{6}\ \rm M_{\odot}$) only form at $z\lesssim 12$, and the DF timescale for Pop~III BBHs can be large. The different trends in the MRDs of field and NSC mergers can be understood through the delay time ($t_{\rm delay}$) distribution, shown in Fig.~\ref{fd_tgw}. Here the delay time is defined as the time taken from the initial formation of the \textit{stellar} binary to the final BBH merger (or NSC infall), which is simply $t_{\rm delay}=\tau_{\rm B}+t_{\rm GW}$ for BBHs that evolve in isolation (Sec.~\ref{sec:bevo}). The delay time distribution of field mergers is approximately log-flat down to $t_{\rm delay}\sim 5\ \rm Myr$. However, it takes at least $\sim 200\ \rm Myr$ for Pop~III BBHs to fall into NSCs in the first place. For BBH mergers in NSCs, we have $t_{\rm delay}\gtrsim 300\ \rm Myr$, and the delay time distribution is almost uniform (i.e., $dN/d\log t_{\rm delay}\propto t_{\rm delay}$) at $t_{\rm delay}\gtrsim 800$~Myr. In Fig.~\ref{fd_tgw}, we also show the distribution of the DH timescale, i.e., the time taken from NSC infall to the final merger, which turns out to be nearly uniform below $\sim 5$~Gyr. 

The predicted local MRD of Pop~III BBH mergers is $\sim 0.06\ \rm yr^{-1}\ Gpc^{-3}$, which only counts for $0.31_{-0.13}^{+0.27}\%$ of the value inferred from observations \citep{Abbott2021}. The Pop~III MRD is overwhelmed by those from Pop~I/II and PBH mergers at low $z$ but its importance increases beyond cosmic noon from $z\sim 2$ to $z\sim 15$. The Pop~III MRD exceeds that of Pop~I/II mergers in GCs at $z\gtrsim 6$ and that of all Pop~I/II mergers at $z\gtrsim 13$, although it remains below the PBH MRD by at least a factor of $\sim 4$, since the PBH rate increases monotonically towards higher $z$. This indicates that although we are more likely to detect Pop~III mergers at higher $z$ for $z\sim 2-15$, distinguishing them from PBH mergers can be a challenge \citep[see, e.g.,][]{Franciolini2022pbh}.

Given the (intrinsic) MRD $\dot{n}(z)$, the all-sky merger rate of Pop~III BBHs observed at $z=0$ as a function of the horizon redshift $z_{\rm horizon}$ can be written as \citep{Hartwig2016} 
\begin{align}
    \dot{\mathcal{N}}(z<z_{\rm horizon})=\int_{0}^{z_{\rm horizon}}dz\frac{\dot{n}(z)}{(1+z)}\frac{dV}{dz}\ ,\label{rate}
\end{align}
where $dV/dz=4\pi cd_{\rm c}^{2}(z)/H(z)$ is the co-moving volume element, given the co-moving distance $d_{\rm c}(z)=\int_{0}^{z}dz'[c/H(z')]$ and Hubble parameter $H(z)$. One can apply a weight to each source using its detection probability above a certain signal-to-noise ratio (SNR) to calculate the \textit{detection} rate $\mathcal{\dot{N}}_{\rm X}$ by a given detector X with finite sensitivity using the same formalism. In this case, $\dot{n}(z)$ should be replaced with the MRD of \textit{detectable} sources, i.e., $\dot{n}_{\rm X}(z_{j})=\sum_{i=1}^{N_{j}}p_{{\rm det},i}$ for a redshift bin $z_j$ given the detection probabilities $p_{{\rm det},i}$ of the $N_{j}$ mergers within the bin. In the calculation of the \textit{all-sky} merger rate $\mathcal{\dot{N}}$, we simply use $p_{{\rm det},i}=1$. Interestingly, this turns out to be a good approximation for our Pop~III BBHs observed by ET. 
According to the results of $p_{{\rm det},i}$ from \textsc{gwtoolbox} \citep{Yi2022a,Yi2022b}, ET can detect most ($\gtrsim 90\%$) Pop~III BBH mergers with $\rm SNR>8$. 
Similarly, most Pop~I/II and PBH mergers in the reference model from \citet{Franciolini2022} are detectable by ET with $\rm SNR>8$ as well given the large horizon redshift of ET at their mass scales \citep[see their fig.~8]{DeLuca2021}. In both cases (Pop~III and Pop~I/II + PBH), the contribution of sources at very high redshifts is very small. Therefore, the Pop~III contribution to the all-sky BBH merger rate within $z<20$ is a good indicator of the chance of finding Pop~III sources in observations of BBH mergers by ET. 

\begin{figure}
    \centering
    \includegraphics[width=1\columnwidth]{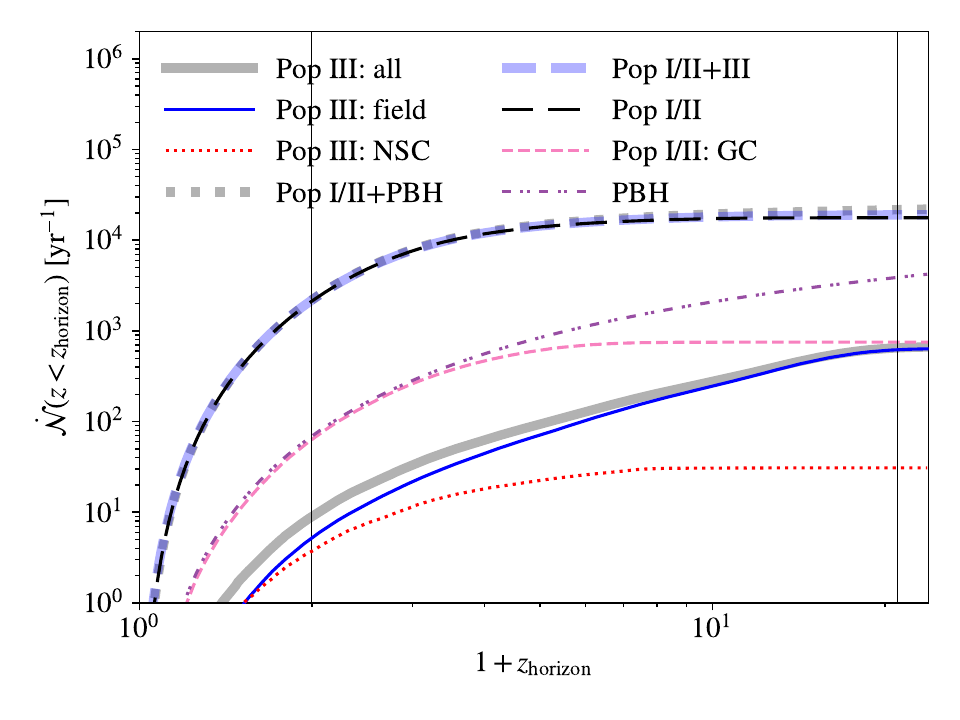}
    \vspace{-20pt}
    \caption{All-sky merger rate (for an observer at $z=0$) of Pop~III BBHs as a function of horizon redshift in the fiducial model LOG1\_obs. The notations of merger populations are the same as in Fig.~\ref{fd_dndz}. 
    The vertical lines label the reference horizon redshifts $z_{\rm horizon}=1$ and 20. }
    \label{fd_event}
\end{figure}

Fig.~\ref{fd_event} (see also Table~\ref{tobs}) shows the results of $\mathcal{\dot{N}}$ for our Pop~III BBHs and the Pop~I/II and PBH mergers from \citet{Franciolini2022}, where 300 bins are used for $z\sim 0-30$ with $\Delta z_{j}=0.1$ to integrate Eq.~\ref{rate} numerically. 
We predict an all-sky merger rate of Pop~III BBHs $\dot{\mathcal{N}}\sim 650\ (9)\ \rm yr^{-1}$ within $z<20\ (1)$, which counts for $\sim 3\ (0.4)\%$ of the total merger rate dominated by Pop~I/II mergers. 
NSC mergers contribute $\sim 5$\% and 42\% of the total Pop~III BBH merger rates for $z_{\rm horizon}=20$ and 1. The contribution of NSC mergers saturates at $z_{\rm horizon}\sim 7$. 
However, NSC mergers dominate ($\gtrsim 99.7$\%) the merger rate of massive BBHs containing at least one IMBH ($m_{1}>100\ \rm M_\odot$) $\mathcal{N}_{\rm IMBH}=11.6\ (1.34)\ \rm yr^{-1}$ within $z<7\ (1)$. 
The Pop~III BBH all-sky merger rate remains below that of PBHs by a factor of $\sim 10$ for $z_{\rm horizon}\lesssim 20$, and becomes comparable to the rate for Pop~I/II BBHs in GCs for $z_{\rm horizon}\gtrsim 20$. Here, the PBH merger rate keeps increasing with $z_{\rm horizon}$, while the GC merger rate saturates around $z_{\rm horizon}\sim 5$. Considering the facts that in our model universe, Pop~III stars only make up $\sim 5\times 10^{-5}$ of the total mass 
density of stars ever formed, and that Pop~III stars are significantly outnumbered (by a factor of $\sim 1000$) by the PBHs (with a characteristic mass of $\sim 35\ \rm M_\odot$) that make up $\sim 2\times 10^{-4}$ of dark matter \citep[see their fig.~4]{Franciolini2022} to achieve merger rates $\sim 6-9$ times higher than those of Pop~III BBHs within $z\lesssim 20$, the contribution by Pop~III stars to the BBH merger rate is already quite large, which reflects the high efficiency of Pop~III stars at forming BBH mergers. 


\subsubsection{Stochastic gravitational wave background}
\label{sec:fd_sgwb}

Next, we drive the contribution to SGWB by Pop~III BBH mergers. Given $N$ mergers predicted by \textsc{a-sloth} in our effective simulation volume $\tilde{V}_{\rm com}=50 V_{\rm com}$, the SGWB (energy density spectrum) can be characterized by the dimensionless parameter \citep[see, e.g.,][]{Abbott2018,Perigois2021,Braglia2021,Martinovic2022,Lehoucq2023}:
\begin{align}
    \Omega_{\rm GW}(\nu)&=\frac{\nu}{c^{2}\rho_{\rm c}}\int d\theta \int_{0}^{\infty}dz\frac{\dot{n}(z)p(\theta,z)}{(1+z)H(z)}\frac{dE_{\rm GW}(\theta)}{d\nu_{\rm s}}\notag\\
    &=\frac{\nu}{c^{2}\rho_{\rm c}}\lim_{\Delta z_{j}\rightarrow 0}\sum_{j=1}^{N_{z}}\sum_{i=1}^{N_{j}}\frac{(\Delta z_{j}/\Delta t_{j})\tilde{V}_{\rm com}^{-1}}{(1+z_{j})H(z_{j})}\frac{dE_{{\rm GW},i}}{d\nu_{\rm s}}\notag\\
    &=\frac{\nu}{c^{2}\rho_{\rm c}}\sum_{i=1}^{N}\frac{1}{\tilde{V}_{\rm com}}\frac{dE_{{\rm GW},i}}{d\nu_{\rm s}}\ ,\label{ogw}
\end{align}
in which $\nu$ is the observer-frame frequency, $\rho_{\rm c}=3H_{0}^{2}/(8\pi G)$ is the critical density, $H(z)$ is the Hubble parameter at redshift $z$, $dE_{\rm GW}/d\nu_{\rm s}$ is the source-frame GW energy spectrum of an individual source evaluated at the source-frame frequency $\nu_{\rm s}=(1+z)\nu$, and the factor $(1+z)$ in the denominator captures time dilation due to cosmic expansion. In the first line, $\dot{n}(z)$ denotes the \textit{source-frame} co-moving merger rate density, and the source properties are captured by the parameter $\theta$ and its probability distribution $p(\theta,z)$. In the second line, we re-write the integration in its discrete form (with $\int dz\int d\theta \dot{n}(z)p(\theta,z)=\lim_{\Delta z_{j}\rightarrow 0} \sum_{j=1}^{N_{z}} \Delta z_{j}\sum_{i=1}^{N_{j}}\Delta t_{j}^{-1}\tilde{V}_{\rm com}^{-1}$) as the summation over $N_{z}$ redshift bins ($z_{j}$, $\Delta z_{j}$) each containing $N_{j}$ mergers at $z_{i}\in [z_{j}-0.5\Delta z_{j},z_{j}+0.5\Delta z_{j})$, whose GW spectra $dE_{{\rm GW},i}/d\nu_{\rm s}$ are added up, where $\nu_{\rm s}=(1+z_{i})\nu$, and $\Delta t_{j}$ is the corresponding cosmic age evolution at redshift bin $j$. Finally, since $dz/dt=(1+z)H(z)$, we arrive at the last line with $\lim_{\Delta z_{j}\rightarrow 0}(\Delta z_{j}/\Delta t_{j})=(1+z_{j})H(z_{j})$. We use the phenomenological single-source energy spectrum model (for mergers with circular orbits) from \citet{Ajith2011} to calculate $dE_{{\rm GW},i}/d\nu_{\rm s}$, assuming random inclinations and a fixed effective spin\footnote{The effect of BH spins is small for the total energy of GWs emitted in the inspiral phase \citep{Zhou2023,Lehoucq2023}.} $\chi_{\rm eff}=0.06$ for all sources, which is around the peak of the $\chi_{\rm eff}$ distribution, inferred from the BBH merger population detected by the LVK network \citep[e.g.,][]{Callister2023}. The ignorance of eccentricities in our calculation is likely an oversimplification for some NSC mergers considering that large eccentricities can arise in dense star clusters and AGN discs \citep[e.g.,][]{Hoang2018,Mapelli2021hierarchical,Zhang2021,Chattopadhyay2023,Dall'Amico2023,Trani2023,ArcaSedda2024iii,Fabj2024}. We plan to take into account the effects of eccentricities on the GW signals in future work, which can reduce $\Omega_{\rm GW}$ at low frequencies\footnote{In this paper, we are mostly interested in the frequency range $\nu\sim 1-200$~Hz accessible by ET, where most Pop~III BBH mergers have their peak GW emission. Since we adopt a conservative coefficient $\kappa=0.01$ for eccentricity enhancement by {binary-single} encounters, the fraction of Pop~III BBH mergers in NSCs with large eccentricities ($e\gtrsim 0.1$) at $\nu\gtrsim 1$~Hz is expected to be small ($\lesssim 1$\%).} \citep[see, e.g.,][]{Chen2017,Buskirk2023,Xuan2024,Islam2024,Raidal2024}. 
To the first order, given the intrinsic chirp mass $m_{\rm c}=(m_{1}m_{2})^{3/5}/(m_{1}+m_{2})^{1/5}$ of a BBH merger at redshift $z$, its contribution to the SGWB $\delta\Omega_{\rm GW}=(c^{2}\rho_{\rm c}\tilde{V}_{\rm com})^{-1}\nu(dE_{\rm GW}/d\nu_{\rm s})$ peaks at an observer-frame frequency $\nu_{\rm p}\propto 1/[(1+z)m_{\rm c}]$ with $\delta\Omega_{\rm GW}(\nu_{\rm p})\propto m_{\rm c}/(1+z)$ and follows $\delta\Omega_{\rm GW}\propto m_{\rm c}^{5/3}\nu^{2/3}(1+z)^{-1/3}$ in the low-frequency inspiral regime ($\nu\ll \nu_{\rm p}$). This indicates that the SGWB is mostly sensitive to massive mergers at low $z$.

\begin{figure}
    \centering
    \includegraphics[width=1\columnwidth]{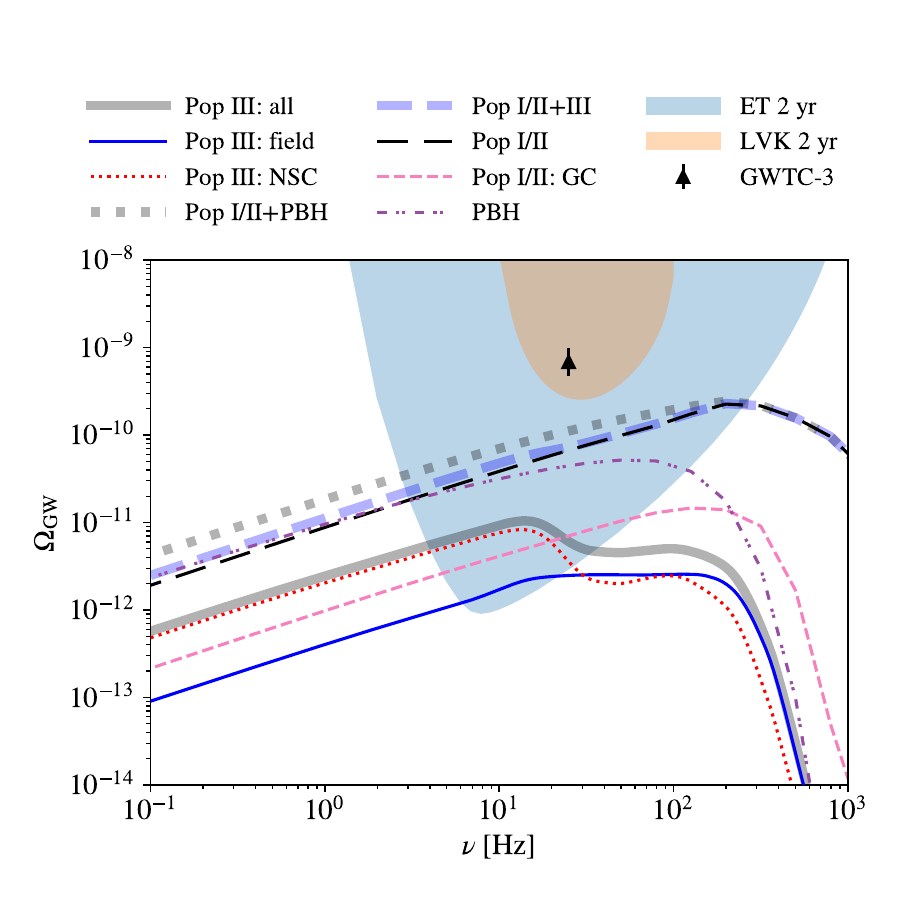}
    \vspace{-20pt}
    \caption{SGWB (dimensionless energy density spectrum) contributed by Pop~III BBH mergers in the fiducial model LOG1\_obs, derived with the phenomenological single-source energy spectrum model in \citet{Ajith2011} assuming a universal fixed effective spin $\chi_{\rm eff}=0.06$. Our Pop~III results are {compared and combined} with those for Pop~I/II and PBH mergers from \citet[see their fig.~4]{Bavera2022} with the same notations of merger populations as in Fig.~\ref{fd_dndz}. 
    The blue and orange shaded regions denote the PI sensitivity curves of 2-year observations by the LVK network and ET for a signal-to-noise ratio $\rm SNR>2$, calculated by \citet[see their appendix C]{Bavera2022} using the public code \textsc{schNell} \citep{Alonso2020}. 
    Here, for LVK, we consider the detector configuration including LIGO, Virgo, and KAGRA \textit{at design sensitivity, assuming no cross-correlations between detectors}. The triangle shows the total SGWB energy density from compact object mergers $\Omega_{\rm GW}(f=25\ {\rm Hz})=6.9_{-2.1}^{+3.0}\times 10^{-10}$ estimated by population analysis through GWTC-3 \citep{Abbott2023}. }
    \label{fd_ogw}
\end{figure}

Our results are shown in Fig.~\ref{fd_ogw} (see also Table~\ref{tobs}), compared with the SGWB energy spectra for Pop~I/II and PBH mergers from \citet[]{Bavera2022}. 
We find that although NSC mergers only account for $\sim 18\%$ of the number of Pop~III mergers across cosmic history, they dominate the SGWB from Pop~III BBH mergers for $\nu\lesssim 20\ \rm Hz$ (contributing $\sim 80$\% of the total energy spectrum at $\nu\lesssim 10$~Hz). The Pop~III SGWB has a peak of $\Omega_{\rm GW}\sim 10^{-10}$ at $\nu=14$~Hz and converges to the standard power-law $\Omega_{\rm GW}\propto\nu^{2/3}$ form for BBH inspiral at $\nu\lesssim 10$~Hz. The energy spectrum is almost flat with $\Omega_{\rm GW}\sim 5\times 10^{-11}$ at $\nu\sim 30-100~\rm Hz$ with similar contributions from field and NSC mergers. The field mergers produce a flat energy spectrum for $\nu\sim 10-200$~Hz, while the SGWB from NSC mergers shows a small secondary peak at $\nu\sim 100$~Hz.  
The maximum ratio between the SGWB of our Pop~III mergers and the total SGWB of Pop~I/II and PBH mergers from \citet{Bavera2022} is $f_{\rm SGWB}^{\rm PopIII}=13.5\%$ achieved in the low-frequency regime ($\nu\lesssim 10$~Hz). Compared with the results for sub-components in \citet{Bavera2022}, the SGWB from our Pop~III BBH mergers exceeds that from Pop~I/II BBH mergers in GCs at $\nu\lesssim 20$~Hz, but remains below the SGWB from PBHs. In Fig.~\ref{fd_ogw} we also show the total SGWB (including contributions from mergers involving neutron stars) $\Omega_{\rm GW}(f=25\ {\rm Hz})=6.9_{-2.1}^{+3.0}\times 10^{-10}$ inferred from GWTC-3 events {(\citealp{Abbott2023}, see also \citealp{Renzini2024})}. 
Actually, the total SGWB estimated by \citet{Abbott2023} is higher than the total SGWB of Pop~I/II and PBH mergers from \citet{Bavera2022} by a factor of $\sim 5-8$ in the frequency range $\nu\sim 1-200$~Hz that we are mostly concerned with. Therefore, if we take the results from \citet{Abbott2023} as the reference, we arrive at a conservative estimate of the maximum contribution of Pop~III BBH mergers to the total SGWB as $\sim 3$\%. 
These features of the Pop~III SWGB, in particular the dominance of NSC mergers at low frequencies, can be interpreted with the different merger histories and mass distributions of field and NSC mergers (see below). 

\subsubsection{Mass distribution}
\label{sec:fd_mass}

\begin{figure}
    \centering
    \includegraphics[width=1\columnwidth]{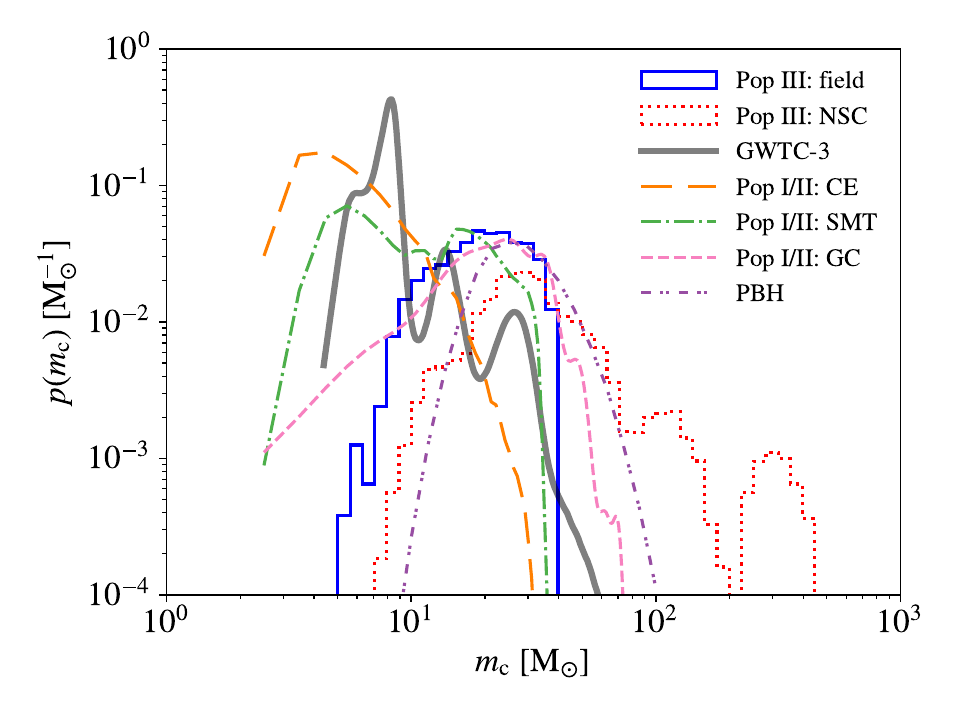}
    \vspace{-20pt}
    \caption{Intrinsic chirp mass distributions of Pop~III BBH mergers in the fiducial model LOG1\_obs, compared with those of Pop~I/II mergers formed via CE (long-dashed curve), SMT (dash-dotted curve), and in GCs (dashed curve), as well as PBH mergers (dash-dot-dotted curve) from the best-fit model in \citet{Franciolini2022} adopted by \citet[see their fig.~1]{Bavera2022}. The results for Pop~III BBH mergers in galaxy fields and NSCs are shown with the solid and dotted contours, respectively. 
    We also plot the intrinsic chirp mass distribution inferred from GWTC-3 events using the flexible mixture model framework \citep[see their fig.~2]{Abbott2023} as the thick solid curve.}
    \label{fd_mchirp}
\end{figure}

In Fig.~\ref{fd_mchirp}, we plot the intrinsic chirp distributions of Pop~III BBH mergers in galaxy fields and NSCs, compared with the results for Pop~I/II mergers formed by CE, SMT, and in GCs, as well as PBH mergers from the best-fit model in \citet{Franciolini2022} adopted by \citet[see their fig.~1]{Bavera2022}, and the distribution inferred from observations \citep[see their fig.~2]{Abbott2023}. The chirp mass distribution of Pop~III BBH mergers in galaxy fields peaks around $m_{\rm c}\sim 20\ \rm M_\odot$ and has a sharp cutoff at $m_{\rm c}\sim 40\ \rm M_\odot$. The number of mergers also declines rapidly for $m_{\rm c}\lesssim 10\ \rm M_\odot$. In between ($m_{\rm c}\sim 10-40\ \rm M_\odot$), the distribution of Pop~III field mergers is similar to that of Pop~I/II BBH mergers in GCs. However, the latter is broader, shaped by repeated mergers of Pop~I/II BHs (smaller than those from Pop~III stars) via $N$-body dynamics in GCs. 
In galaxy fields, Pop~III BBH mergers produced mostly by IBSE are generally more massive than their Pop~I/II counterparts (from CE and SMT during IBSE) due to the massive and compact nature of Pop~III stars. The Pop~I/II mergers involve significant fractions of low-mass ($m_{\rm c}\lesssim 10\ \rm M_\odot$) mergers that are rare in our case. Compared with field mergers, Pop~III BBH mergers in NSCs are even more massive. Their chirp mass distribution has three peaks at $m_{\rm c}\sim 30$, 100 and $300\ \rm M_\odot$. Interestingly, the location and shape of the first peak are similar to the (quasi-log-normal) chirp mass distribution of PBH mergers, but the peak of Pop~III mergers is broader, which renders the second peak at $m_{\rm c}\sim 100\ \rm M_{\odot}$ insignificant. However, there is a (narrow) gap at $m_{\rm c}\sim 200\ \rm M_\odot$ between the second and third peaks. 

As mentioned in Sec.~\ref{sec:fd_sgwb}, the peak contribution from a BBH merger to the SGWB satisfies $\delta\Omega_{\rm GW}(\nu_{\rm p})\propto m_{\rm c}/(1+z)$ with the peak location $\nu_{\rm p}$ inversely proportional to $m_{\rm c}(1+z)$ to the first order. The strong dependence of $\delta\Omega_{\rm GW}(\nu_{\rm p})$ on $m_{\rm c}$ indicates that it is the (low-$z$) massive NSC mergers with $m_{\rm c}\gtrsim 60\ \rm M_\odot$ around the second and third peaks in the chirp mass distribution that produce the primary peak of the SGWB at $\nu=14$~Hz in Fig.~\ref{fd_ogw}, while less massive NSC mergers shape the secondary SGWB peak at $\nu\sim 100$~Hz. On the other hand, due to the redshift dependence, Pop~III BBH mergers in galaxy fields (with $m_{\rm c}\sim 10-40\ \rm M_{\odot}$) make their peak contributions to the SGWB in a frequency range $\nu\sim 10-200$~Hz much broader than their chirp mass range given their extremely broad redshift distribution with the MRD rapidly rising from $z=0$ to $z\sim 15$ (Fig.~\ref{fd_dndz}). This 
explains the broad, flat peak of the SGWB spectrum from Pop~III field mergers at $\nu\sim 10-200$~Hz. 



\begin{figure}
    \centering
    \includegraphics[width=1\columnwidth]{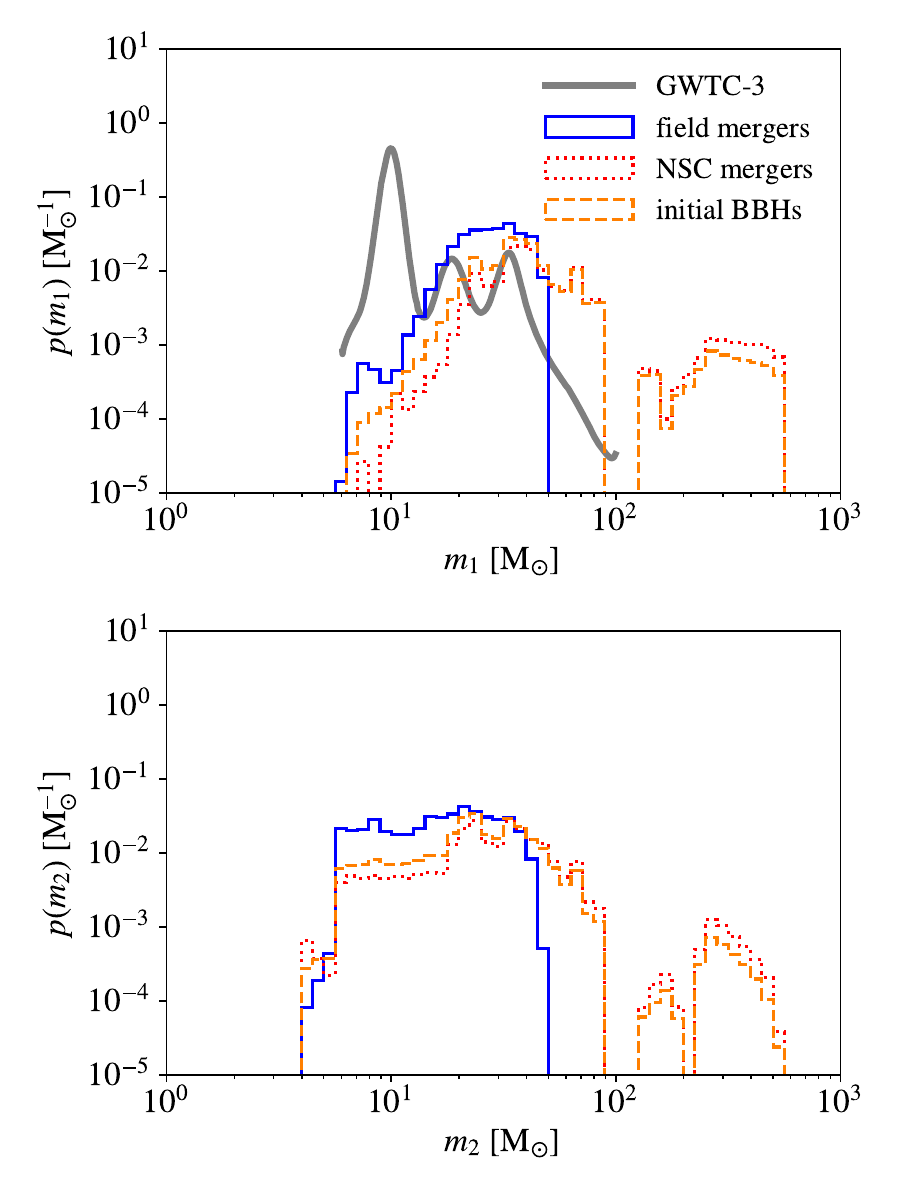}
    \vspace{-20pt}
    \caption{Intrinsic primary (top) and secondary (bottom) mass distributions of Pop~III BBH mergers in galaxy fields (solid) and NSCs (dotted) and initial BBHs (dashed) in the fiducial model LOG1\_obs. Here initial BBHs denote all Pop~III BBHs sampled in \textsc{a-sloth} (including those that do not merge at $z>0$). 
    In the top panel, we also plot the primary mass distribution inferred from the 69 confident BBH mergers in GWTC-3 \citep[see their fig.~1]{Edelman2023} as the thick solid curve. }
    \label{fd_m}
\end{figure}

\begin{figure}
    \centering
    \includegraphics[width=1\columnwidth]{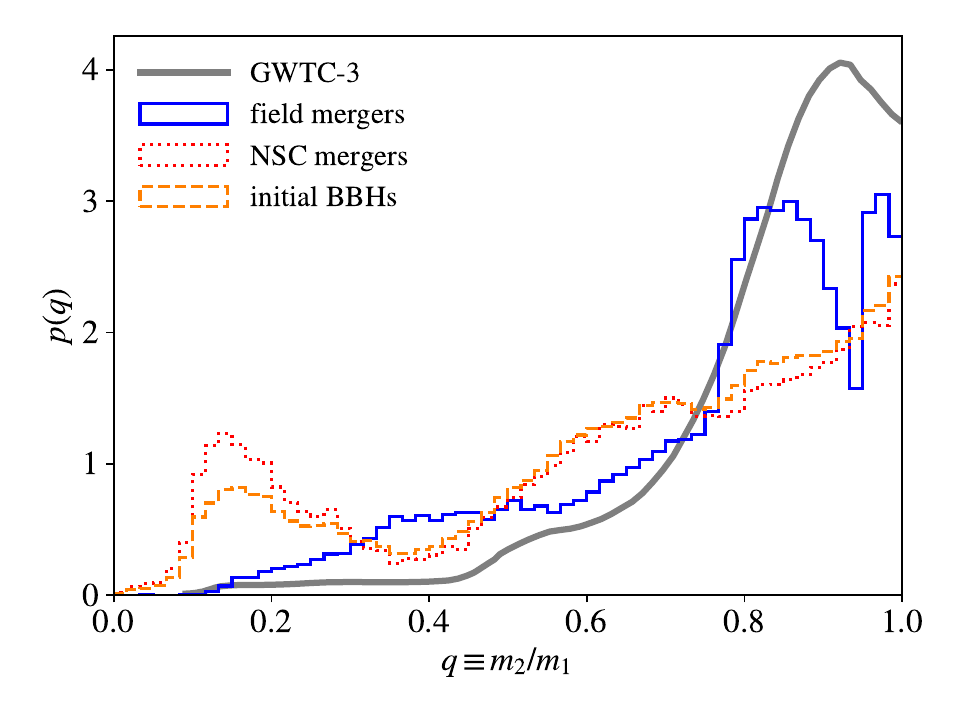}
    \vspace{-20pt}
    \caption{Intrinsic mass ratio distributions of Pop~III BBH mergers in galaxy fields (solid) and NSCs (dotted) and initial BBHs (dashed) in the fiducial model LOG1\_obs. The mass ratio distribution inferred from the 69 confident BBH mergers in GWTC-3 \citep[see their fig.~2]{Edelman2023} is shown with the thick solid curve.}
    \label{fd_q}
\end{figure}

To better understand the chirp mass distributions of Pop~III BBH mergers from different channels, we plot the underlying primary and secondary mass distributions in Fig.~\ref{fd_m} as well as the mass ratio distributions in Fig.~\ref{fd_q}, where we also show the results for initial Pop~III BBHs (including those that do not merger at $z>0$). The mean values of $m_{\rm c}$, $m_{1}$, and $q$ are summarized in Table~\ref{tmass}. 
In galaxy fields, almost all 
Pop~III BBH mergers come from BHs below $50\ \rm M_\odot$. This upper bound is even significantly lower than the lower edge of the Pop~III single-star PISN mass gap ($86\ \rm M_\odot$). In fact, only $\sim 0.31\ (0.14)$\% of field mergers have $m_{1}>50\ (100)\ \rm M_\odot$, among which $51\ (41)\%$ are ejected from NSCs, while $59.2\ (35.5)\%$ of NSC mergers have $m_{1}>50\ (100)\ \rm M_\odot$. The mass ratio distribution of field mergers is dominated by nearly equal-mass mergers with $q\gtrsim 0.8$, similar to the distribution inferred from GWTC-3 events \citep{Edelman2023}. 

However, NSC mergers involve (both primary and secondary) BHs below, above, and also inside the PISN mass gap $\sim 86 - 242\ \rm M_\odot$ expected from the \textit{single-star} evolution models adopted in \textsc{sevn} \citep{Costa2023}. The BHs inside the mass gap are relatively rare and only cover the mass range $\sim 130-242\ \rm M_\odot$ so another narrower gap around $86-130\ \rm M_\odot$ exists in the mass distribution of BHs in binaries, which can be regarded as a stricter definition of the Pop~III PISN mass gap. This feature is also seen in other BPS studies \citep[e.g.,][]{Tanikawa2021mrd,Tanikawa2022}. 
It is the mass gap that results in the three peaks in the chirp mass distribution of NSC mergers (dotted contour in Fig.~\ref{fd_mchirp}): (1) The low-mass peak around $m_{\rm c}\sim 30\ \rm M_\odot$ is made of two BHs below the gap. (2) The intermediate-mass peak at $m_{\rm c}\sim 100\ \rm M_\odot$ involves one BH below the gap and another above the gap. (3) The massive peak at $m_{\rm c}\sim 300\ \rm M_\odot$ is produced by BHs above the gap. Here the (2) mergers between one BH below and another above the mass gap produce a peak at $q\sim 0.15$ in the mass ratio distribution. The mass and mass ratio distributions of BHs in NSC mergers are similar to those for all initial BBHs. The difference is that for NSC mergers, the mass distribution is tilted to the massive end, and the peak at $q\sim 0.15$ in the mass ratio distribution is more obvious. The reason is that more massive BBHs are more likely to fall into NSCs by DF (see Eqs.~\ref{tdf1} and \ref{tdf2}). 
{Such distinct features of the NSC-DH channel with respect to IBSE, i.e., large fractions of mergers with low mass ratios and massive BHs (inside/above the mass gap), are also seen in the \textit{in-situ} dynamical channel for BBH mergers in massive Pop~III clusters \citep[][]{Wang2022,Liu2023sc,Mestichelli2024}.}

\begin{table}
\caption{Summary statistics of BH masses in Pop~III BBH mergers for the 18 models listed in Table~\ref{runlist}. The second, third and last columns show the average chirp mass $m_{\rm c}$, primary mass $m_{1}$ and mass ratio $q$ for BBH mergers in galaxy fields/NSCs. }
    \centering
    \begin{tabular}{cccccccccccc}
    \hline
    Model & $\bar{m}_{\rm c}^{\rm field/NSC}$ & $\bar{m}_{\rm 1}^{\rm field/NSC}$ & $\bar{q}^{\rm field/NSC}$ \\
    & $\rm [M_{\odot}]$ & $\rm [M_{\odot}]$ \\
    \hline
LOG1\_obs & 23.4/97.2 & 31.3/160 & 0.75/0.64 \\
TOP1\_obs & 25.3/169 & 33.8/282 & 0.76/0.58 \\
KRO1\_obs & 20.4/38.4 & 28.1/56.7 & 0.71/0.71 \\
LOG5\_obs & 25.8/131 & 37/242 & 0.69/0.55 \\
TOP5\_obs & 27.9/185 & 41.7/345 & 0.68/0.48 \\
KRO5\_obs & 24.4/58.3 & 34.6/94.3 & 0.7/0.68 \\
\hline
LOG1\_full & 23.4/94.2 & 31.3/154 & 0.75/0.64 \\
TOP1\_full & 25.2/165 & 33.5/276 & 0.77/0.58 \\
KRO1\_full & 20.4/37.5 & 28/55.4 & 0.71/0.7 \\
LOG5\_full & 25.6/129 & 36.5/238 & 0.69/0.55 \\
TOP5\_full & 26.2/182 & 38.1/341 & 0.68/0.48 \\
KRO5\_full & 24.3/57.3 & 34.4/92.2 & 0.7/0.69 \\
\hline
LOG1\_low & 23.4/107 & 31.3/174 & 0.75/0.64 \\
TOP1\_low & 25.2/180 & 33.6/295 & 0.76/0.58 \\
KRO1\_low & 20.4/40.5 & 28.1/59.9 & 0.71/0.71 \\
LOG5\_low & 25.6/143 & 36.6/261 & 0.69/0.54 \\
TOP5\_low & 26.3/193 & 38.3/357 & 0.68/0.48 \\
KRO5\_low & 24.2/63.4 & 34.4/104 & 0.7/0.68 \\
    \hline
    \end{tabular}
    \label{tmass}
\end{table}

\subsubsection{Progenitor and host system properties} 
\label{sec:fd_progenitor}

\begin{figure}
    \centering
    \includegraphics[width=1\columnwidth]{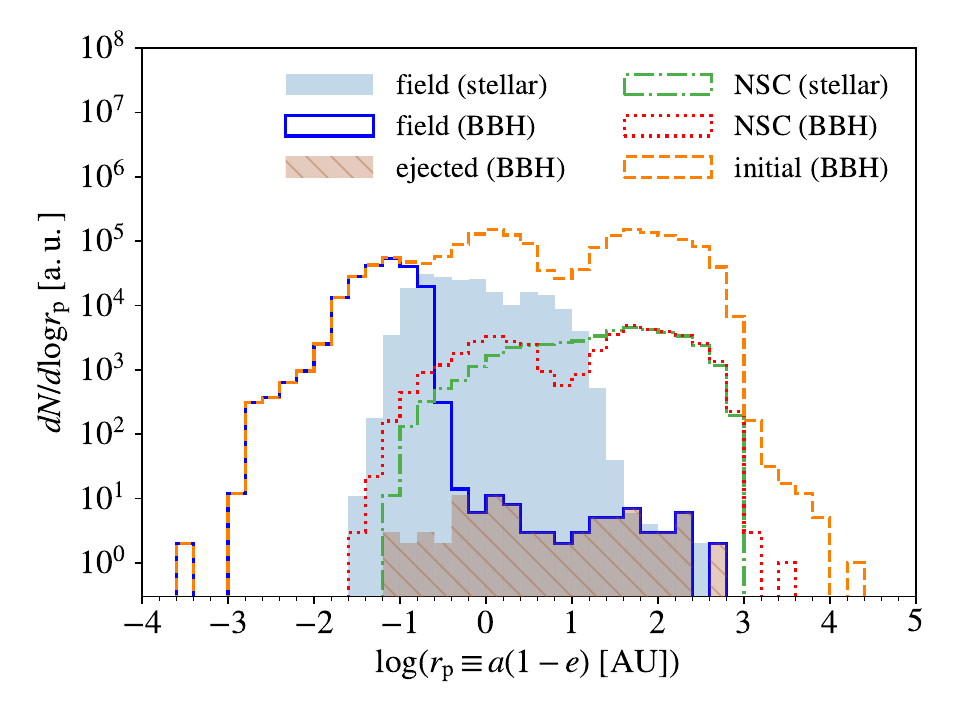}
    \vspace{-20pt}
    \caption{Initial pericenter separation distribution for the progenitors of Pop~III BBH mergers in the fiducial model LOG1\_obs. Shaded region: (ZAMS) stellar progenitors of field mergers. Dash-dotted contour: stellar progenitors of NSC mergers. Solid contour: initial BBHs that become field mergers. Dotted contour: initial BBHs that become NSC mergers. Dashed contour: all initial BBHs sampled in \textsc{a-sloth}. Hatched shaded region: initial BBHs that once fall into NSCs but are eventually ejected and merge in galaxy fields, which make up a tiny fraction ($\sim 0.04\%$) of field mergers.}
    \label{fd_adis}
\end{figure}

The difference between the IBSE and NSC-DH channels in the mass distribution of BHs can be explained by the facts that (i) only close 
BBHs can merge in isolation within a Hubble time while the NSC-DH channel can also drive initially wide BBHs to merge, and (ii) the masses and initial separations of BHs are correlated as a result of binary stellar evolution. To illustrate the first point (i), we plot the initial pericenter separation $r_{\rm p}$ distributions of the progenitors of Pop~III BBH mergers in Fig.~\ref{fd_adis}, for both initial stellar binaries and BBHs. It can be seen that field mergers mostly originate from BBHs with initial pericenter separations $r_{\rm p}\equiv a(1-e)\lesssim 0.16$~AU, which are mainly produced by close stellar binaries with $r_{\rm p}\lesssim 10$~AU initially. However, NSC mergers mostly come from initial BBHs with $r_{\rm p}\sim 0.1-10^{3}$~AU, and the shape of the $r_{\rm p}$ distribution for such NSC merger progenitors is similar to that of all BBHs for $r_{\rm p}\gtrsim 0.1$~AU. 
The corresponding progenitor stellar binaries also have a broad $r_{\rm p}$ distribution extending to $r_{\rm p}\sim 10^{3}$~AU. Interestingly, there is a tiny fraction (0.04\%) of field mergers from initially wide BBHs. These binaries have once fallen into NSCs and been processed by DH, but are ejected later and eventually merge in galaxy fields. 
In general, the progenitor BBHs of field mergers and NSC mergers are approximately separated by a critical initial pericenter separation $r_{\rm p,crit}\sim 0.1$~AU, which is a common feature in all cases considered in this work (Table~\ref{runlist}). Below this critical separation, Pop~III BBHs typically merge within $~\rm 1$~Gyr, so there is usually not enough time for them to sink into NSCs by DF. Above the critical separation, due to the strong dependence of merger timescale on initial separation for isolated evolution (Eq.~\ref{tgw_approx}), most BBHs cannot merge within a Hubble time in isolation. 

\begin{figure}
    \centering
    \includegraphics[width=1\columnwidth]{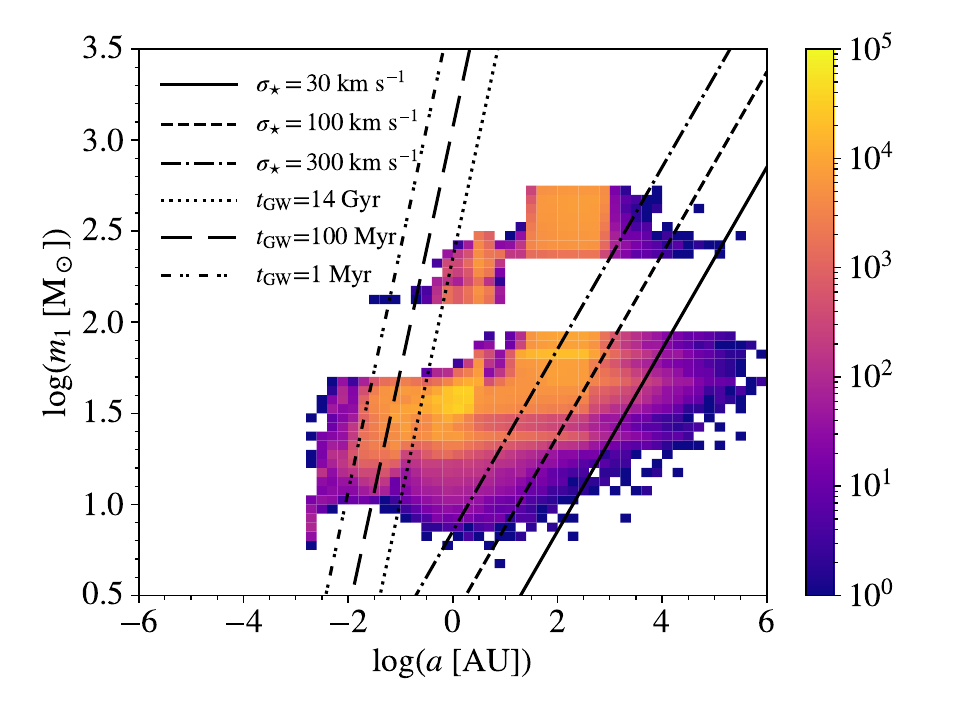}
    \vspace{-20pt}
    \caption{Distribution of Pop~III BBHs at formation in the $\log m_{1}-\log a$ space for the LOG1 simulation by \textsc{sevn} from \citet{Costa2023} with the close IBS model. To estimate which BBHs can merge by GW emission in galaxy fields, we plot the primary mass as a function of initial separation given the merger timescales $t_{\rm GW}=14000$ (dotted), 100 (long-dashed), and 1~Myr (dash-dot-dotted) for \textit{equal-mass BBHs on initially circular orbits}. Similarly, for the NSC-DH channel, we derive the minimum primary mass required for a BH binary to be hard in NSCs with $\sigma_{\star}=30$ (solid), 100 (dashed), and $300\ \rm km\ s^{-1}$ (dash-dotted) given the criterion $a<a_{\rm HDB}=Gm_{1}^{2}/\left[m_{\star,\rm SC}\sigma_{\star}^{2}\right]$.}
    \label{log1_m1_a}
\end{figure}

For the second point (ii), Fig.~\ref{log1_m1_a} shows the distribution of all Pop~III BBHs at formation in the $\log m_{1}-\log a$ space from the LOG1 \textsc{sevn} catalogue used in our fiducial model, where we also demonstrate the criteria for equal-mass BBHs on initially circular orbits to (a) merge in galaxy fields by GW emission for a given merger timescale $t_{\rm GW}$, and (b) survive {binary-single} encounters and merge by DH in NSCs with a given velocity dispersion $\sigma_{\star}$. On the left side of the dotted line for $t_{\rm GW}=14$~Gyr, we have $m_{1}\lesssim 50\ \rm M_\odot$ for most BBHs that can merge within a Hubble time in isolation. Such BBHs originate from close ($r_{\rm p}\lesssim 10$~AU) binaries of stars with ZAMS masses below $242\ \rm M_\odot$, which lose their envelopes through mass transfer and/or CE phases, so the resulting naked cores can only produce BHs below $\sim 50\ \rm M_\odot$ \citep{Iorio2023}. Close binaries of more massive stars tend to merge during unstable mass transfer when the donor leaves the MS as a red super-giant while the accretor \textit{still on the MS} also fills the Roche lobe due to significant expansion\footnote{If massive Pop~III stars remain compact during MS (by, e.g., suppressed convective overshooting or chemically homogeneous evolution), some of these close binaries can undergo stable mass transfer without stellar mergers and produce massive BBHs (with BHs above $130\ \rm M_\odot$) merging within a Hubble time \citep[][]{Tanikawa2021,Tanikawa2021mrd,Tanikawa2022,Hijikawa2021,Santoliquido2023,Mestichelli2024}. This indicates that the fates of massive binary stars and the efficiency of forming massive BBH mergers via IBSE are sensitive to the modelling of single star evolution. 
} \citep{Costa2023}. 
Initially wider ($r_{\rm p}\gtrsim 10$~AU) binaries undergo less and even negligible mass loss during binary stellar evolution and produce more massive BHs up to $86\ (550)\ \rm M_\odot$ from stars with ZAMS masses below (above) $242 \rm M_\odot$.

However, the resulting BBHs are mostly too wide ($a\gtrsim 0.3$~AU) to merge within a Hubble time in isolation. 
Nevertheless, the majority of such massive BBHs with wide orbits are still hard binaries in typical host NSCs with $\sigma_{\star}=30\ \rm km\ s^{-1}$ (solid line), so a significant fraction ($\sim 45-64$\%) of them can merge within a Hubble time by DH. 
In particular, BBHs with at least one BH inside or above the PISN mass gap $\sim 86-242\ \rm M_\odot$ originate from massive stellar binaries with $m_{\star,1}\gtrsim 242\ \rm M_\odot$ and form two groups in the $\log m_{1}-\log a$ space. The first group with $m_{1}\sim 130-320\ \rm M_\odot$ (including all BHs inside the PISN mass gap) and $a\sim 0.1-30$~AU is produced by stellar binaries with initial pericenter separations of $r_{\rm p}\sim 5-30$~AU, in which the primary star loses 
the H envelope by CE evolution and stable mass transfer during the red super-giant phase before collapsing into a BH directly. The second group with $m_{1}\sim 242-550\ \rm M_\odot$ (above the PISN mass gap) and $a\gtrsim 30$~AU comes from initially wider ($r_{\rm p}\gtrsim 30$~AU) stellar binaries in which mass transfer/loss via Roche lobe overflow never happens, 
so the primary star collapses entirely into a BH (as mass loss by stellar winds is negligible in the Pop~III tracks adopted by \textsc{sevn}). 
Almost all BBHs with BHs inside or above the PISN mass gap are massive enough to merge by DH even in massive NSCs with $\sigma_{\star}=300\ \rm km\ s^{-1}$ (dash-dotted line).

Beyond distinct progenitors, the two evolution channels also produce Pop~III BBH mergers in very different environments. For conciseness, here we briefly summarize the host halo/galaxy/NSC properties of Pop~III BBH mergers in our fiducial model and comment on how they vary with model parameters. 
Most ($\gtrsim 90\%$) field mergers occur in small haloes with $M_{\rm h}\lesssim 5\times 10^{9}\ \rm M_\odot$ at $z\gtrsim 0.8$, which host low-mass, metal-poor galaxies with $M_{\star}\lesssim 1.3\times 10^{6}\ \rm M_\odot$ and $Z\lesssim 0.22\ \rm Z_\odot$. The typical (median) host system has $M_{\rm h}\sim 9.4\times 10^{7}\ \rm M_\odot$, $M_{\star}\sim 3.5\times 10^{3}\ \rm M_\odot$, $Z\sim 0.005\ \rm Z_\odot$, and $z\sim 6.8$. In fact, $\sim 10$\% of Pop~III BBHs merge before the formation of the second generation of stars (i.e, $M_{\star}=0$). Therefore, direct observations of the host galaxies of Pop~III BBH mergers from IBSE are very challenging. 
On the contrary, the majority ($\sim 80\%$) of NSC mergers happen in more massive structures with $M_{\rm h}\sim 6.7\times 10^{9}-1.1\times 10^{11}\ \rm M_\odot$, $M_{\star}\sim 1.3\times 10^{6}-8\times 10^{8}\ \rm M_\odot$, $Z\sim 0.017-0.48\ \rm Z_\odot$, and $M_{\rm SC}\sim 9\times 10^{4}-3.4\times 10^{9}\ \rm M_\odot$ at $z\sim 0.1-3$. The typical host system has $M_{\rm h}\sim 2\times 10^{10}\ \rm M_\odot$, $M_{\star}\sim 1.5\times 10^{7}\ \rm M_\odot$, $Z\sim 0.075\ \rm Z_\odot$, $M_{\rm SC}\sim 4.5\times 10^{5}\ \rm M_\odot$, and $z\sim 0.7$. Nevertheless, the host galaxies of our Pop~III BBH mergers in NSCs are generally less massive than the galaxies hosting NSCs in local observations \citep[see their fig.~12]{Neumayer2020nuclear}, especially for Late-Type NSC-host galaxies that mainly have $M_{\star}\gtrsim 10^{8}\ \rm M_\odot$. This indicates that the NSC-DH channel favours low-mass, metal-poor, compact (Early-Type) galaxies in small haloes where the inspiral of Pop~III BBHs into NSCs by DF is efficient. We also find that among the host NSCs of Pop~III BBHs, each NSC typically only swallows a few Pop~III BBHs throughout its lifetime, so the total mass (as well as binding energy) carried by the Pop~III BBHs is minor (up to a few percent) compared with that of the NSC itself. This justifies our ignorance of the heating effect by DH of Pop~III BBHs on NSC evolution. 

In general, the host system properties of field mergers are insensitive to model parameters, although more massive systems at lower redshifts are slightly favoured by initially wider binaries with more bottom-heavy IMFs. However, increasing the abundances of Pop~III BBHs and NSCs will shift NSC mergers to less massive systems with lower metallicities at higher redshifts. For instance, in the model LOG5\_full with the highest abundances of Pop~III BBHs and NSCs, the typical host system has $M_{\rm h}\sim 10^{10}\ \rm M_\odot$, $M_{\star}\sim 3.6\times 10^{6}\ \rm M_\odot$, $Z\sim 0.05\ \rm Z_\odot$, $M_{\rm SC}\sim 2.8\times 10^{5}\ \rm M_\odot$, and $z\sim 1.1$. In the opposite case of KRO1\_low, we have $M_{\rm h}\sim 3.7\times 10^{10}\ \rm M_\odot$, $M_{\star}\sim 4.9\times 10^{7}\ \rm M_\odot$, $Z\sim 0.17\ \rm Z_\odot$, $M_{\rm SC}\sim 8\times 10^{5}\ \rm M_\odot$, and $z\sim 0.3$. These trends can be explained by the distinct merger histories in different models, which are discussed below. 

\subsection{Exploration of parameter space}
\label{sec:comp}

The general features of Pop~III BBH mergers from the IBSE and NSC-DH channels described above hold in all the 18 models considered in this paper. However, the detailed properties of BBH mergers and the relative importance of the two channels do vary greatly with the underlying assumptions on (1) Pop~III IMF, (2) IBS, and (3) occupation fraction of high-$z$ NSCs. In this subsection, we first discuss the general trends in the BBH formation and merger efficiencies (Sec.~\ref{sec:epsilon}). Then we illustrate the effects of each aspect on the merger history, SGWB, and mass distribution (Sec.~\ref{sec:ibs}-\ref{sec:focc}). 

\subsubsection{BBH formation and merger efficiencies}
\label{sec:epsilon}

\begin{figure}
    \includegraphics[width=1\columnwidth]{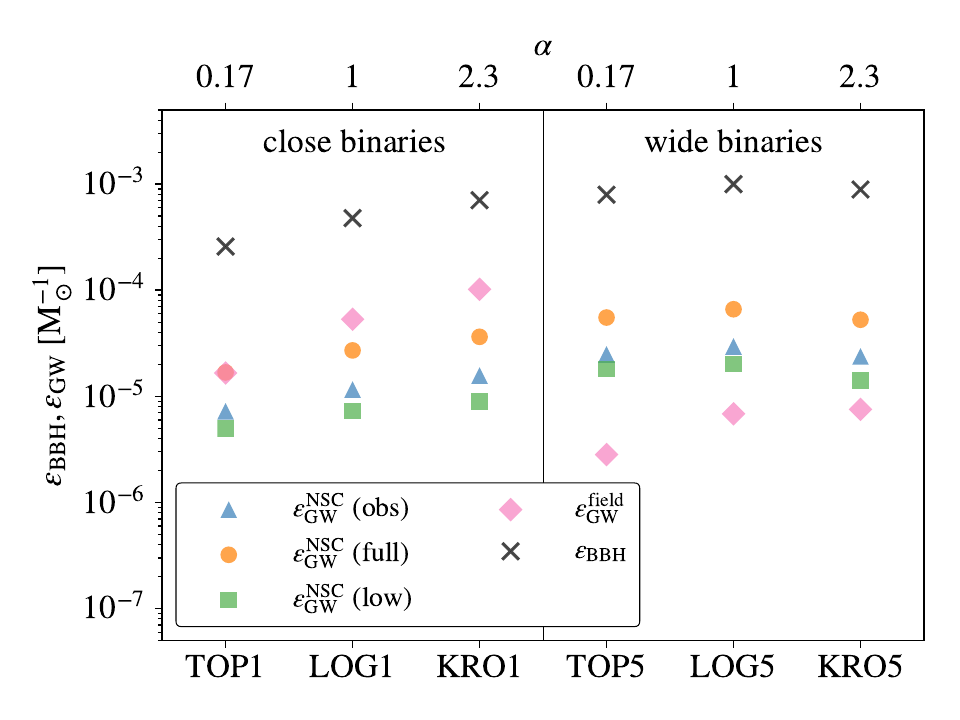}
    \vspace{-20pt}
    \caption{Pop~III BBH formation and merger efficiencies from the 18 models listed in Table~\ref{runlist}. The BBH formation efficiency (crosses) is independent of the NSC parameters, so we only show 6 data points for the 6 input \textsc{sevn} catalogues. The efficiency of field mergers (diamonds) is insensitive to the NSC parameters, and we only show the results for the fiducial NSC model (obs). For NSC mergers, we show the results for the obs, full, and low NSC models (Sec.~\ref{sec:nsc_form}) with triangles, circles, and squares, respectively. The left (right) section of the plot shows the results for the close (wide) IBS model (Sec.~\ref{sec:sample}). Within each section, the IMF becomes more bottom-heavy (with higher $\alpha$) from left to right. We have $\epsilon_{\rm GW}^{\rm field}\simeq \epsilon_{\rm GW}^{\rm NSC}$ in the model TOP1\_full, causing the two data points to overlap. }
    \label{epsilon_imf}
\end{figure}

\begin{figure}
    \centering
    \includegraphics[width=1\columnwidth]{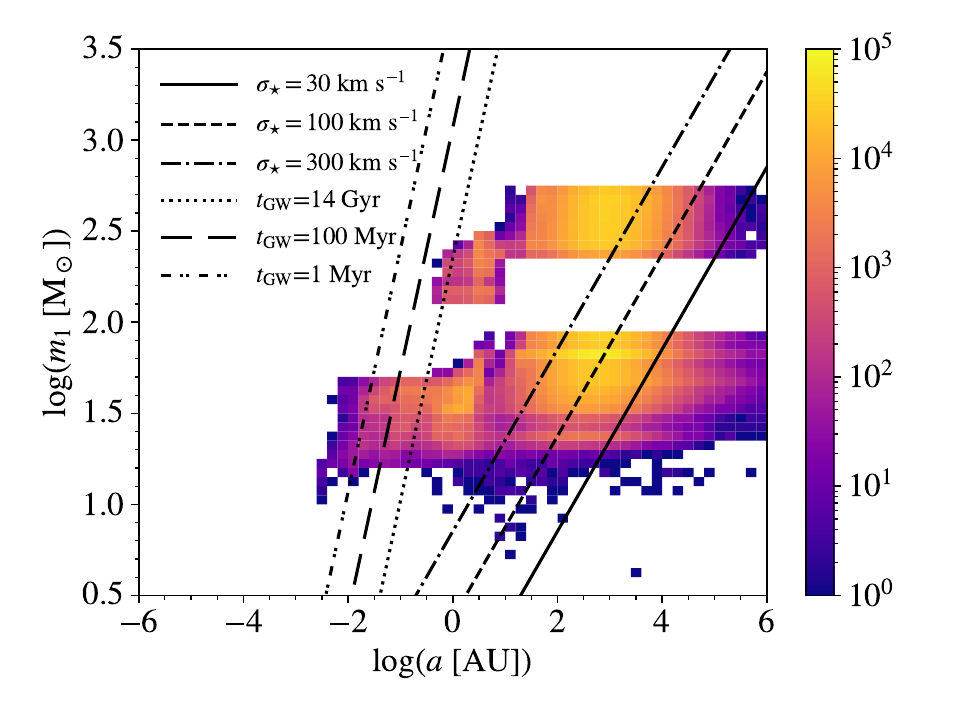}
    \vspace{-20pt}
    \caption{Same as Fig.~\ref{log1_m1_a} but for the LOG5 simulation by \textsc{sevn} from \citet{Costa2023} with the wide IBS model. }
    \label{log5_m1_a}
\end{figure}

The BBH formation and merger efficiencies of the 18 simulations are summarized in Table~\ref{runlist} and visualized in Fig.~\ref{epsilon_imf}. 
Under the wide IBS model in which close binary interactions are rare, $\epsilon_{\rm BBH}$ is not very sensitive to the IMF, and the fiducial log-flat IMF shows the highest $\epsilon_{\rm BBH}$. This is produced by two competing effects: When individual stars become more massive (with increasing $\alpha$), the number of stars/binaries is simply smaller for a fixed total stellar mass, but meanwhile, the fraction of stars massive enough to form BHs is larger. However, under the close IBS model where binary interactions are important, $\epsilon_{\rm BBH}$ becomes higher when the IMF is more bottom-heavy (with higher $\alpha$). The reason is that more massive stars are more likely to merge during close binary interactions due to their larger radii, so a larger fraction of stellar binaries that would form BBHs without interactions will be lost in mergers for a larger $\alpha$. Similarly, when there are more wide binaries in the IBS model, mergers of massive stars are suppressed, resulting in higher $\epsilon_{\rm BBH}$ and more massive BBHs, as shown in Fig.~\ref{log5_m1_a}. 

For Pop~III BBH mergers in galaxy fields, the merger efficiency $\epsilon_{\rm GW}^{\rm field}$ is sensitive to the NSC parameters since only a small ($\sim 3-12\%$) fraction of Pop~III BBHs fall into NSCs. Therefore, we focus on the results of $\epsilon_{\rm GW}^{\rm field}$ for the fiducial NSC model (obs). We find that $\epsilon_{\rm GW}^{\rm field}$ always increases with $\alpha$, and the trend is stronger with the close IBS model. The evolution of $\epsilon_{\rm GW}^{\rm field}$ with $\alpha$ is also stronger compared with the case of $\epsilon_{\rm BBH}$, as the fraction $f_{\rm GW}^{\rm field}$ of BBHs in galaxy fields that merge at $z>0$ increases with $\alpha$. These trends can be explained by the fact that field mergers mostly originate from interacting close ($r_{\rm p}\lesssim 10$~AU) binaries of relatively small stars ($\lesssim 240\ \rm M_\odot$) while close binaries of massive stars will merge, as discussed in Sec.~\ref{sec:fd_progenitor} (see Fig.~\ref{log1_m1_a}). 
For the same reason, with a fixed IMF, $\epsilon_{\rm GW}^{\rm field}$ is significantly reduced (by a factor of $\sim 6-14$) when we switch from the close IBS model to that dominated by wide binaries. The reduction is stronger when the IMF is more bottom-heavy because more massive stars are more likely to interact given a fixed separation. 

For NSC mergers, the variation of merger efficiency $\epsilon_{\rm GW}^{\rm NSC}$ with IMF and IBSE is very similar to that of $\epsilon_{\rm BBH}$. Indeed, if the probability for a BBH to merge in a NSC at $z>0$ is constant, $\epsilon_{\rm GW}^{\rm NSC}$ should be proportional to $\epsilon_{\rm BBH}$. In fact, the probability is slightly higher for more massive BBHs that sink into NSCs more efficiently by DF and require less hardening to merge rapidly via GW emission. As a result, on top of the trend driven by $\epsilon_{\rm BBH}$, $\epsilon_{\rm GW}^{\rm NSC}$ is further enhanced by a more top-heavy IMF or IBS with more wide binaries, which tend to produce more massive BBHs. This point is reflected in the weak evolution of the fraction $f_{\rm GW}^{\rm NSC}$ of BBHs in NSCs that merge at $z>0$ and the fraction $f_{\rm infall}$ of BBHs that fall into NSCs (see the last two columns of Table~\ref{runlist}) with IMF and IBS. As expected, $f_{\rm infall}$ and thus $\epsilon_{\rm GW}^{\rm NSC}$ increase when the occupation fraction of NSCs is higher. 
Combing these distinct features of field and NSC mergers, we find that the fraction $f_{\rm NSC}$ of mergers in NSCs across cosmic history is higher 
when the IMF is more top-heavy, the initial condition includes more wide stellar binaries, and high-$z$ NSCs are more abundant. 
The most important factor here 
is IBS: For the close IBS model, field mergers are more abundant than NSC mergers in most cases with $f_{\rm NSC}\sim 8-51$\% except for the TOP1\_full model in which the two populations of mergers have almost identical numbers. However, for the wide IBS model, NSC mergers always dominate with $f_{\rm NSC}\sim 65-95$\%. Besides, the BBH merger efficiency of the NSC-DH channel is less sensitive to the IMF and IBS compared with that of the IBSE channel. 


\subsubsection{Effects of initial binary statistics}
\label{sec:ibs}

\begin{figure}
    \centering
    \includegraphics[width=1\columnwidth]{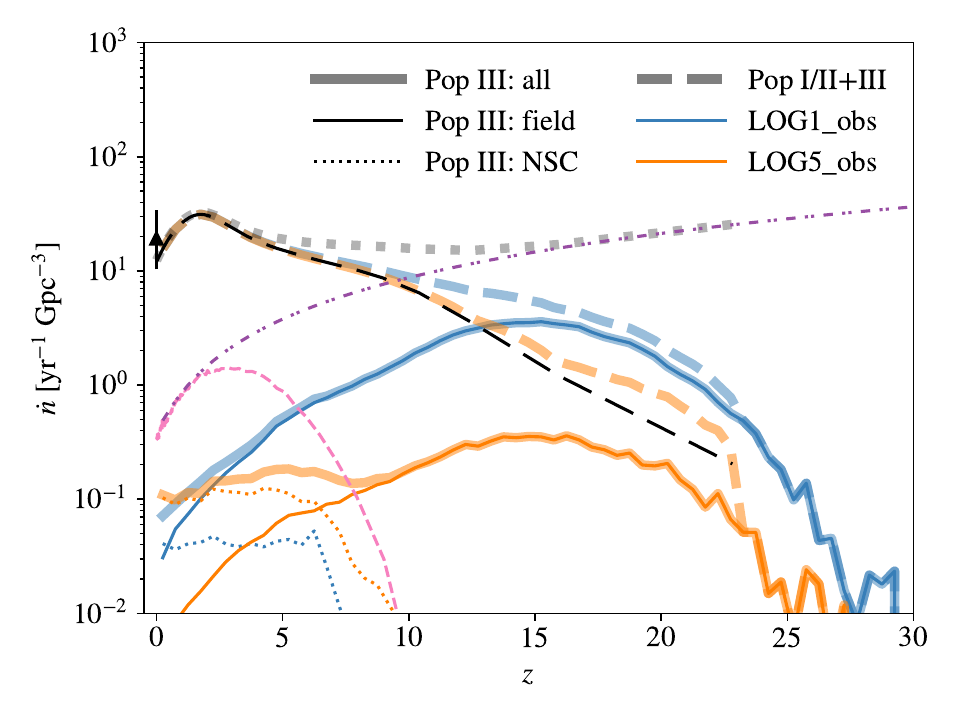}
    \vspace{-20pt}
    \caption{Same as Fig.~\ref{fd_dndz} but comparing the MRDs for two IBS models dominated by close (blue) and wide (orange) binaries assuming the log-flat IMF and fiducial NSC occupation fraction.}
    \label{bs_dndz}
\end{figure}

To illustrate the effects of IBS, as an example, we compare the results from LOG1\_obs (i.e., the fiducial model) and LOG5\_obs with the close and wide IBS models (see Sec.~\ref{sec:sample}), respectively, where the IMF and NSC models are fixed to the fiducial choices. 
Consistent with the trends seen in merger efficiencies $\epsilon_{\rm GW}$ (Fig.~\ref{epsilon_imf}), the MRDs of field and NSC mergers are reduced and increased when there are more wide binaries in the initial condition (LOG5\_obs), respectively, as shown in Fig.~\ref{bs_dndz}. As a result, NSC mergers dominate the MRD at $z\lesssim 6$ with the wide IBS model. 
Similarly, the reduction (enhancement) of the MRD of field (NSC) mergers is weaker (stronger) when the IMF is more bottom-heavy. 
The MRD redshift evolution for NSC mergers is not significantly affected by IBS, which holds for other choices of IMFs and NSC models. This indicates that the NSC-DH channel is mainly sensitive to the abundances of Pop~III BBHs rather than their initial properties, which is also reflected in the correlation between $\epsilon_{\rm GW}^{\rm NSC}$ and $\epsilon_{\rm BBH}$. 
For field mergers, the MRD evolution is insensitive to IBS at $z\gtrsim 5$, which always shows a peak at $z\sim 15$ closely following the peak of the Pop~III SFRD (Fig.~\ref{sfrd}), since this regime is dominated by mergers with short delay times. However, at lower redshifts when Pop~III star formation has terminated, and the MRD is dominated by mergers of long delay times, the decrease of MRD with decreasing redshift is slower for the wide IBS model compared with the case of the close IBS model. The reason is that the wide IBS model produces a larger fraction of BBHs with long merger timescales ($t_{\rm GW}\gtrsim 300$~Myr) 
\citep[see fig.~11 in][]{Costa2023}. 
This effect is stronger when the IMF is more top-heavy. 
In fact, the close and wide IBS models prefer different binary evolution pathways to BBH mergers in isolation, which produce different delay time distributions \citep[see fig. 12 and 17 in][]{Costa2023}. The former favours stable mass transfer during early evolutionary stages, while the latter is dominated by CE evolution of red giants with large radii ($\sim 1-10$~AU). The reader is referred to \citet{Costa2023} and \citet{Santoliquido2023} for in-depth discussions on the roles played by different binary evolution pathways in shaping the properties of BBH mergers via IBSE.

\begin{figure}
    \centering
    \includegraphics[width=1\columnwidth]{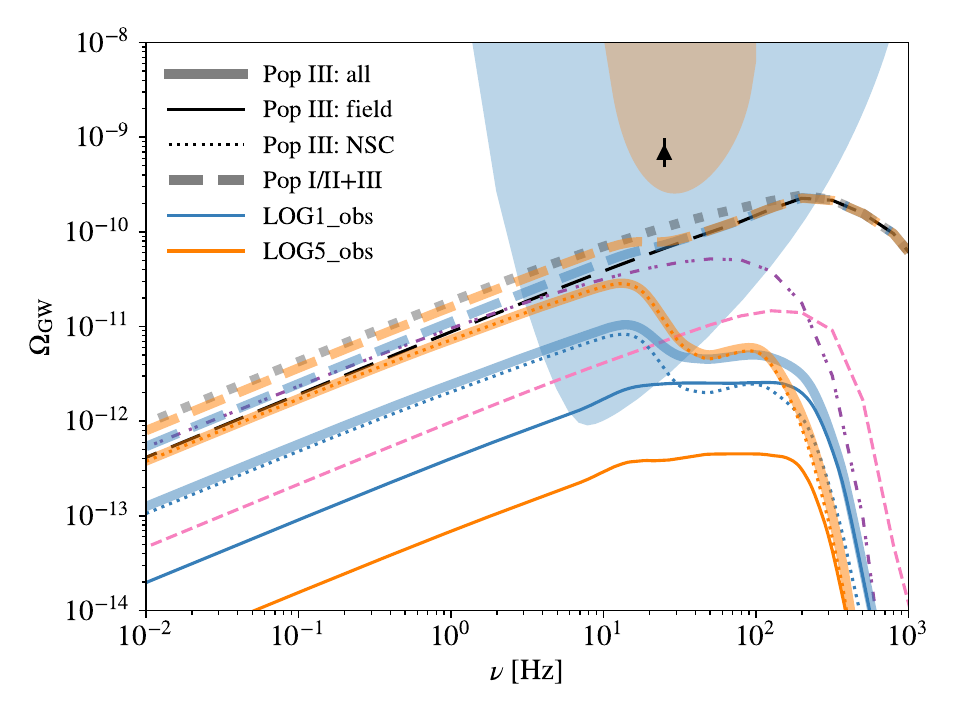}
    \vspace{-20pt}
    \caption{Same as Fig.~\ref{fd_ogw} but comparing the energy spectra of SGWB for the two IBS models dominated by close (blue) and wide (orange) binaries assuming the log-flat IMF and fiducial NSC occupation fraction.}
    \label{bs_ogw}
\end{figure}

Fig.~\ref{bs_ogw} shows the contributions of Pop~III BBH mergers to the SGWB for the two IBS models. Compared with LOG1\_obs with the close IBS model, the contribution from field (NSC) mergers is lower (higher) for LOG5\_obs with the wide IBS model, where NSC mergers completely dominate the Pop~III SGWB at $\nu\lesssim 200$~Hz. 
This dominance of NSC mergers is a common feature of the wide IBS model regardless of the chosen IMF and NSC parameters. This can be understood by the fact that the SGWB (observed at $z=0$) captures the summation of the GW energies from all mergers across cosmic history and more massive mergers at lower $z$ contribute higher energies to the SGWB (Sec.~\ref{sec:fd_sgwb}). 
As mentioned in the last subsection and shown in Fig.~\ref{bs_dndz}, the merger efficiency and local MRD of NSC mergers are always larger than those of field mergers under the wide IBS model. Moreover, NSC mergers are more massive than field mergers, and thus release more energies in GWs (Sec.~\ref{sec:fd_mass}). Therefore, in LOG5\_obs, the contribution from NSC mergers is higher than that from field mergers by about 100 times in the Pop~III SGWB, being as strong as the SGWB from PBH mergers predicted by \citet{Bavera2022} at $\nu\lesssim 10$~Hz. {In this case, Pop~III NSC mergers produce a significant flattening feature around $\nu\sim 10-20\ \rm Hz$ deviating from the canonical shape $\Omega_{\rm GW}\propto f^{2/3}$ in the total SGWB from stellar BBH mergers (Pop~I/II+III).}

\begin{figure}
    \centering
    \includegraphics[width=1\columnwidth]{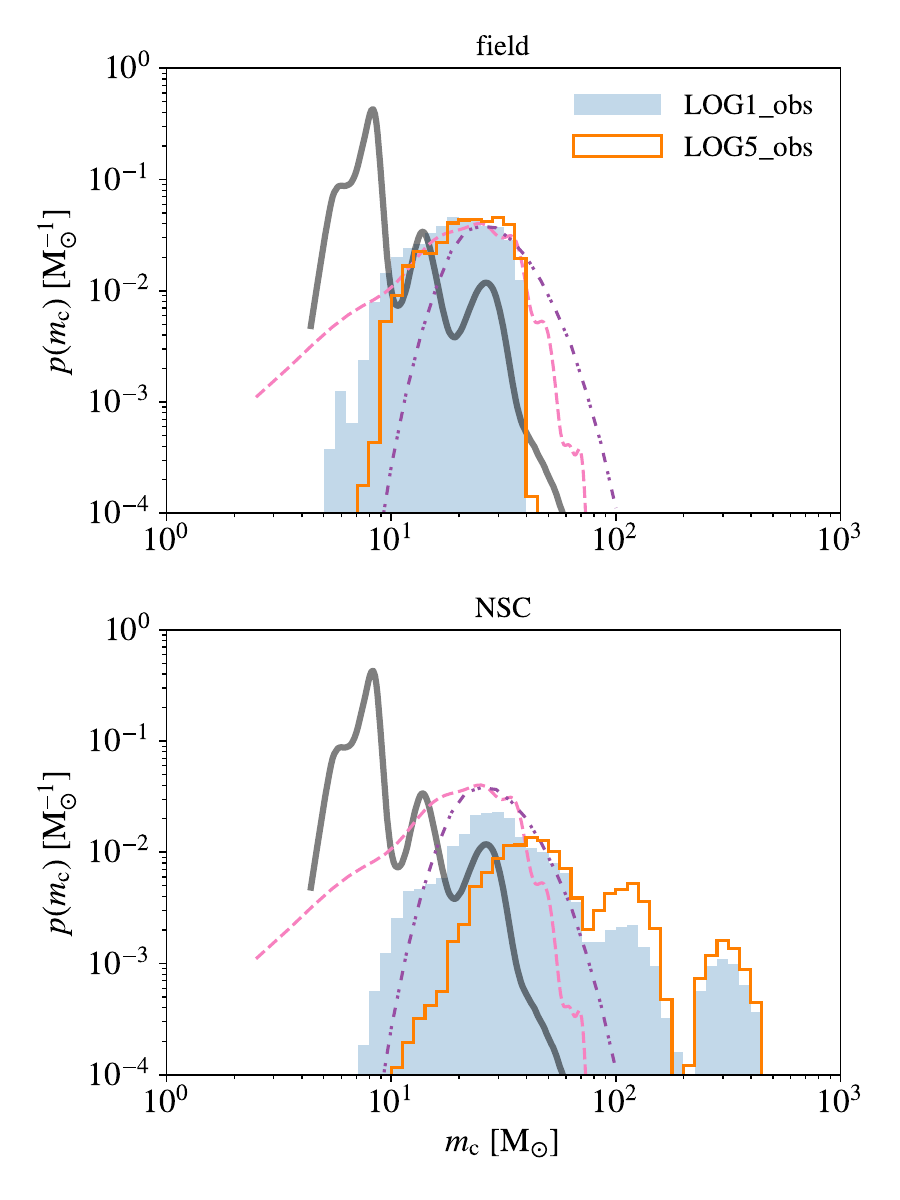}
    \vspace{-20pt}
    \caption{Same as Fig.~\ref{fd_mchirp}, but comparing the chirp mass distributions of Pop~III BBH mergers in galaxy fields (middle) and NSCs (bottom) for the two IBS models dominated by close (blue shaded region) and wide (orange contour) binaries assuming the log-flat IMF and fiducial NSC occupation fraction. Here, we only show the results for Pop~I/II BBH mergers in GCs (dashed curve) and PBH mergers (dash-dot-dotted curve) from the best-fit model in \citet{Franciolini2022} adopted by \citet[see their fig.~1]{Bavera2022}, and the intrinsic chirp mass distribution (thick solid curve) inferred from GWTC-3 events from \citep[see their fig.~2]{Abbott2023}.}
    \label{bs_mchirp}
\end{figure}

Finally, we compare the chirp mass distributions between the two models in Fig.~\ref{bs_mchirp} (see also Table~\ref{tmass}). We find that both field and NSC mergers become more massive with the wide IBS model, and the effect is stronger for NSC mergers. The average chirp mass increases by 10\% and 35\% for field and NSC mergers given the wide IBS model compared with the case of the close IBS model. 
On the one hand, mass loss and stellar mergers during close binary interactions are suppressed in the wide IBS model, which enhances the formation of massive BBHs (see Fig.~\ref{log1_m1_a} and \ref{log5_m1_a}) and thus greatly increases the masses of NSC mergers (whose mass distribution closely correlates with the mass distribution of all BBHs at formation, see Fig.~\ref{fd_m}). Besides, the fraction of NSC mergers with small mass ratios $q\lesssim 0.4$ is higher in the wide IBS model. 
On the other hand, since close binary interactions are required to produce BBHs that can merge within a Hubble time in isolation, field mergers mostly involve low-mass BHs ($\lesssim 50\ \rm M_\odot$) regardless of the IBS model. Therefore, changing from the close IBS model to the wide IBS model only shifts the primary mass distribution of field mergers to the massive end by a few solar masses (but still limited by $m_{1}\lesssim 50\ \rm M_\odot$), leading to slightly larger chirp masses.







\subsubsection{Effects of IMF}
\label{sec:imf}

\begin{figure}
    \centering
    \includegraphics[width=1\columnwidth]{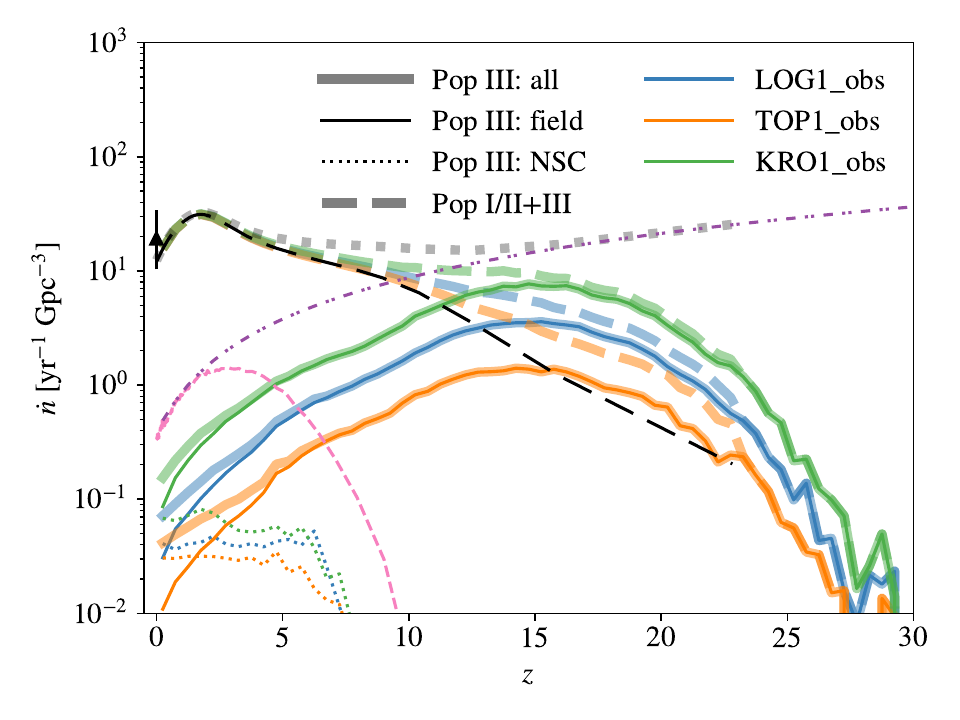}
    \vspace{-20pt}
    \caption{Same as Fig.~\ref{fd_dndz}, but comparing the MRDs for the three IMFs with power-law slopes of $\alpha=1$ (blue) $0.17$ (orange) and 2.3 (green), assuming the close IBS and fiducial NSC models. }
    \label{imf_dndz}
\end{figure}

\begin{figure}
    \centering
    \includegraphics[width=1\columnwidth]{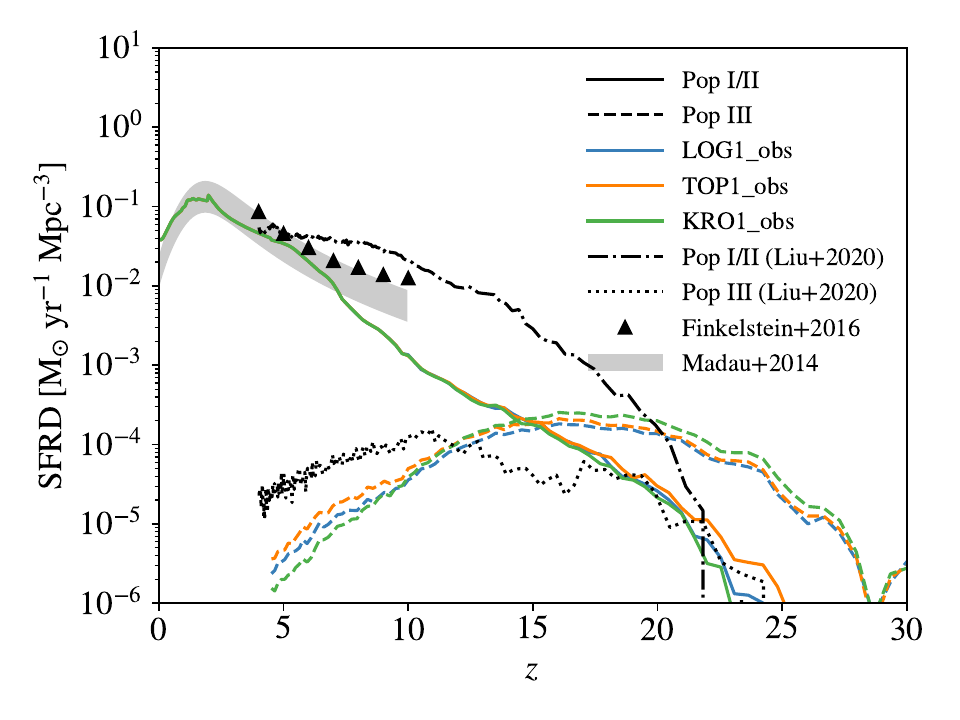}
    \vspace{-20pt}
    \caption{Same as Fig.~\ref{sfrd} but comparing the SFRDs of Pop~III (dashed) and Pop I/II (solid) stars from LOG1\_obs (blue), TOP\_obs (orange), and KRO1\_obs (green) with Pop~III IMF slopes of $\alpha=1$, $0.17$, and 2.3 under the close IBS model and the fiducial NSC model. }
    \label{imf_sfrd}
\end{figure}

Next, we look into the effects of IMF variations using the results from LOG1\_obs, TOP\_obs, and KRO1\_obs with IMF slopes of $\alpha=1$, $0.17$, and 2.3 (see Sec.~\ref{sec:sample}), respectively, where we adopt the close IBS model and fiducial NSC occupation fraction. As shown in Fig.~\ref{imf_dndz}, when the IMF is more bottom-heavy (with a larger $\alpha$), the MRD of field (NSC) mergers is significantly (mildly) increased. For both field and NSC mergers, the MRD redshift evolution is insensitive to the IMF, while the normalization of the MRD curve is approximately proportional to the merger efficiency (Fig.~\ref{epsilon_imf}), whose dependence on the IMF is explained in Sec.~\ref{sec:epsilon}. These trends reflect that the IMF has minor effects on the delay time distributions for both field and NSC mergers, and that the Pop~III formation history does not vary greatly with $\alpha$, as shown in Fig.~\ref{imf_sfrd}. {In fact, the (co-moving) stellar mass density of Pop~III stars ever formed in our simulation box is $\rho_{\star,\rm III}\simeq 4.6$, $5.6$ and $5.7\times 10^{4}\ \rm M_\odot\ Mpc^{-3}$ for $\alpha=1$, 0.17, and 2.3, respectively. The small variation of the Pop~III star formation history with $\alpha$ justifies our simple modelling of Pop~III stellar feedback (Sec.~\ref{sec:sample}). Our $\rho_{\star,\rm III}$ values are generally consistent with the predictions from cosmological simulations \citep[e.g.,][]{Johnson2013,Smith2015,Xu2016,Sarmento2017,Liu2020gw,Liu2020did,Venditti2023,Incatasciato2023} and other semi-analytical models \citep[e.g.,][]{Dayal2020,Visbal2020,Hegde2023,Ventura2024,Feathers2024} and the upper limit $\sim 10^{5-6}\ \rm M_\odot\ Mpc^{-3}$ placed by reionization \citep{Visbal2015,Inayoshi2016}.}

\begin{figure}
    \centering
    \includegraphics[width=1\columnwidth]{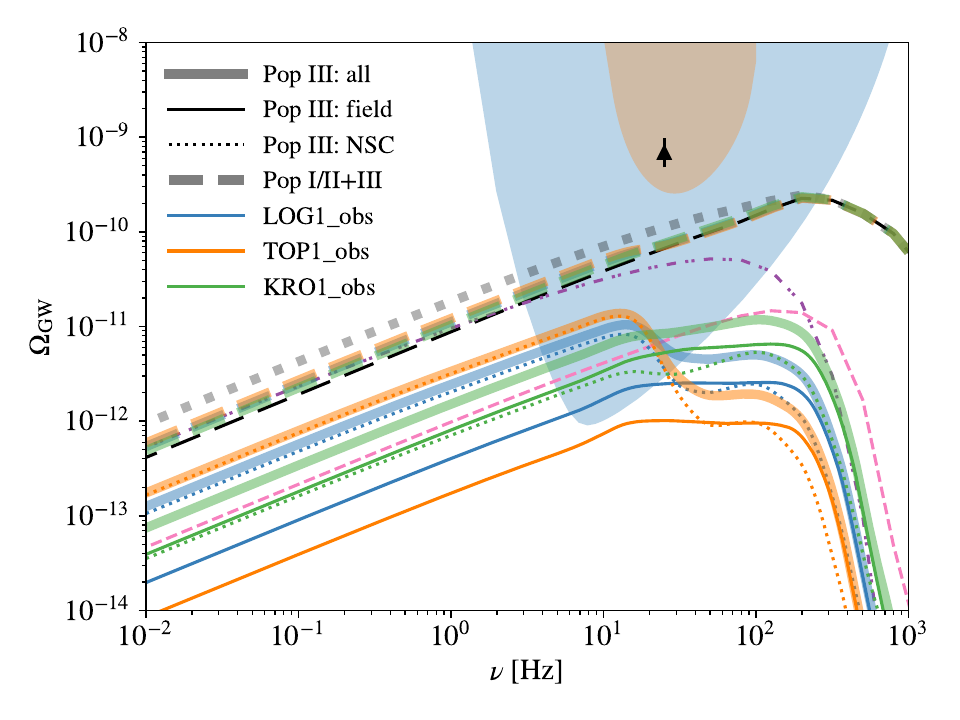}
    \vspace{-20pt}
    \caption{Same as Fig.~\ref{fd_ogw}, but comparing the energy spectra of SGWB for the three IMFs with power-law slopes of $\alpha=1$ (blue) $0.17$ (orange) and 2.3 (green), under the close IBS model and the fiducial NSC model.}
    \label{imf_ogw}
\end{figure}

\begin{figure}
    \centering
    \includegraphics[width=1\columnwidth]{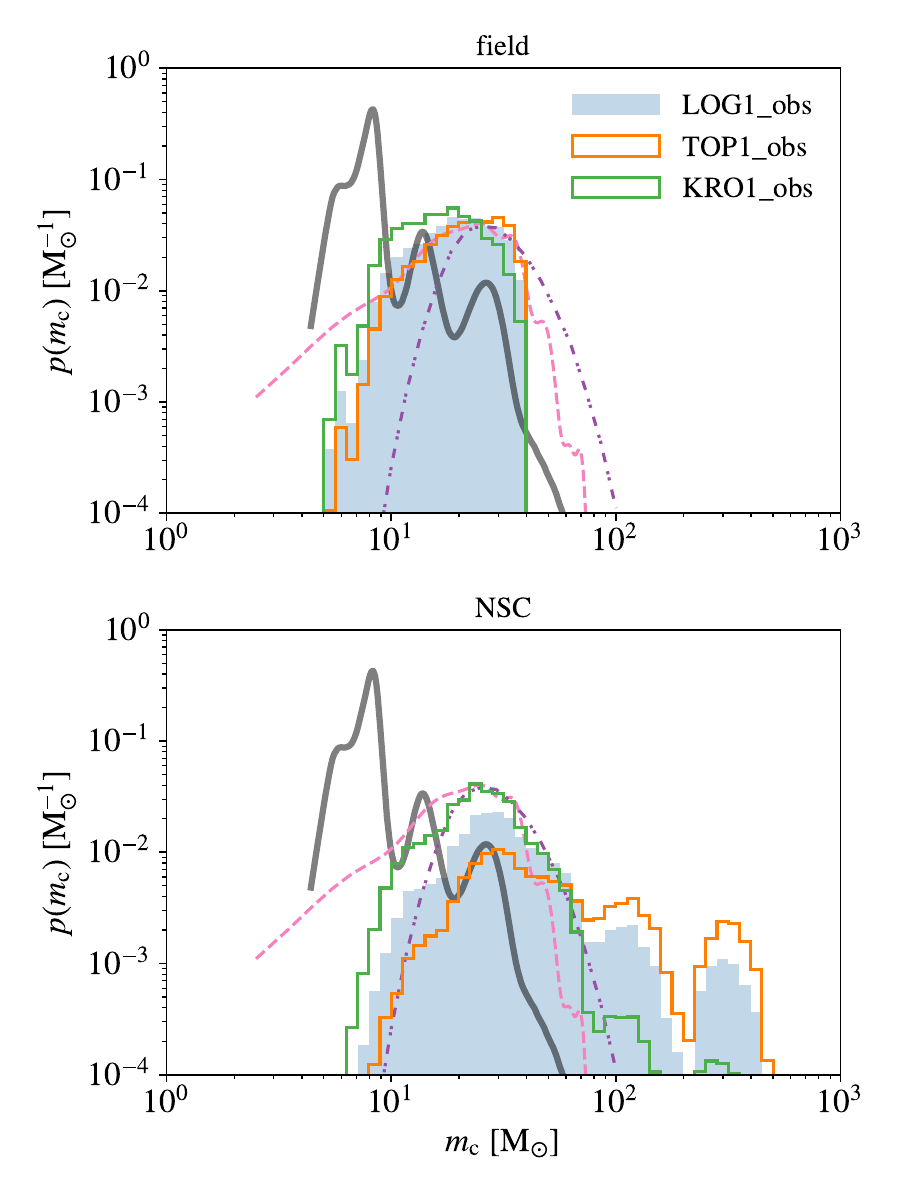}
    \vspace{-20pt}
    \caption{Same as Fig.~\ref{fd_mchirp} and Fig.~\ref{bs_mchirp}, but comparing the chirp mass distributions of Pop~III BBH mergers in galaxy fields (middle) and NSCs (bottom) for the three IMFs with power-law slopes of $\alpha=1$ (blue shaded region) $0.17$ (orange contour) and 2.3 (green contour), again under the close IBS and fiducial NSC models.}
    \label{imf_mchirp}
\end{figure}

The SGWB contributed by Pop~III BBH mergers in galaxy fields also increases with $\alpha$, as shown in Fig.~\ref{imf_ogw}. The shape of the energy spectrum at $\nu\sim 15-200$~Hz varies slightly with the IMF. Compared with the flat spectrum for $\alpha=1$, $\Omega_{\rm GW}$ slowly increases (decreases) with $\nu$ for $\alpha=2.3$ (0.17). The reason is that BBH mergers become slightly more massive when the IMF is more top-heavy (see Fig.~\ref{imf_mchirp} and Table~\ref{tmass}), which enhances GW emission at lower frequencies. However, for NSC mergers, the SGWB increases with $\alpha$ at $\nu\gtrsim 30$~Hz, but decreases with $\alpha$ at $\nu\lesssim 30$~Hz. Here, the high-frequency regime is dominated by low-mass mergers with $m_{\rm c}\lesssim 60\ \rm M_\odot$ whose fraction increases with $\alpha$ as shown in Fig.~\ref{imf_mchirp}. Since the merger efficiency and local MRD of NSC mergers also increase with $\alpha$ (Fig.~\ref{epsilon_imf} and \ref{imf_dndz}), it is reasonable that the SGWB from such low-mass mergers is stronger. On the other hand, the low-frequency part of the SGWB is produced by massive mergers with $m_{\rm c}\gtrsim 60\ \rm M_\odot$ whose fraction is reduced when the IMF is more bottom-heavy (see Fig.~\ref{imf_mchirp} and Table~\ref{tmass}). This reduction overcomes the increase of merger number with $\alpha$ and drives the decrease of SGWB with $\alpha$. 

\subsubsection{Effects of NSC occupation fraction}
\label{sec:focc}

\begin{figure}
    \centering
    \includegraphics[width=1\columnwidth]{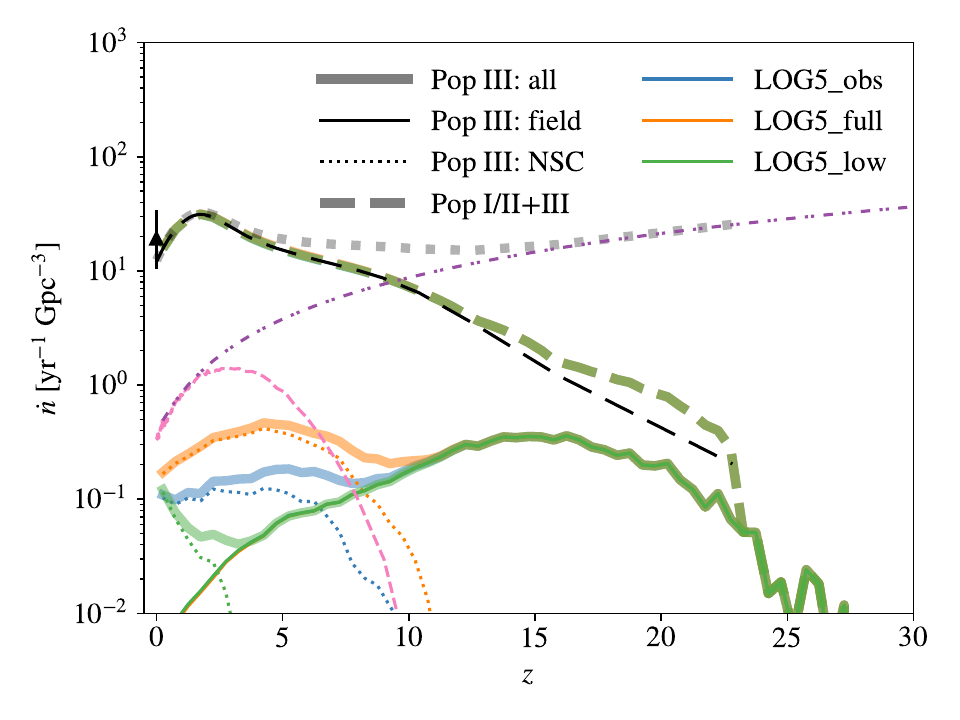}
    \vspace{-20pt}
    \caption{Same as Fig.~\ref{fd_dndz} but comparing the MRDs for the fiducial (blue), optimistic (orange), and conservative (green) models of NSC occupation fraction under the wide IBS model and log-flat IMF.}
    \label{focc_dndz}
\end{figure}

\begin{figure}
    \centering
    \includegraphics[width=1\columnwidth]{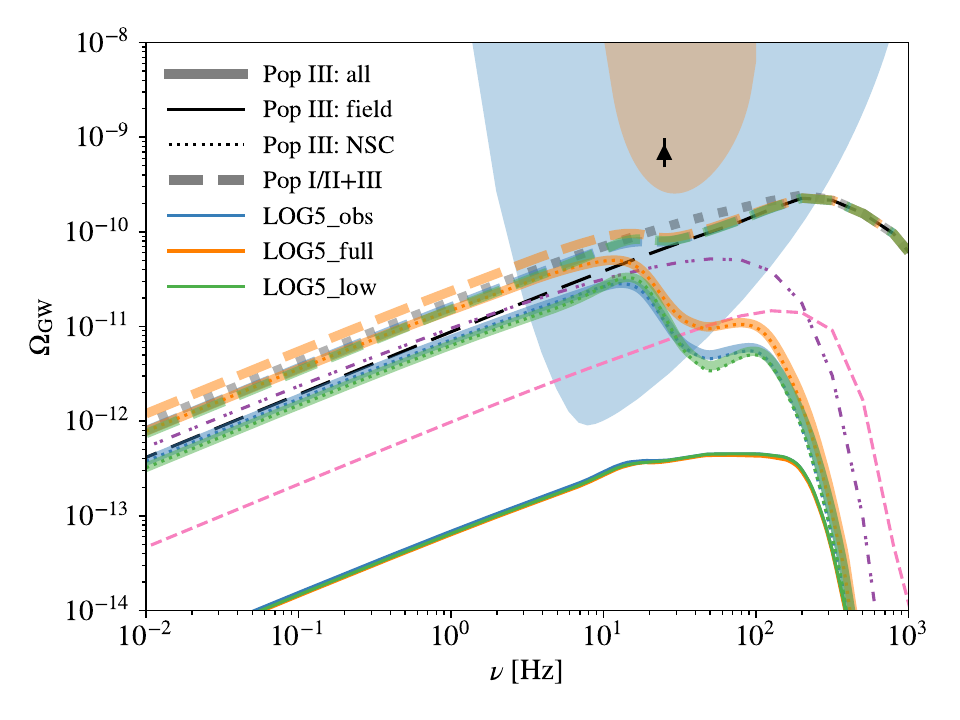}
    \vspace{-20pt}
    \caption{Same as Fig.~\ref{fd_ogw} but comparing the energy spectra of SGWB for the fiducial (blue), optimistic (orange), and conservative (green) models of NSC occupation fraction under the wide IBS model and log-flat IMF.}
    \label{focc_ogw}
\end{figure}

Finally, we consider the effects of the NSC occupation fraction in Fig.~\ref{focc_dndz}-\ref{focc_ogw} with the wide IBS model and log-flat IMF. We find that the properties of field mergers are hardly affected by the NSC parameters, which is reasonable as only a small ($\lesssim 12$\%) fraction of Pop~III BBHs ever fall into NSCs. Therefore, we focus on NSC mergers in this subsection. 
The chirp mass distributions of NSC mergers are very similar in the three NSC models and not shown here for conciseness (see Table~\ref{tmass} for the average chirp masses). 
As expected, the MRD of NSC mergers is enhanced when more galaxies host NSCs. Unlike the fiducial NSC model where the MRD of NSC mergers flattens at $z\lesssim 4$, the MRD of NSC mergers shows a peak at $z\sim 4$ in the optimistic NSC model (where every galaxy with $M_{\star}>10^{6}\ \rm M_\odot$ hosts a NSC). The MRD keeps increasing rapidly towards lower redshifts in the conservative NSC model, producing a local MRD that is slightly higher than that in the fiducial NSC model. 
The difference between the fiducial and conservative NSC models in the SGWB is very small, which can be understood with their similar merger efficiencies and local MRDs, although the peak value of the SGWB in the conservative NSC model is slightly ($\sim 16\%$) higher than that in the fiducial NSC model due to the slightly higher local MRD. 
The SGWB is stronger in the optimistic NSC model as expected from its higher merger efficiency and local MRD. In this case (LOG5\_full), the Pop~III SGWB becomes comparable ($\sim 84$\%) to the SGWB from Pop~I/II and PBH mergers predicted by \citet{Bavera2021} at $\nu\lesssim 10$~Hz{, which may leave a detectable feature in the total SGWB \citep{Perigois2021,Martinovic2022}}. 


\subsection{Observational perspective}
\label{sec:et}

\begin{figure}
    \centering
    \includegraphics[width=1\columnwidth]{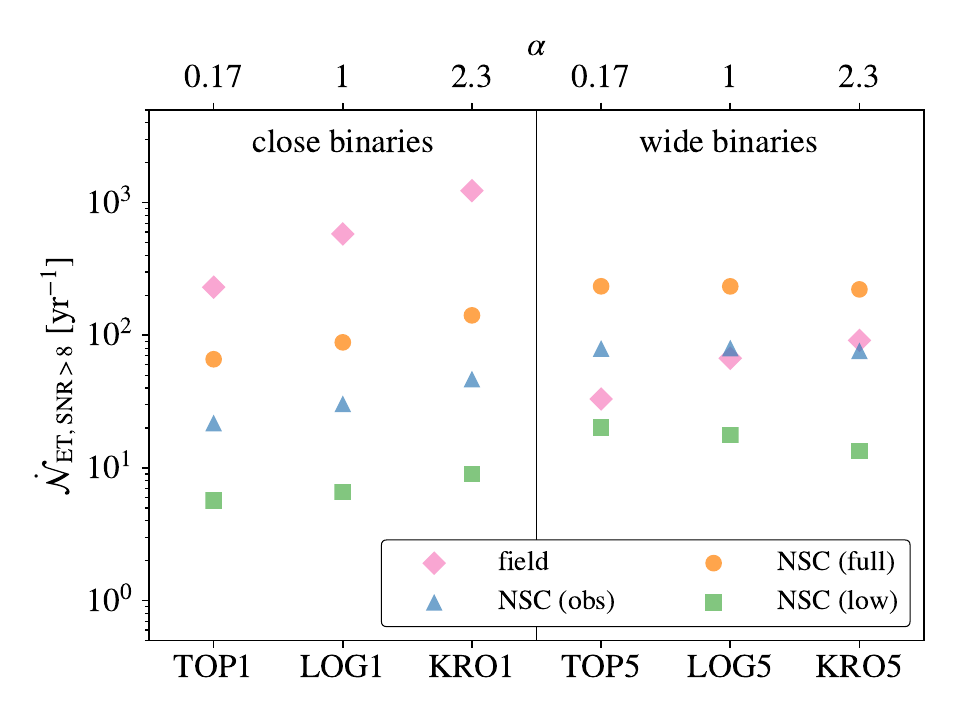}
    \caption{Detection rates of Pop~III BBH mergers by ET with $\rm SNR>8$ for the 18 models listed in Table~\ref{runlist}. The detection rates of field mergers (diamonds) are insensitive to the NSC parameters, and we only show the results for the fiducial NSC model (obs). For NSC mergers, we show the detection rates for the obs, full, and low NSC models (Sec.~\ref{sec:nsc_form}) with triangles, circles, and squares, respectively. The left (right) section of the plot shows the results for the close (wide) IBS model (Sec.~\ref{sec:sample}). Within each section, the IMF becomes more bottom-heavy (with higher $\alpha$) from left to right. }
    \label{ndet_et}
\end{figure}

We estimate the detection rates $\dot{\mathcal{N}}_{\rm ET,SNR>8}$ of Pop~III BBH mergers by ET for all the 18 models (Table~\ref{runlist}) given the detection probability of each source with $\rm SNR>8$ derived from the \texttt{python} package \textsc{gwtoolbox} \citep{Yi2022a,Yi2022b}, using the formalism of Eq.~\ref{rate}. For simplicity, we do not distinguish high-$z$ and low-$z$ sources but sum up all events detectable by ET. We also do not consider eccentricities in this calculation as in the case of SGWB (Sec.~\ref{sec:fd_sgwb}). 
The results are summarized in Fig.~\ref{ndet_et}, where we only show the detection rates of field mergers with the fiducial NSC model since the results for the other two NSC models are almost identical. As discussed in Sec.~\ref{sec:fd_mrd}, most ($\gtrsim 90$\%) BBH mergers with $z<20$ will be detectable by ET thanks to its high sensitivity across a broad frequency range, so the all-sky merger rate $\dot{\mathcal{N}}(z<20)$ within $z<20$ is a good approximation of the ET detection rate (for $\rm SNR>8$). Therefore, we derive the fraction of $\dot{\mathcal{N}}(z<20)$ contributed by Pop~III BBHs based on the reference rates for Pop~I/II and PBH mergers from \citet{Franciolini2022} to estimate the chance of finding Pop~III BBHs by ET. The results are summarized in Table~\ref{tobs}. 

In general, the ET detection rate of all Pop~III BBH mergers is $\dot{\mathcal{N}}_{\rm ET,SNR>8}\sim 50-1370\ \rm yr^{-1}$, and we have $\dot{\mathcal{N}}_{\rm ET,SNR>8}\sim 30-1230\ \rm yr^{-1}$ and $\dot{\mathcal{N}}_{\rm ET,SNR>8}\sim 6-230\ \rm yr^{-1}$ for field and NSC mergers, respectively. The highest rate is achieved by KRO1\_full, in which case Pop~III BBHs contribute $\sim 6.5$\% of the all-sky merger rate within $z<20$. 
The dependence of detection rates on model parameters follows similar trends as those seen in merger efficiencies (Fig.~\ref{epsilon_imf}), which are explained in Sec.~\ref{sec:epsilon}. However, here the effects of NSC parameters are stronger, and field mergers become more important. The reason is that the detection rate is more sensitive to the high-$z$ MRD than the merger efficiency. IBS still plays the most important role in determining the relative importance of the NSC-DH and IBSE channels. Given the close IBS model, field mergers always dominate the total detection rate, as the contribution of NSC mergers remains low ($\sim 0.7-22$\%). With the wide IBS model, the NSC mergers account for $\sim 13-88$\% of the total detection rate. Their contributions are only higher than those of field mergers in 5 of the 9 models with relatively high NSC occupation fractions and large $\alpha$, although they always dominate the total number of mergers across cosmic history with $f_{\rm NSC}\sim 65-95\%$. 

\begin{figure}
    \centering
    \includegraphics[width=1\columnwidth]{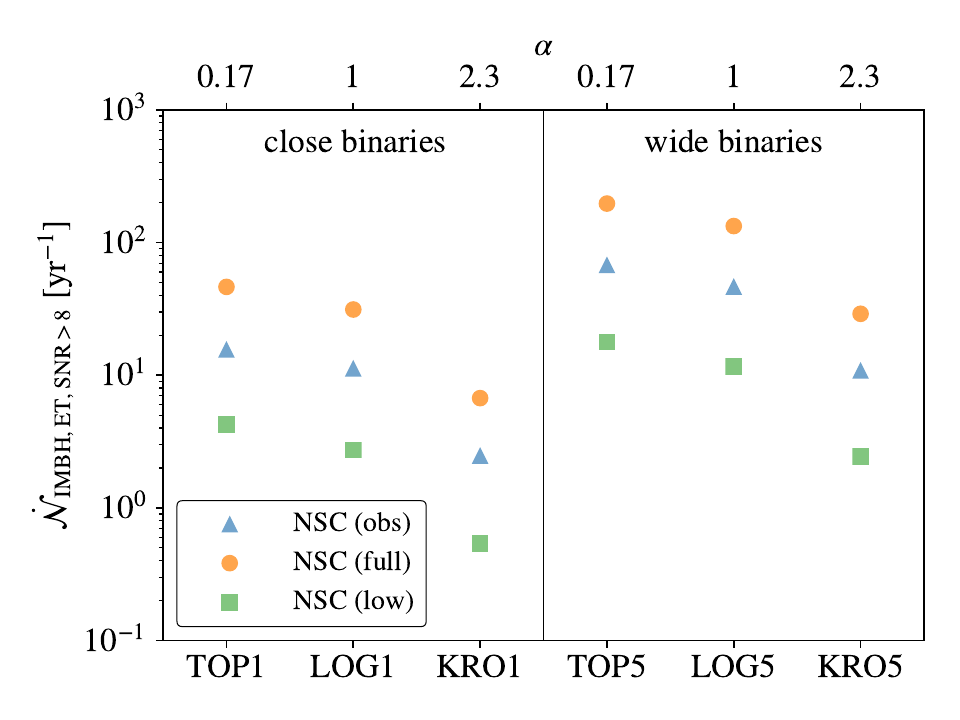}
    \caption{Same as Fig.~\ref{ndet_et} but for the detection rates of Pop~III BBH mergers including at least one IMBH ($m_{1}>100\ \rm M_\odot$) in NSCs. The detection rates of such mergers in galaxy fields remain below $0.1\ \rm yr^{-1}$ and are therefore not shown here.}
    \label{ndet_imbh_et}
\end{figure}

As discussed in Sec.~\ref{sec:fd_mass}, an important feature of the NSC-DH channel is the ability to produce Pop~III BBH mergers with massive ($\gtrsim 50\ \rm M_\odot$) BHs, especially those above $130\ \rm M_\odot$. In light of this, we further look into the ET detection rates $\dot{\mathcal{N}}_{\rm ET,IMBH,SNR>8}$ of Pop~III mergers with at least one IMBH ($m_{1}>100\ \rm M_\odot$). We find that such mergers are extremely rare in galaxy fields, with ET detection rates below $0.1\ \rm yr^{-1}$ in all the 18 models considered here. However, we have $\dot{\mathcal{N}}_{\rm ET,IMBH,SNR>8}\sim 0.5-200\ \rm yr^{-1}$ for NSC mergers as shown in Fig.~\ref{ndet_imbh_et}. The completeness of detection of such mergers by ET is also very high ($\gtrsim 95$\%). 
Here, the detection rate is sensitive to all model parameters on Pop~III IMF, IBS, and occupation fraction of high-$z$ NSCs, implying that BBH mergers with IMBHs\footnote{Because of PISNe, all IMBHs produced by Pop~III stars in our models are above $130\ \rm M_\odot$ (see Fig.~\ref{fd_m}), which is a common feature of the adopted BPS catalogues from \textsc{sevn} \citep{Costa2023}, and also found in other BPS studies \citep{Tanikawa2021,Tanikawa2022,Hijikawa2021}.} of $\sim 100-1000\ \rm M_\odot$ are valuable for constraining the properties of Pop~III binaries and high-$z$ SCs with GW observations. 

{We obtain similar detection rates for CEO\footnote{We use the CEO second stage (wideband) sensitivity curve \citep[see fig.~2 of][]{Abbott2017} from \url{https://dcc.ligo.org/LIGO-P1600143/public} (\texttt{curve\_data.txt}).}: $\dot{\mathcal{N}}_{\rm CEO,SNR>8y}\sim 30-1270\ \rm yr^{-1}$ for field mergers, $\dot{\mathcal{N}}_{\rm CEO,SNR>8}\sim 6-210\ \rm yr^{-1}$ for NSC mergers, and $\dot{\mathcal{N}}_{\rm CEO,SNR>8}\sim 0.5-120\ \rm yr^{-1}$ for mergers involving IMBHs. The main difference here is that the detection probability of massive sources by CEO is lower at $z\gtrsim 2$, because CEO is less sensitive than ET at low frequencies \citep[$\nu\lesssim 10\ \rm Hz$, see fig.~8 of][]{DeLuca2021}. The completeness of detection of NSC mergers (with IMBHs) can be as low as $\sim 70\ (60)\%$ for CEO (in TOP5\_full), while it remains $\gtrsim 95\%$ for ET.}

Finally, we estimate the detection rate of Pop~III BBH mergers by the LVK network during the O4 run with $\rm SNR>8$ as $\dot{\mathcal{N}}_{\rm LVK,SNR>8}\sim 0.9-9\ \rm yr^{-1}$. 
This small detection rate is consistent with the small local MRD $\dot{n}(z=0)\sim 0.034-0.17\ \rm yr^{-1}\ Gpc^{-3}$ that is much lower than the observed local MRD $\dot{n}_{\rm obs}(z=0)=19.3_{-9}^{+15.1}\ \rm yr^{-1}\ Gpc^{-3}$ \citep{Abbott2021}. 
This indicates that Pop~III mergers are overwhelmed by other populations of BBH mergers in the low-$z$ Universe (see also Table~\ref{tobs}). Although Pop~III stars can still be important for the most massive mergers, such events are rare in LVK observations. 
For instance, the LVK O4 detection rate of Pop~III mergers with IMBHs (in NSCs) is $\dot{\mathcal{N}}_{\rm LVK,IMBH,SNR>8}\sim 0.04-1.5\ \rm yr^{-1}$, and the expected number of events detected in 18 months\footnote{\url{https://observing.docs.ligo.org/plan/}} of observations is only $\sim 0.07-2.2$. 
Considering the typical horizon redshift $z_{\rm horizon}\sim 1$ of the LVK network during O3 for BBH mergers with total masses $\sim 100-300\ \rm M_\odot$ \citep[see fig.~8 of][]{DeLuca2021}, we derive the MRD of Pop~III mergers with IMBHs for $z<1$ as $\dot{n}_{\rm IMBH}(z<1)\sim 0.003-0.15\ \rm yr^{-1}\ Gpc^{-3}$, which is lower than the upper limits derived from O3 data under most conditions \citep[see their fig.~2]{Abbott2022}. 
For Pop~III BBH mergers with both BHs in the standard PISN mass gap $50-130\ \rm M_\odot$ like GW190521 \citep[detected at $z=0.82_{-0.34}^{+0.28}\lesssim 1$,][]{Abbott2020}, we predict the MRD to be $\dot{n}_{\rm gap}(z<1)\sim 0.0013-0.034\ \rm yr^{-1}\ Gpc^{-3}$ for $z<1$, also consistent with the rate $0.08_{-0.07}^{+0.19}\ \rm yr^{-1}\ Gpc^{-3}$ estimated for GW190521-like mergers from observations \citep{Abbott2022}. In fact, we find that GW190521 can be marginally explained by our Pop~III BBH mergers (in NSCs) assuming the log-flat or bottom-heavy IMF and the wide IBS model. In general, the results from all the 18 models are consistent with current observations of BBH mergers, but it will be challenging to derive strong constrains on Pop~III BBH mergers from current and upcoming observations by the LVK network that is only sensitive to low-$z$ events among which Pop~III mergers are expected to be sub-dominant (unless Pop~III star formation is much more efficient than that modelled by \textsc{a-sloth} in our case). 
This highlights the importance of the 3rd-generation GW detectors like ET and CEO in observing Pop~III BBH mergers at high $z$.

\section{Summary and Discussion}
\label{sec:dis}

We study the roles played by two evolution channels in producing binary black hole (BBH) mergers from Population~III (Pop~III) stars using the public semi-analytical model \textsc{a-sloth} \citep{Hartwig2022,Magg2022} applied to the halo merger trees produced by the cosmological simulation from \citet{Ishiyama2016}. The first channel considers isolated binary stellar evolution (IBSE) in which interactions of close ($\lesssim 10$~AU) binary stars shrink binary orbits to form tight ($\lesssim 0.1$~AU) BBHs that can merge within a Hubble time by gravitational wave (GW) emission in galaxy fields (i.e., effectively in isolation). 
In the second channel, Pop~III BBHs fall into nuclear star cluster (NSCs) by dynamical friction (DF) and are subsequently driven to merge via dynamical hardening (DH) by {binary-single} encounters, which also works for initially wide (up to $10^{5}$~AU) Pop~III BBHs given their massive nature. We combine the binary population synthesis (BPS) results of \textsc{sevn} \citep{Costa2023} with customized routines for the formation, dynamics, and internal orbital evolution of Pop~III BBHs in galaxy fields and NSCs (Fig.~\ref{bbh_process}), the formation and evolution of NSCs (Fig.~\ref{nsc_process}), as well as halo growth and galaxy evolution at lower redshifts ($z\lesssim 5$, Sec.~\ref{sec:extra}) in \textsc{a-sloth} based on the earlier work by \citet{Liu2021}. 
We explore 18 models (Table~\ref{runlist}) under different assumptions on Pop~III initial mass function (IMF), initial binary statistics (IBS, see Sec.~\ref{sec:sample}), and occupation fraction of high-$z$ NSCs (see Sec.~\ref{sec:nsc_form}). 
Our main findings are summarized as follows. 

\begin{enumerate}
    \item Although only a small fraction ($3-12\%$) of Pop~III BBHs fall into NSCs, a significant fraction ($\sim 45-64\%$) of the BBHs in NSCs merge at $z>0$. 
    The NSC-DH channel can be as efficient as the classical IBSE channel for producing Pop~III BBH mergers at $z\lesssim 8$, while the IBSE channel always dominates the merger rate density at $z\gtrsim 10$, producing a peak at $z\sim 15$ that closely follows the peak of Pop~III star formation. The NSC-DH channel contributes $f_{\rm NSC}\sim 8-95\%$ of Pop III BBH mergers across cosmic history, and is generally more important at lower redshifts, accounting for $\mathcal{F}_{\rm NSC}\sim 34-98$\% ($0.7-88$\%) of the all-sky merger rates of Pop~III BBH mergers within $z<1$ (20). Higher contributions from the NSC-DH channel are achieved by initially wider binary stars, more top-heavy IMFs, and higher occupation fractions of NSCs. 
    \item The most important factor that determines the relative importance of the two channels is IBS. When the IBS is dominated by close binaries, the NSC-DH channel produces $f_{\rm NSC}\sim 8-51$\% of all Pop~III mergers across cosmic history and explains $\mathcal{F}_{\rm NSC}\sim 0.7-22$\% of the all-sky merger rate within $z<20$. However, when the IBS is dominated by wide binaries, we have $f_{\rm NSC}\sim 65-95$\% and $\mathcal{F}_{\rm NSC}\sim 12-88$\% for $z<20$. 
    These outcomes are mainly driven by the strong dependence of the IBSE channel on IBS. The NSC-DH channel is relatively insensitive to the initial properties of Pop~III binary stars (i.e., IMF and IBS). 
    \item Pop~III BBH mergers in NSCs are more massive than those from IBSE. The latter mostly involve BHs below $\sim 50\ \rm M_\odot$ from the collapse of naked cores of stellar progenitors that have relatively low initial masses ($\lesssim 240\ \rm M_\odot$) and lose their hydrogen envelopes during binary interactions. {As pointed out in \citet{Mestichelli2024}, binaries of more massive stars that can potentially form more massive BHs either merge during unstable mass transfer when they are initially too close or do not produce tight enough BBHs that can merge within a Hubble time in isolation when they are initially too far apart.} 
    However, the NSC-DH channel can produce BBH mergers from wide ($\gtrsim 0.1$~AU) binaries of more massive BHs (in the mass ranges $50-86\ \rm M_\odot$ and $130-550\ \rm M_\odot$) whose stellar progenitors start on wide ($\gtrsim 5$~AU) orbits and experience less and even negligible mass loss during binary stellar evolution. For instance, a significant fraction ($\sim 4-84\%$) of Pop~III BBH mergers in NSCs involve at least one intermediate mass BH (IMBH) above $100\ \rm M_\odot$, while such mergers are extremely rare from IBSE. 
    \item We find that the Einstein Telescope \citep[ET,][]{Punturo2010,Maggiore2020} is able to detect most ($\gtrsim 90$\%) Pop~III BBH mergers in our models with a signal-to-noise ratio (SNR) above 8. The total detection rate is estimated as $\dot{\mathcal{N}}_{\rm ET,SNR>8}\sim 50-1370\ \rm yr^{-1}$, and we have $\dot{\mathcal{N}}_{\rm ET,SNR>8}\sim 30-1230\ \rm yr^{-1}$ and $\dot{\mathcal{N}}_{\rm ET,SNR>8}\sim 6-230\ \rm yr^{-1}$ for the IBSE and NSC-DH channels, respectively. Moreover, ET will detect $\sim 0.5-200$ Pop~III BBH mergers involving at least one IMBH ($m_{1}>100\ \rm M_\odot$) per year from the NSC-DH channel (mostly at $z\lesssim 10$). The detection rate of such mergers from IBSE is below $0.1\ \rm yr^{-1}$. 
    In our simulations, Pop~III stars only make up a tiny fraction ($\lesssim 10^{-4}$) of the total mass budget of stars ever formed in the universe. However, we estimate that Pop~III BBH mergers will make up a much larger fraction ($\sim 0.3-6.5$\%) of the BBH mergers that the 3rd-generation GW detectors will observe, given the conservative predictions on the merger rates of BBHs from metal-enriched Population I/II (Pop~I/II) stars and primordial BHs (PBHs) by \citet{Franciolini2022} as the reference. 
    \item Our results are consistent with current observations of BBH mergers by the LVK network, especially for the merger rates of IMBHs and BHs with masses in the standard PISN mass gap ($50-130\ \rm M_\odot$) like those in GW190521 \citep{Abbott2020,Abbott2022}. However, it will be challenging to identify and characterize Pop~III mergers with observations of the LVK network that are mostly sensitive to the low-$z$ regime ($z\lesssim 1$) where Pop~III mergers in our models are completely sub-dominant (accounting for $\lesssim 1\%$ of the total merger rate). Even for massive BBH mergers with IMBHs for which Pop~III stars are expected to make higher contributions, the estimated detection rate of such mergers from Pop~III remnants (mostly in NSCs) during the O4 run of the LVK network is very low ($\dot{\mathcal{N}}_{\rm LVK,IMBH,SNR>8}\sim 0.04-1.5\ \rm yr^{-1}$). This indicates that the 3rd-generation detectors like ET are required to efficiently constrain Pop~III star formation (and high-$z$ SCs) with GW observations \citep{Iwaya2023,Franciolini2024,Santoliquido2024}. 
    
    \item The stochastic GW background (SGWB) produced by Pop~III BBH mergers has a peak value of $\Omega_{\rm GW}\sim 10^{-11}-8\times 10^{-11}$ around observer-frame frequencies $\nu_{\rm peak}\sim 10-100$~Hz. Mergers in NSCs dominate the Pop~III SGWB at $\nu\lesssim 20$~Hz in most cases due to their massive nature\footnote{Interestingly, our predictions on the SGWB from Pop~III BBH mergers at $\nu\lesssim 20$~Hz (dominated by the NSC-DH channel in most cases) are similar to previous results \citep{Perigois2021,Martinovic2022} that only consider the IBSE channel and assume much higher Pop~III star formation rates. This indicates that significant degeneracy exists in the large parameter space of Pop~III BBH mergers to explain a single observable.}. 
    The SGWB from Pop~III BBH mergers is equal to $\sim2-32\%$ of the total SGWB of compact object mergers inferred from observations \citep{Abbott2023} at $\nu\lesssim 10\ \rm Hz$. Therefore, it is very difficult to observe such a sub-dominant signal from Pop~III BBHs in the total SGWB. However, the contribution of Pop~III mergers can be larger in the residual background observed by the 3rd-generation detectors like ET where individual detected sources can be properly subtracted \citep{Zhong2024}, under certain conditions for the other BBH merger populations \citep{Perigois2021,Martinovic2022,Kouvatsos2024}. 
    Moreover, if the cosmic star formation rate density of Pop~III stars is higher than that predicted by \textsc{a-sloth} in our case by a factor of a few, {Pop~III BBHs in NSCs can dominate the SGWB at $\nu\lesssim 10\ \rm Hz$. Considering the cumulative Pop~III stellar mass density $\rho_{\star,\rm III}\sim 5\times 10^4\ \rm M_\odot\ Mpc^{-3}$ found in our simulations, such enhancement of Pop~III star formation is still allowed by the constraints $\rho_{\star,\rm III}\lesssim 10^{5-6}\ \rm M_\odot\ Mpc^{-3}$ from reionization under conservative assumptions for the production/escape of ionizing photons from Pop~III stars \citep{Visbal2015,Inayoshi2016}.}
\end{enumerate}

In general, the IBSE and NSC-DH channels produce Pop~III BBH mergers of distinct properties and shape the Pop~III contributions to the GW signals from compact object mergers in different ways. In particular, an important feature of the NSC-DH channel is that it can produce mergers with IMBHs ($>100\ \rm M_\odot$) and BHs in the (standard) PISN mass gap ($50-130\ \rm M_\odot$) efficiently. {This is consistent with previous studies showing that such massive BBH mergers form efficiently via dynamical processes in SCs \citep[][]{Rodriguez2015,Rodriguez2019,DiCarlo2020,fragione2020repeated,Fragione2020gw,Fragione2022,ArcaSedda2020finger,Arca-Sedda2021,Khan2021,Gerosa2021,Kimball2021,Liu2021gw,Mapelli2021hierarchical,Mapelli2022,Anagnostou2022,Wang2022,Bruel2023,Chattopadhyay2023,Dall'Amico2023,Fragione2023,Liu2023sc,Mestichelli2024}}, which can also provides seeds for supermassive BHs \citep[e.g.,][]{Askar2022,Rose2022,Kritos2022,Kritos2023,Rantala2024}.

Our results show strong, complex dependence on the largely unknown properties of Pop~III binary stars and high-$z$ NSC clusters. 
In addition to the uncertain aspects explored in the current paper, there are other uncertainties/caveats in our modelling that may have significant effects on our results. Below we discuss the ones known to date in the hope that they provide directions for future work.


\begin{itemize}
    \item[(1)] The properties of Pop~III BBHs at birth are crucial inputs in our model, which not only depend on initial conditions but are also highly sensitive to the parameters/algorithms governing binary interactions and single stellar evolution \citep[see, e.g.,][]{Iorio2023}. Our conclusions are based on the BPS results from \citet{Costa2023} using \textsc{sevn} with a specific set of parameters, and thus can be altered if other BPS results are considered. For instance, the finding that the IBSE channel hardly produces mergers with BHs above $100\ \rm M_\odot$ results from stellar mergers caused by significant expansion of massive Pop~III stars during the MS phase in the stellar evolution tracks adopted by \textsc{sevn} (see Sec.~\ref{sec:fd_mass}). However, as shown in \citet[][]{Tanikawa2021mrd}, if massive Pop~III stars remain compact during MS evolution, massive BBH mergers (involving BHs above $130\ \rm M_\odot$) can still form via IBSE. 
    \item[(2)] Our modelling of the dynamics of Pop~III BBHs in host galaxies and during halo mergers under spherical symmetry is highly idealized. The adopted assumptions on the galaxy size and structure \citep{Arca-Sedda2014} are based on local observations and may not hold at high redshifts. Besides, we assume that Pop~III BBHs are always on their own, i.e., not embedded by larger structures (e.g., satellite galaxies, bound clusters of Pop~III stars and remnants) that can sink into galaxy centres more efficiently by dynamical friction, so the efficiency of the NSC-DH channel may be underestimated. Similarly, we assume that the smaller galaxies hosting NSCs during galaxy mergers are immediately destroyed, leaving behind naked NSCs. 
    \item[(3)] We use a phenomenological model of NSCs with adiabatic evolution based on empirical scaling relations from local observations \citep{Neumayer2020nuclear}. These relations may not be valid at high redshifts and cannot capture dynamical NSC assembly by mergers of globular clusters and young star clusters and in-situ star formation, which may affect the evolution of BBHs in NSCs. Moreover, we only consider dynamical hardening, softening/disruption and ejection by {binary-single} encounters to follow the internal orbital evolution of BBHs in NSCs, ignoring higher-order processes (e.g., relativistic phase space diffusion, tides-driven eccentricity excitation, and Kozai–Lidov mechanism) 
    involving tidal fields, general relativity effects, and interactions with other components (e.g., single/binary Pop~I/II BHs, central massive BHs and discs of active galactic nuclei) in NSCs (see Sec.~\ref{sec:bevo}). 
    \item[(4)] The merger trees adopted in our simulations only cover the high-$z$ regime ($z\gtrsim 4.5$) where the limited simulation volume is cosmologically representative. At lower redshifts, we use a simple galaxy evolution model to keep track of Pop~III BBHs, calibrated to reproduce the observed star formation rate density within a factor of 2. This model only considers smooth halo growth and star formation without taking into account halo/galaxy mergers which are important for the infall of BBHs into NSCs. How exactly this simplification affects our results is unclear and related to the assumptions on the dynamics of BBHs and NSCs during galaxy mergers.
    \item[(5)] We use a stochastic model for the formation of Pop~III BBHs that is not self-consistently connected to the stellar feedback routine, since we do not use the same population of binary and single stars to model stellar feedback, which is instead derived from a separate population of single stars. Besides, we use the fiducial star formation and stellar feedback parameters of \textsc{a-sloth} (calibrated against observations) in all our simulations. These parameters may need re-calibrations to be consistent with observations when we change the IMF and include binary stars. In general, the cosmic star formation rate density of Pop~III stars is poorly constrained without direct observations, and current theoretical predictions have a scatter of $\sim 1.5$~dex \citep[see, e.g.,][]{Liu2020did,Hartwig2022,Klessen2023}. Considering other Pop~III star formation histories will change the merger histories of Pop~III BBHs, especially for the IBSE model \citep{Santoliquido2023}.
    
\end{itemize}

{In this paper, we focus on the merger rate, SGWB, and mass distribution of Pop~III BBH mergers, whereas the spins of merging BHs (with respect to the orbital vector) can also provide clues for the formation/evolution channels of BBHs \citep[e.g.,][]{Biscoveanu2022,Callister2022,Mould2022,Adamcewicz2023,Pierra2024,Guo2024}. We plan to investigate the spin distributions of Pop~III BBH mergers from different channels in future work. Here we briefly comment on the physics involved. In the standard pathway of Pop~III star cluster formation (in minihaloes of $M_{\rm h}\sim 10^{5-6}\ \rm M_\odot$) via disc fragmentation \citep{Liu2021binary,Klessen2023}, the initial stellar spins and binary orbital vectors are expected to be aligned as they both inherit the angular momentum of the same star-forming disc. On the other hand, Pop~III binary stars with miss-aligned spins can form through non-standard pathways with turbulent fragmentation and mergers of sub-clusters in more massive haloes \citep[under the influence of peculiar environmental effects such as baryon-dark matter streaming motion,][]{Hirano2018sc,Hirano2023}. The stellar spins are further regulated by stellar winds, tidal effects, and mass transfer during (binary) stellar evolution leading to diverse outcomes \citep[e.g., alignment, tilt, and flipping,][]{Kinugawa2020,Tanikawa2021mrd,Stegmann2021}. Next, the angular momentum transport in stellar interiors (before and during the collapse) and SN natal kicks determine the BH spin magnitudes and orientations, which are still highly uncertain in current theoretical models \citep[e.g.,][]{Jimenez-Forteza2017,Qin2018,Qin2019,Fuller2019,Fuller2019apj,Bavera2020,Belczynski2020spin,Olejak2021spin,Ghodla2023}. 
Finally, if the BBHs reside in dense star clusters, the BH spins and orbit plane can also be altered by dynamical interactions, among which (fly-by) binary-single encounters 
typically cause small tilts ($\sim 15^{\circ}$) of the orbit plane while exchanges tend to result in isotropic orientations \citep[e.g.,][]{Bouffanais2019,Trani2021,Banerjee2023,Dall'Amico2024}.

In conclusion,} our results indicate that valuable information on the initial properties of Pop~III binary stars and the evolution channels/environments of BBHs is encoded in the GWs from Pop~III BBH mergers. 
The former can be fundamentally different from the observed properties of Pop~I/II binaries in the local Universe, and capture the elusive interplay between fragmentation, (proto)stellar feedback, competitive accretion, migration, gravitational scatters, and mergers of protostars during Pop~III star (cluster) formation, as well as {the dynamics of Pop~III star clusters \citep[][]{Sakurai2017,Hirano2018,Chon2019,Sugimura2020,Sugimura2023,Liu2021binary,Wang2022,Riaz2022,Park2022,Park2024,Franchini2023,Liu2023sc,Mestichelli2024}}. The latter are governed by key processes in galaxy evolution at Cosmic Dawn, such as star cluster formation and galactic dynamics which have been increasingly revealed by JWST observations \citep[e.g.,][]{Baggen2023,Ferreira2023,Kartaltepe2023,Langeroodi2023,Ito2023,Ormerod2024,Ji2024,Adamo2024}. Although distinguishing Pop~III BBHs from other populations of mergers on an individual basis 
is non-trivial \citep{Franciolini2022pbh,Costa2023}, identifying a subgroup of BBH mergers of Pop~III origins from a large enough sample of mergers is feasible and promising given the unique features of Pop~III BBHs and the non-proportionally high contributions to GW signals by Pop~III stars \citep{Tanikawa2022,Iwaya2023,Franciolini2024,Santoliquido2024}. In the next decades, the 3rd-generation GW detectors like ET and CEO \citep{Reitze2019,Evans2023} will observe thousands of events per year reaching $z\sim 30$. The ‘gravitational-wave archaeology’ in synergy with other observations of Cosmic Dawn will provide us new insights into the first stars, star clusters, and galaxies in the early Universe. 



\section*{Acknowledgements}
{The authors thank the referee, Tomoya Kinugawa\textsuperscript{\href{https://orcid.org/0000-0002-3033-4576}{\includegraphics[width=2.5mm]{orcid.png}}}, for insightful comments that helped improve our paper.} We thank Michela Mapelli\textsuperscript{\href{https://orcid.org/0000-0001-8799-2548}{\includegraphics[width=2.5mm]{orcid.png}}} for providing the input \textsc{sevn} catalogues underlying \citet{Costa2023} and valuable discussions. We thank Giuliano Iorio\textsuperscript{\href{https://orcid.org/0000-0003-0293-503X}{\includegraphics[width=2.5mm]{orcid.png}}} for developing and making \textsc{sevn} available. We thank Benedetta Mestichelli\textsuperscript{\href{https://orcid.org/0009-0002-1705-4729}{\includegraphics[width=2.5mm]{orcid.png}}} for useful discussions on Pop~III binary stellar evolution. We thank Mattis Magg\textsuperscript{\href{https://orcid.org/0000-0002-9022-5136}{\includegraphics[width=2.5mm]{orcid.png}}} and Li-Hsin Chen\textsuperscript{\href{https://orcid.org/0000-0003-0475-1947}{\includegraphics[width=2.5mm]{orcid.png}}} for their help in code development with \textsc{a-sloth}. We thank Simone S. Bavera\textsuperscript{\href{https://orcid.org/0000-0002-3439-0321}{\includegraphics[width=2.5mm]{orcid.png}}} for providing the data underlying \citet{Bavera2022}. We thank Sebastian Ljung for feedback on the BBH and NSC modules in \textsc{a-sloth}. The authors acknowledge the Texas Advanced Computing Center (TACC) for providing HPC resources under FRONTERA allocation AST22003. BL is supported by the Royal Society University Research Fellowship. NS gratefully acknowledges the support of the Research Foundation - Flanders (FWO Vlaanderen) grant 1290123N. GC acknowledges support from the Agence Nationale de la Recherche grant POPSYCLE number ANR-19-CE31-0022. FS acknowledges financial support from the AHEAD2020 project (grant agreement n. 871158). GC and FS acknowledge financial support from the European Research Council for the ERC Consolidator grant DEMOBLACK, under contract no. 770017. RSK acknowledges financial support from the European Research Council via the ERC Synergy Grant `ECOGAL' (project ID 855130),  from the German Excellence Strategy via the Heidelberg Cluster of Excellence (EXC 2181 - 390900948) `STRUCTURES', and from the German Ministry for Economic Affairs and Climate Action in project `MAINN' (funding ID 50OO2206). RSK thanks for computing resources provided by the Ministry of Science, Research and the Arts (MWK) of the State of Baden-W\"{u}rttemberg through bwHPC and the German Science Foundation (DFG) through grants INST 35/1134-1 FUGG and 35/1597-1 FUGG, and also for data storage at SDS@hd funded through grants INST 35/1314-1 FUGG and INST 35/1503-1 FUGG.
This work excessively used the public packages \texttt{numpy} \citep{vanderWalt2011}, \texttt{matplotlib} \citep{Hunter2007}, and \texttt{scipy} \citep{2020SciPy-NMeth}. The authors wish to express their gratitude to the developers of these packages and to those who maintain them.


\section*{Data availability}
\textsc{a-sloth} is publicly available at \url{https://gitlab.com/thartwig/asloth}. 
The version used in this paper (including the new modules for compact object mergers and NSCs) is public at \url{https://gitlab.com/Treibeis/a-sloth-cob}. 
\textsc{sevn} is publicly available at \url{https://gitlab.com/sevncodes/sevn}. 
The data underlying this paper will be shared on reasonable request to the corresponding author. 



\bibliographystyle{mnras}
\bibliography{ref} 


\appendix

\bsp	
\label{lastpage}
\end{document}